\documentclass[manuscript,screen]{acmart}
\AtBeginDocument{%
  }

\usepackage{enumitem}
\setlist[itemize]{
    leftmargin=*,
    nosep
}

\usepackage{booktabs}
\usepackage{multirow}
\usepackage{graphicx}
\usepackage{enumitem}
\usepackage{tabularx}
\usepackage{longtable}
\usepackage{pifont}
\usepackage{fontawesome}

\usepackage[most]{tcolorbox}

\usepackage{tikz}
\usetikzlibrary{calc}
\usepackage{forest}       
\useforestlibrary{edges}
\usepackage{xcolor}       
\usepackage{adjustbox} 

\definecolor{taxfill-L0}{RGB}{220, 230, 250}
\definecolor{taxdraw-L0}{RGB}{54, 110, 210}
\definecolor{taxfill-L1}{RGB}{255, 230, 230}
\definecolor{taxdraw-L1}{RGB}{200, 100, 100}
\definecolor{taxfill-L2}{RGB}{230, 240, 255}
\definecolor{taxdraw-L2}{RGB}{100, 140, 200}
\definecolor{taxfill-L3}{RGB}{245, 245, 245}
\definecolor{taxdraw-L3}{RGB}{160, 160, 160}
\definecolor{taxfill-R1}{RGB}{255, 230, 230}
\definecolor{taxdraw-R1}{RGB}{200, 100, 100}
\definecolor{taxfill-R2}{RGB}{255, 230, 230}
\definecolor{taxdraw-R2}{RGB}{200, 100, 100}
\definecolor{taxfill-R3}{RGB}{245, 245, 245}
\definecolor{taxdraw-R3}{RGB}{160, 160, 160}

\definecolor{color1}{RGB}{192, 0, 0}        
\definecolor{color2}{RGB}{0, 112, 192}      
\definecolor{color3}{RGB}{237, 125, 49}     
\definecolor{color4}{RGB}{0, 146, 70}       
\definecolor{color5}{RGB}{128, 0, 64}       
\definecolor{color6}{RGB}{109, 120, 40}     
\definecolor{color7}{RGB}{112, 48, 160}     
\definecolor{color8}{RGB}{0, 132, 132}      

\NewDocumentCommand{\heng}
{ mO{} }{\textcolor{red}{\textsuperscript{\textit{Heng}}\textsf{\textbf{\small[#1]}}}}

\newtcolorbox{analysisbox}{
  boxrule=0pt,
  toprule=0pt,
  bottomrule=0pt,
  leftrule=1pt,
  rightrule=1pt,
  arc=0pt,
  left=6pt,
  right=6pt,
  top=6pt,
  bottom=6pt
}

\newcommand{\ie}{\emph{i.e., }}
\newcommand{\eg}{\emph{e.g., }}

\newcommand{\cf}{\emph{cf. }}

\setcopyright{acmlicensed}
\copyrightyear{2018}
\acmYear{2018}
\acmDOI{XXXXXXX.XXXXXXX}
\acmConference[Conference acronym 'XX]{Make sure to enter the correct
  conference title from your rights confirmation email}{June 03--05,
  2018}{Woodstock, NY}
\acmISBN{978-1-4503-XXXX-X/2018/06}

\begin{document}

\title{Trustworthy Recommendation in the Era of Large Language Models: Opportunities and Challenges}


\author{Bohao Wang}
\orcid{0009-0006-8264-3182}
\email{bohao.wang@zju.edu.cn}
\author{Yu Cui}
\orcid{0009-0001-6203-3022}
\email{cuiyu23@zju.edu.cn}
\author{Zhenxiang Xu}
\orcid{0009-0004-3678-1141}
\email{zhenxiangxu@zju.edu.cn}
\affiliation{
    \institution{Zhejiang University}
    \city{Hangzhou}
    \country{China}
}

\author{Jujia Zhao}
\orcid{0009-0003-6951-7593}
\email{zhao.jujia.0913@gmail.com}
\affiliation{
    \institution{Leiden University}
    \city{Leiden}
    \country{The Netherlands}
}

\author{Chenxiao Fan}
\orcid{0009-0009-2509-7092}
\email{simonfan@mail.ustc.edu.cn}
\author{Jizhi Zhang}
\orcid{0000-0002-0251-465X}
\email{cdzhangjizhi@mail.ustc.edu.cn}
\affiliation{
    \institution{University of Science and Technology of China}
    \city{Hefei}
    \country{China}
}

\author{Weiqin Yang}
\orcid{0000-0002-5750-5515}
\email{tinysnow@zju.edu.cn}
\author{Shengjia Zhang}
\orcid{0009-0004-0209-2276}
\email{shengjia.zhang@zju.edu.cn}
\author{Sirui Chen}
\orcid{0009-0006-5652-7970}
\email{chenthree@zju.edu.cn}
\affiliation{
    \institution{Zhejiang University}
    \city{Hangzhou}
    \country{China}
}

\author{Yang Zhang}
\orcid{0000-0002-7863-5183}
\email{zyang1580@gmail.com}
\author{Xiaoyan Zhao}
\orcid{0000-0001-6001-1260}
\email{xzhao@se.cuhk.edu.hk}
\affiliation{
    \institution{National University of Singapore}
    \country{Singapore}
}

\author{Wenjie Wang}
\orcid{0000-0002-5199-1428}
\email{wenjiewang96@gmail.com}
\author{Chongming Gao}
\authornote{Corresponding author.}
\orcid{0000-0002-5187-9196}
\email{chongminggao@ustc.edu.cn}
\author{Fuli Feng}
\orcid{0000-0002-5828-9842}
\email{fulifeng93@gmail.com}
\author{Xiangnan He}
\orcid{0000-0001-8472-7992}
\email{xiangnanhe@gmail.com}
\affiliation{
    \institution{University of Science and Technology of China}
    \city{Hefei}
    \country{China}
}

\author{Jiawei Chen}
\authornotemark[1]
\orcid{0000-0002-4752-2629}
\email{sleepyhunt@zju.edu.cn}
\affiliation{
    \institution{Zhejiang University}
    \city{Hangzhou}
    \country{China}
}

\renewcommand{\shortauthors}{Bohao Wang et al.}

\begin{abstract}
    The field of recommender systems (RS) is currently undergoing two profound paradigm shifts. From the perspective of objectives, the goal has shifted beyond mere recommendation accuracy to comprehensive trustworthiness, encompassing multiple dimensions such as robustness, fairness, and privacy preservation.  From a technical perspective, Large Language Models (LLMs) have been extensively integrated into RS, reshaping the foundations of recommendation through richer semantic understanding, stronger intent reasoning, and more flexible user interactions. The convergence of these two shifts prompts a timely and pivotal question: how does the integration of LLMs reshape the landscape of trustworthy recommendation?

    In this work, we present a systematic review of trustworthy LLM-empowered recommendation. By comprehensively analyzing over 200 recent studies, we reveal that the introduction of LLMs acts as a double-edged sword. While their advanced mechanisms and user-friendly interfaces offer unprecedented opportunities to enhance trustworthiness, they simultaneously introduce new risks, such as novel forms of bias and hallucination-induced issues. To characterize this dual impact, we systematically identify 13 opportunities and 18 challenges across six fundamental dimensions of trustworthiness, and accordingly organize the existing literature into a novel taxonomy. We also provide a comprehensive review of commonly used datasets and evaluation metrics to facilitate empirical validation. Finally, we identify critical open challenges and outline future directions, hoping to inspire future research on this emerging topic.

\end{abstract}

\ccsdesc[500]{Information systems~Recommender systems}

\keywords{Recommendation System, Large Language Model, Trustworthiness}


\maketitle

\section{Introduction}
Recommender systems (RS) have become fundamental components of modern online platforms, helping users navigate vast information spaces and discover relevant content~\cite{resnick1997recommender}. Today, RS are widely deployed across e-commerce, social media, and lifestyle applications~\cite{raza2022news}, and have even permeated high-stakes domains such as healthcare~\cite{tran2021recommender} and education~\cite{urdaneta2021recommendation}. As key mediators of information exposure, RS guide users' access to digital resources, not only driving immediate choices but also subtly shaping human cognition through prolonged interactions~\cite{chaney2018algorithmic}.

\textbf{Importance of trustworthy RS.} Given this pivotal role and profound societal impact, the design objectives of RS have increasingly expanded beyond mere recommendation accuracy to encompass broader concerns of trustworthiness~\cite{chen2023bias,ge2024survey,wang2024trustworthy,fan2022comprehensive}. Formally, a \textit{trustworthy recommender system} is defined as a reliable system that fundamentally aligns with human values and is inherently deserving of user trust~\cite{ge2024survey}. Recent studies have explored multiple aspects of trustworthiness, such as robustness against perturbations~\cite{zhang2025robust}, fairness across diverse users and items~\cite{chen2023bias}, explainability in decision-making processes~\cite{zhang2020explainable}, and the preservation of user privacy~\cite{jeckmans2012privacy}. Failures in trustworthiness not only degrade the user experience and compromise commercial ecosystems~\cite{lam2004shilling}, but also trigger severe societal risks, such as the dissemination of misinformation~\cite{oro72186}, the formation of echo chambers~\cite{ge2020understanding}, and algorithmic discrimination~\cite{chen2023bias}. Consequently, building trustworthy RS has become an active research frontier in both academia and industry, and has drawn significant attention from governments and regulatory bodies~\cite{messmer2023auditing}.

\textbf{Emergence of LLM-empowered techniques.} From a technical perspective, RS are currently undergoing a profound paradigm shift.  Traditional RS are primarily founded on the principle of collaborative filtering, aiming to mine and model the correlations among users, items, and contexts from historical interactions~\cite{he2020lightgcn, ma2008sorec, kang2018self}. Recently, the advent of large language models (LLMs) has begun to fundamentally reshape the technical foundations of recommendation. Driven by their powerful language understanding, reasoning capabilities, and extensive world knowledge, LLMs have been extensively explored for integration into RS~\cite{zhu2025recommender}. Specifically, LLMs have been utilized as backbone models~\cite{wang2025msl}, comprehensive knowledge bases~\cite{ren2024enhancing}, semantic encoders~\cite{liu2025llmemb}, auxiliary reasoners~\cite{wang2024can} or conversational agents~\cite{gao2023chat}, enabling RS to better capture semantic user intent~\cite{bao2023tallrec}, integrate heterogeneous information~\cite{wang2025empowering}, and perform more flexible interaction modeling~\cite{chang2024conversational}. Beyond academic exploration, LLM-empowered RS have increasingly been deployed in industrial applications, such as ByteDance's HLLM~\cite{chen2024hllm}, KuaiShou's OneRec~\cite{deng2025onerec}, and Amazon's Rufus~\footnote{\url{https://www.aboutamazon.com/news/retail/amazon-rufus}}, yielding significant gains in practical recommendation effectiveness.

\textbf{Necessity of this survey.} With the rise of LLM-empowered recommendation techniques, ensuring trustworthiness in this new paradigm has attracted increasing attention. Recent years have witnessed a rapid surge of research efforts in this emerging topic. Despite this flourishing body of work, the existing literature remains highly fragmented, often focusing on a specific subproblem within a single dimension of trustworthiness rather than providing a unified and systematic overview. As a result, researchers and practitioners face difficulties in developing a structured understanding of this rapidly evolving field and in identifying how LLMs reshape the objectives, risks, and methodologies of trustworthy recommendation. Therefore, we believe it is both timely and imperative to present a comprehensive survey of LLM-empowered trustworthy recommendation.

\textbf{Contributions.} In this work, we provide a systematic review of this growing research area, focusing primarily on how the integration of LLMs changes the landscape of trustworthy recommendation. By comprehensively analyzing over 200 works, we reveal that the introduction of LLMs acts as a double-edged sword, presenting both unprecedented opportunities and emerging challenges. Specifically, we systematically summarize 13 opportunities and 18 challenges across six fundamental dimensions of trustworthiness (\cf Figure~\ref{fig:taxonomy}): robustness~\cite{zhang2025robust}, bias and fairness~\cite{chen2023bias}, explainability~\cite{zhang2020explainable}, factuality~\cite{huang2025survey}, controllability~\cite{ge2024survey}, and privacy~\cite{jeckmans2012privacy}. 
On the one hand, LLMs provide advanced technical mechanisms and user-friendly interactive interfaces that can enhance these dimensions of trustworthiness. On the other hand, they introduce new risks, including novel forms of bias~\cite{bito2025evaluating}, expanded attack surfaces~\cite{zhang2024stealthy}, hallucination-induced issues~\cite{bao2025bi}, and new pathways for privacy leakage~\cite{carlini2021extracting}. Building on this multi-dimensional perspective, we further organize the existing literature into a novel three-layer taxonomy (\cf Section~\ref{sec:opportunity&challenge}).

Furthermore, we observe that the evaluation protocols in this area remain highly fragmented and often lack alignment. Accordingly, we provide a comprehensive review of commonly used datasets and evaluation metrics, enabling readers to quickly grasp the experimental settings and facilitating future empirical validation (\cf Section~\ref{sec:evaluation}).

\begin{figure*}[t]
    \centering
    \includegraphics[width=0.95\linewidth]{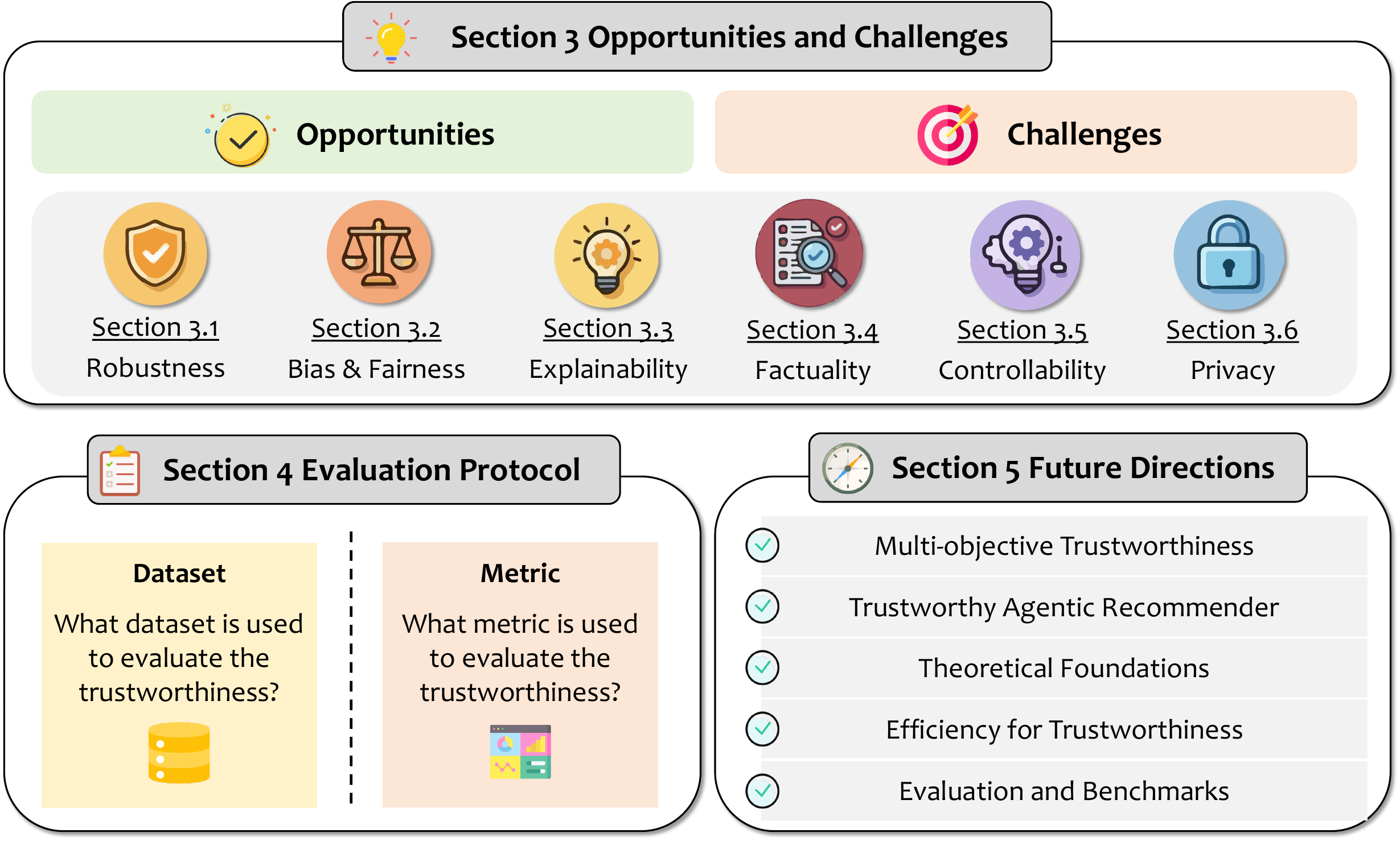} 
    \caption{Overview of this survey structure.}
    \label{fig:overview}
\end{figure*}

Finally, we identify five open problems in this area that deserve further exploration, including the trade-offs between multiple dimensions of trustworthiness and recommendation utility, trustworthiness in the novel agentic recommendation paradigm, the theoretical understanding of trustworthiness, the efficiency concerns in integrating LLMs, and the need for more comprehensive evaluation protocols. We hope this discussion will draw the community's attention to these critical issues and inspire future research on this topic (\cf Section~\ref{sec:future}).

\section{Preliminary}
\label{sec:preliminary}
In this section, we present the technical background of LLM-empowered recommender systems and introduce the concept of trustworthiness in recommendation.

\begin{figure*}[t]
    \centering
    \includegraphics[width=0.95\linewidth]{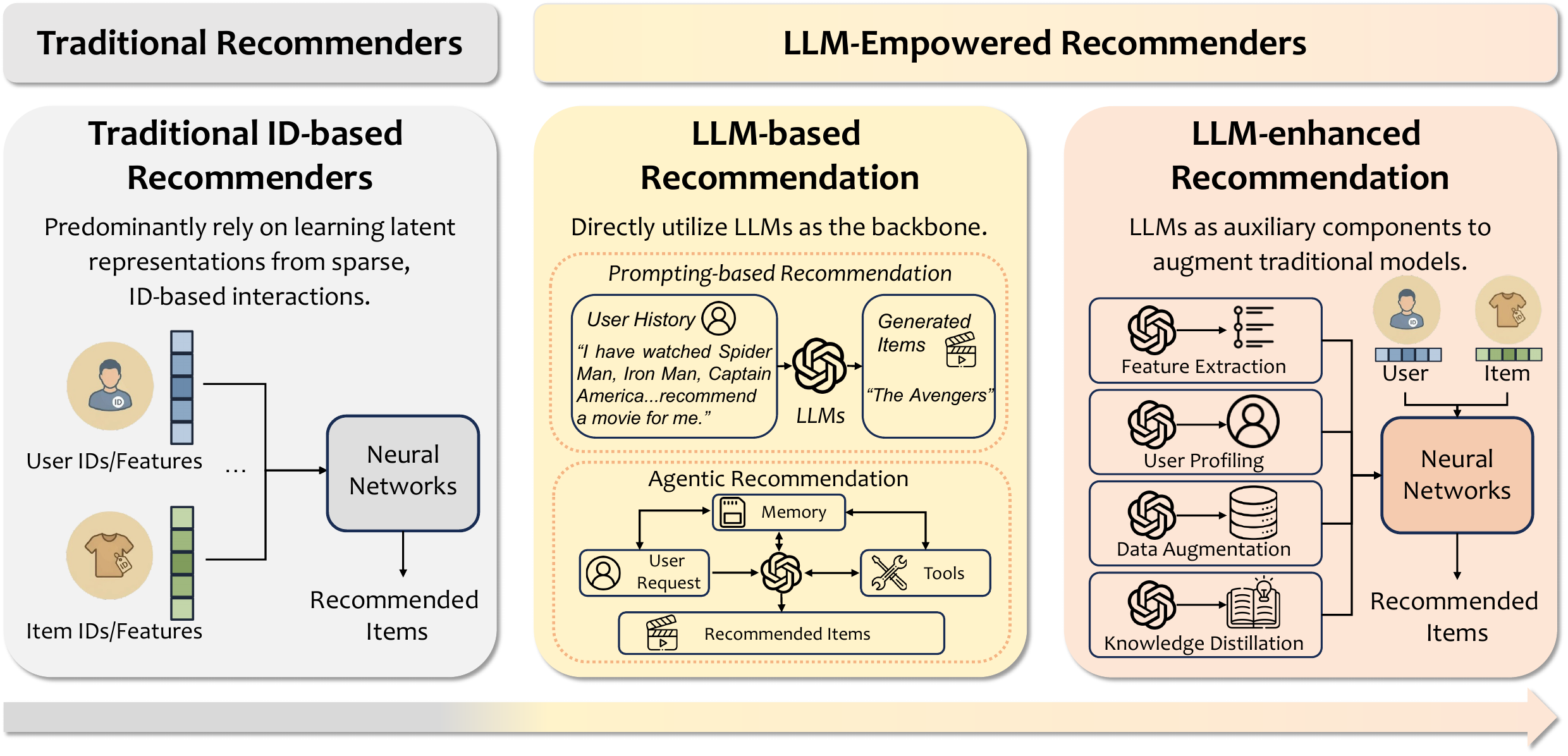} 
    \caption{Illustration of the paradigms of traditional recommender systems and LLM-empowered recommenders.}
    \label{fig:LLMRS}
\end{figure*}

\subsection{LLM-Empowered Recommendation}
Traditional ID-based recommender systems aim to predict, rank, or retrieve items of interest from a large-scale corpus. Most existing methods follow a model-based paradigm, representing users and items as unique IDs and learning their latent representations from collaborative signals to estimate user-item preferences~\cite{sun2019bert4rec,kang2018self,he2020lightgcn,hidasi2015session,tang2018personalized}.
While effective, these methods often struggle to capture rich semantic context, reason over complex user intents, and generalize to zero-shot scenarios. LLMs have opened new opportunities to mitigate these limitations~\cite{wang2024enhanced,sun2024large,wang2025msl,bao2023tallrec,liao2024llara,yu2025fashiondpo}, owing to their strong reasoning, semantic understanding, generation capabilities, and broad open-world knowledge~\cite{adcock2026llama,liu2024deepseek,ke2025survey,miao2026unidetect}. This has given rise to a new paradigm known as \emph{LLM-empowered recommendation}. According to the role of LLMs in the recommendation pipeline, existing studies can be broadly divided into two paradigms, as illustrated in Figure~\ref{fig:LLMRS}.

\textbf{LLM-based recommenders.} This paradigm directly utilizes LLMs as the backbone to generate recommendation results. In this paradigm, recommendation tasks are often reformulated as language modeling problems, where user preferences, historical behaviors, and candidate items are represented in textual form \cite{bao2023tallrec} or semantic IDs \cite{rajput2023recommender}. The LLM then generates recommendations through \textit{prompting}.
Moreover, LLMs can naturally support conversational recommendation scenarios, where users interact with the system through natural language \cite{chang2024conversational}.
More recently, an emerging line of research has extended LLM-based RS toward \emph{Agentic RS}~\cite{da2026agentic}. In these methods, the LLM is instantiated as an autonomous agent capable of perceiving user requests, maintaining user memory, invoking external tools or retrieval modules, and iteratively refining its decisions through reasoning and reflection~\cite{chen2026memrec,xia2026multi}. This agent-based paradigm enhances the flexibility and adaptability of LLM-based RS.

\textbf{LLM-enhanced recommenders.} 
In this paradigm, LLMs function as auxiliary modules that enhance traditional recommendation models rather than replacing them. They are commonly used to encode textual content such as item descriptions into rich semantic representations, which can be integrated with collaborative signals to improve feature quality and alleviate cold-start issues \cite{liu2024llm}. 
LLMs can also infer user preferences and intentions from historical interactions and textual feedback, enabling more expressive user profiling \cite{ren2024enhancing}. In addition, they support data augmentation by generating synthetic interactions, behavioral rationales, or enriched item attributes to mitigate data sparsity \cite{wei2024llmrec}. Their open-world knowledge can further be distilled into lightweight recommendation models, allowing this hybrid paradigm to combine the strengths of LLMs with the efficiency advantages of traditional RS \cite{cui2024distillation}.

\begin{figure*}[t]
    \centering
    \includegraphics[width=0.95\linewidth]{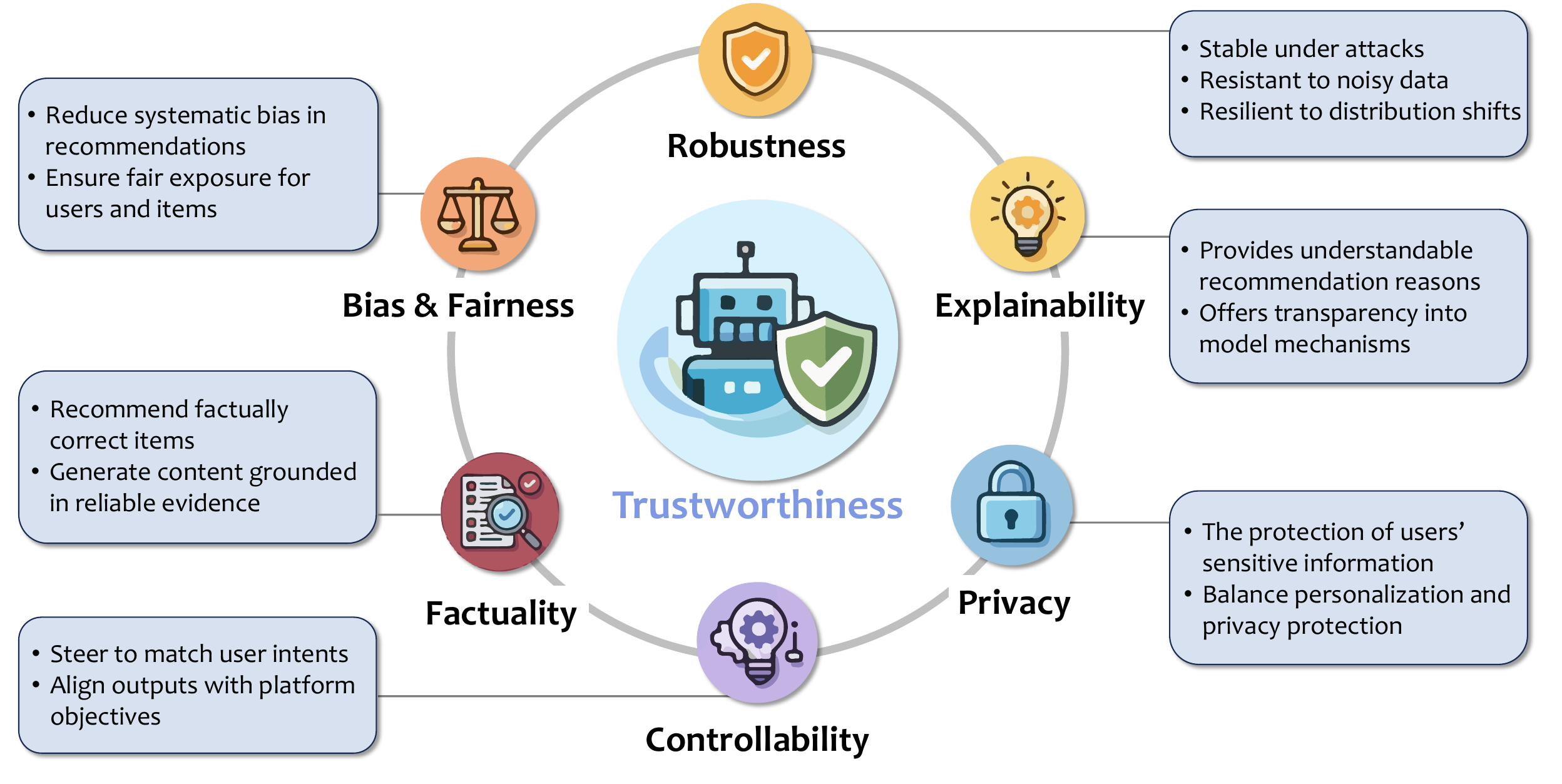} 
    \caption{Six key dimensions of trustworthy recommendation in the era of large language models.}
    \label{fig:trustworthy_dimenson}
\end{figure*}

\subsection{Trustworthy Recommendation}
Trustworthy AI has emerged as a foundational paradigm for the responsible development and deployment of AI systems~\cite{li2023trustworthy}.
According to the \textit{European Commission’s guidelines}\footnote{\url{https://digital-strategy.ec.europa.eu/en/library/ethics-guidelines-trustworthy-ai}}, trustworthy AI rests on three interrelated pillars: it must be lawful, ethical, and robust. Driven by the core ethical principles of human autonomy, harm prevention, fairness, and explicability, this paradigm demands rigorous operationalization throughout the AI lifecycle.
\input{taxonomy/taxonomy}

Trustworthy recommender systems (TRS) instantiate the above principles in personalized information access. 
A trustworthy RS can be defined as a reliable recommendation service that fundamentally aligns its objectives and behaviors with human values and is inherently worthy of user trust~\cite{ge2024survey}. 
This notion goes beyond conventional recommendation utility and further requires the system to mitigate adverse impacts caused by opaque decision-making, biased treatment or exposure, privacy leakage, adversarial manipulation, and insufficient user agency. 
In this survey, we operationalize trustworthy recommendation through six perspectives in the era of LLMs, as illustrated in Figure~\ref{fig:trustworthy_dimenson}: \emph{Robustness}, \emph{Bias and Fairness}, \emph{Explainability}, \emph{Factuality}, \emph{Controllability}, and \emph{Privacy}. 
In the following section, we will introduce each perspective in detail and further discuss the opportunities and challenges brought by LLMs for trustworthy recommendation.

\section{Opportunities and Challenges of LLM-Empowered Trustworthy Recommendation}
\label{sec:opportunity&challenge}

The integration of LLMs into RS signifies a profound paradigm shift, transitioning traditional ID-based pattern matching into knowledge-intensive, reasoning-driven, and highly interactive recommendation paradigms. However, while LLMs significantly augment the capability of RS, they fundamentally reshape the landscape of trustworthy recommendation. 
The introduction of LLMs acts as a double-edged sword: On one hand, empowered by their natural language interaction paradigms and formidable capabilities, \emph{LLMs unlock unprecedented opportunities for trustworthy recommendation}. For instance, they can significantly enhance explainability by generating human-readable natural language justifications~\cite{gao2023chat}, and improve controllability by allowing users to steer and refine recommendations through conversational interfaces~\cite{carroll2025ctrl}. 
On the other hand, \emph{LLMs introduce novel and critical challenges}, such as more complex forms of bias~\cite{bito2025evaluating}, broader attack surfaces~\cite{zhang2024stealthy}, hallucinated recommendations~\cite{bao2025bi}, and increased risks of privacy leakage~\cite{carlini2021extracting}.
To systematically navigate this complex landscape, this section comprehensively examines the dual impact of LLMs on trustworthy recommendation. We delineate the emerging opportunities and pressing challenges across six pivotal dimensions: Robustness (\cf Section~\ref{sec:robustness}), Bias and Fairness (\cf Section~\ref{sec:bias_and_fairness}), Explainability (\cf Section~\ref{sec:explainability}), Factuality (\cf Section~\ref{sec:factuality}), Controllability (\cf Section~\ref{sec:controllability}), and Privacy (\cf Section~\ref{sec:privacy}), as illustrated in Figure~\ref{fig:taxonomy}.

\subsection{Robustness}
\label{sec:robustness}
Robustness refers to the ability of a RS to provide consistent and reliable recommendations under various perturbations and uncertainties~\cite{zhang2025robust}. It mainly encompasses three dimensions: attack robustness~\cite{gunes2014shilling}, which evaluates its resistance to deliberate adversarial manipulations; noise robustness~\cite{wang2025llm4dsr}, which measures the system’s tolerance to noisy or imperfect interaction data; and out-of-distribution (OOD) robustness~\cite{wang2024distributionally}, which assesses the model’s capacity to generalize when the distribution of test data differs from that of the training data.
Robustness is crucial for RS because they operate in dynamic, real-world environments characterized by continuously changing user behaviors, data noise, and potential adversarial threats~\cite{koren2009collaborative}. 

\begin{figure*}[t]
    \centering
    \includegraphics[width=0.95\linewidth]{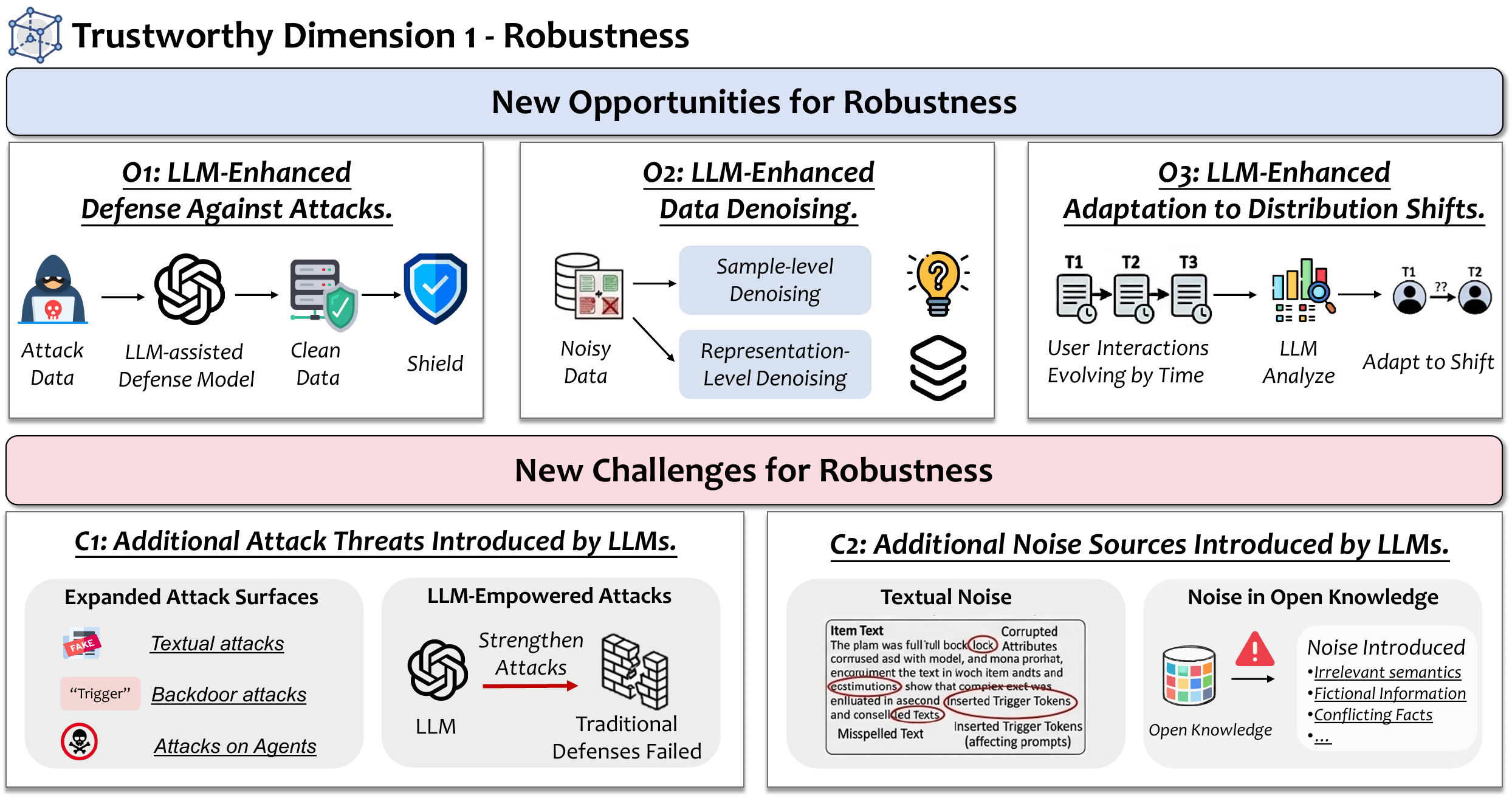} 
    \caption{An illustration of the opportunities and challenges introduced by LLMs for the robustness of recommender systems.}
    \label{fig:robustness}
\end{figure*}

With the emergence of LLMs, the robustness landscape of RS has undergone fundamental changes. On the one hand, LLMs create new opportunities for improving robustness, including enhancing the detection of malicious behaviors, enabling more effective denoising, and strengthening the system’s generalization under distribution shifts. On the other hand, the integration of LLMs also introduces new challenges, such as additional attack surfaces and new sources of noise.
In this section, we summarize both the opportunities and challenges brought by LLMs to the robustness of RS from three perspectives: attacks, noise, and OOD scenarios, as illustrated in Figure~\ref{fig:robustness}.

\subsubsection{New Opportunities for Robustness}
The opportunities that LLMs introduce to the robustness of RS manifest in the following three aspects.

\begin{itemize}
\item \textbf{LLM-enhanced defense against attacks.}
Recommender system attacks refer to deliberate manipulations intended to influence the system's output~\cite{nguyen2024manipulating}. The integration of LLMs into the recommendation pipeline has substantially strengthened the defensive capabilities of RS. 
Existing studies have mainly explored the use of LLMs to defend against two major categories of attacks.

\begin{itemize}
    \item \textit{1) Defending data poisoning.} In this type of attack, attackers inject carefully crafted fraudulent data to manipulate recommendation outcomes, for example by registering multiple fake user accounts and submitting fabricated ratings or reviews to poison the training data, aiming to achieve specific goals such as promoting or demoting target items \cite{huang2021data}.
    Traditional defense methods are often limited by their reliance on heuristic priors or supervised signals derived from labeled fraudulent data \cite{zhang2024lorec, zhang2020gcn, chung2013betap}.
    To overcome these limitations, recent studies have explored leveraging LLMs to assist in identifying fraudulent behaviors \cite{zhang2024lorec, ning2025retrieval} and detecting fake reviews generated by malicious users \cite{nawara2024shilling, li2026semanticshield}. For instance, LoRec~\cite{zhang2024lorec} prompts LLMs to assess users' fraudulence potential and downweights identified suspicious users during training to reduce the impact of poisoning attacks. The open knowledge of LLMs enables them to generalize more effectively across diverse attack patterns, thereby improving the detection of previously unseen fraudulent activities.
    \item \textit{2) Defending textual attacks.} Textual attacks manipulate recommendation outcomes by introducing subtle and malicious modifications to item textual descriptions, aiming to promote or demote target items~\cite{wang2025id}. RS that rely on textual information, particularly LLM-empowered RS, are especially vulnerable to such attacks (\cf Section \ref{sec:challenge_attack}). 
    To defend against these threats, many studies leverage the superior semantic understanding and contextual reasoning capabilities of LLMs~\cite{chen2025llm}. For instance, RewriteDetection \cite{wang2025id} segments item text into two parts and uses LLMs to discriminate whether the remaining content has been maliciously modified by LLMs. P-Scanner \cite{ning2025exploring} applies an LLM-based detection module, fine-tuned on a diverse set of poisoned item titles generated by an augmentation agent, to identify abnormal textual content.
\end{itemize}

\item \textbf{LLM-enhanced data denoising.}
Recommendation data often contain noise from factors such as clickbait \cite{wang2021clicks}, position bias \cite{chen2023bias}, or accidental interactions \cite{lin2023self}, which can impair model training and degrade performance. Traditional denoising approaches struggle due to the lack of accurate noise annotations, often relying on prior knowledge or assumptions that limit generalization \cite{sun2021does, zhang2022hierarchical, zhang2024ssdrec}. To overcome these limitations, numerous studies have explored leveraging the open knowledge or reasoning ability of LLMs to enhance denoising, which can be categorized into two main approaches.

\begin{itemize}
    \item \textit{1) Sample-level denoising.} These studies utilize LLMs to explicitly identify noisy samples within datasets and mitigate their impact during training by removing them or reducing their weights~\cite{song2024large, wang2025llm4dsr, wang2025unleashing, wang2025ruleagent, sun2025llm4rsr}. 
    For instance, LLM4DSR \cite{wang2025llm4dsr} employs a self-supervised denoising task to enable the model to learn noise patterns, and replaces the identified noisy interactions with appropriate alternatives suggested by the LLM. LLaRD \cite{wang2025unleashing} leverages the reasoning capabilities of LLMs, combined with collaborative signals, to generate knowledge that guides the denoising process. 
    However, directly employing LLMs as denoisers incurs substantial computational costs. To improve efficiency, some approaches utilize LLMs as enhancers to augment traditional denoising methods. For instance, LLMHD \cite{song2024large} utilizes LLMs to identify hard samples among potential noisy instances initially detected by traditional denoising methods, thereby reducing the volume of samples requiring LLM-based discrimination. RuleAgent \cite{wang2025ruleagent} circumvents the direct use of LLMs for noise detection by employing an agent-based framework to discover effective denoising rules, addressing the limitation that heuristic rules in traditional methods often lack generalizability across varied scenarios.
    \item \textit{2) Representation-level denoising.} Such methods enhance the model's robustness to noise by incorporating semantic representations generated by LLMs without explicitly identifying the noisy samples \cite{peng2025denoising, ren2024representation}. Specifically, these approaches maximize mutual information to align representations learned from collaborative filtering signals with semantic representations produced by LLMs, which encourages the model to focus on useful shared information during training, thereby implicitly suppressing signals that appear only in noisy interactions and lack semantic interpretability.
\end{itemize}

\item \textbf{LLM-enhanced adaptation to distribution shifts.}
In real-world recommendation scenarios, models are often deployed in dynamic environments where data distributions evolve over time due to changes in user preferences, item popularity, or external events. Such distribution shifts can lead to severe performance degradation when the recommendation model encounters OOD data. 
Several studies \cite{wang2023drdt, ebrat2024lusifer, nguyen2025enhancing} have leveraged LLMs to enhance the robustness of RS against distribution shift \cite{wang2024distributionally, yang2023generic}. 

One representative approach is DRDT \cite{wang2023drdt}, which directly employs LLMs as recommenders and utilizes their reasoning capabilities to achieve greater robustness to dynamic data. Specifically, DRDT instructs the LLM to iteratively generate multi-aspect analyses of high-level user preferences at each time step within the user interaction sequence through chain-of-thought (CoT) prompting \cite{wei2022chain}. By explicitly modeling the temporal dynamics of user preferences, DRDT improves the recommender’s resilience to OOD scenarios.
In contrast, some other methods follow an LLM-enhanced paradigm, leveraging LLMs’ understanding of user preference dynamics to perform data augmentation and thereby improve downstream recommendation models. For instance, Lusifer \cite{ebrat2024lusifer} constructs a simulation environment powered by LLM-based agents to highlight the evolution of user preferences over time. The simulated data generated is subsequently used to train downstream recommender models, enabling them to better adapt to the dynamic nature of user interests. 

\end{itemize}

\subsubsection{New Challenges for Robustness}
Although LLMs can enhance the robustness of RS, they also introduce new vulnerabilities. From the attack perspective, integrating LLMs expands the attack surface of recommendation pipelines and can also be exploited to strengthen existing attack methods. From the noise perspective, LLMs introduce additional sources of noise into the pipeline. Therefore, LLMs play a dual role in robustness: they can serve as powerful defenders, but also become new sources of fragility.
\begin{itemize}
\item \textbf{Additional attack threats introduced by LLMs.}
\label{sec:challenge_attack}
The challenges that LLMs pose to attack robustness are mainly reflected in the following two aspects.

\begin{itemize}
\item  \textit{1) Expanded attack surfaces of LLMs.} 
Compared with traditional RS, LLM-based RS involve a wider variety of input data modalities \cite{liao2024llara} and more complex model architectures and mechanisms \cite{peng2025survey}, thereby substantially expanding the potential attack surface of the recommendation pipeline. Some methods specifically exploit these characteristics to carry out more effective attacks, which can be broadly categorized into three main types according to their attack strategies.

(1) \textit{Textual attacks} \cite{zhang2024stealthy, nazary2025stealthy, wang2025id, filandrianos2025bias, ning2024cheatagent, morris2020textattack} are among the most prominent threats. Since LLM-based RS rely more heavily on textual item content \cite{bao2025bi, liao2024llara, wang2025msl}, attackers can manipulate item descriptions to promote or demote target items. For instance, \citet{filandrianos2025bias} insert pre-designed sentences containing cognitive biases (e.g., ``This is the most popular choice among customers!'') into item descriptions to increase/decrease item exposure. 
(2) \textit{Backdoor attacks} \cite{ning2025exploring, wu2023attacking} constitute another major threat. Due to the high vulnerability of LLMs to backdoor behaviors \cite{ning2025exploring, wu2023attacking}, attackers can inject triggers into textual item descriptions during training, causing the model to recommend trigger-containing target items regardless of user preferences. Such attacks are often highly stealthy and effective even with limited poisoning.
(3) \textit{Attacks on agentic RS} \cite{yang2025drunkagent} further expose vulnerabilities in agent-based recommendation. These systems often rely on memory mechanisms for personalized behavior modeling \cite{peng2025survey}, which can be exploited by adversaries. For example, DrunkAgent \cite{yang2025drunkagent} uses stealthy textual triggers to manipulate memory updates, causing a persistent bias toward target items.

Overall, the integration of LLMs significantly broadens the attack surface of RS by introducing richer input modalities, heavier dependence on textual semantics, and more complex model mechanisms, thereby enabling attacks that are more diverse, stealthy, and persistent.

\item \textit{2) LLM-empowered attack methods.}
Given their ability to generate highly deceptive and sophisticated fake content or interaction data, LLMs are increasingly being exploited by attackers as powerful tools to strengthen attacks against RS. Moreover, traditional defense methods, which primarily rely on predefined rules based on known attack patterns, often struggle to detect the malicious data produced by LLMs \cite{zhang2024lorec, zhang2020gcn, chung2013betap}. Existing studies mainly focus on enhancing three categories of attacks.

(1) One prominent line is \textit{textual attacks} \cite{chiang2023shilling, zhang2024stealthy, filandrianos2025bias, nazary2025stealthy, wang2025id, ning2024cheatagent}, where LLMs generate or rewrite reviews, titles, and descriptions to promote or demote target items. Their strong language generation ability enables more fluent, diverse, and stealthy attack content. Some methods further optimize prompts or mimic platform-specific styles to improve effectiveness \cite{zhang2024stealthy, nazary2025stealthy, filandrianos2025bias, wang2025id}, while CheatAgent \cite{ning2024cheatagent} uses an LLM-based agent to refine prompt injection strategies.
(2) Another direction is \textit{data poisoning attacks} \cite{chiang2023shilling, gu2025llm}. In this setting, LLMs are used to generate fake interactions \cite{gu2025llm} or fabricated reviews \cite{chiang2023shilling}. For instance, Agent4SR \cite{gu2025llm} introduces fake user agents that interact with items in a human-like manner and assign fraudulent ratings to manipulate target item exposure. Compared with conventional poisoning methods, such synthetic data is often more realistic and thus harder to detect.
(3) LLMs have also been applied to \textit{data-free Model Extraction Attacks (MEAs)} \cite{zhao2025llm4mea}. MEAs aim to replicate black-box models by querying them and training a surrogate model on collected input-output pairs, without access to the original training data~\cite{truong2021data}. Traditional methods often rely on randomly sampled synthetic user data, which poorly matches real-world distributions. To improve attack effectiveness, LLM4MEA \cite{zhao2025llm4mea} employs an LLM-driven agent to simulate user behavior and generate more realistic interaction sequences for model extraction.

\end{itemize}

In conclusion, LLMs exacerbate attack-related challenges mainly by expanding the attack surface and serving as more powerful attack instruments. Therefore, future work should urgently address these vulnerabilities by focusing on: 1) developing advanced LLM-assisted detection algorithms capable of identifying LLM-generated fake texts and poisoned interactions; 2) enhancing the architectural security of LLM-based RS, particularly concerning backdoor mitigation and the robust filtering of agent memories; and 3) establishing comprehensive benchmarks and dynamic adversarial defense frameworks that can continuously adapt to the rapidly evolving landscape of LLM-empowered attacks.

\item \textbf{Additional noise sources introduced by LLMs.}
In traditional RS, prior studies have primarily focused on removing noise from users' historical interactions, \ie interactions that do not faithfully reflect users' true preferences \cite{wang2021denoising, lin2023self, zhang2022hierarchical}. 
However, the integration of LLMs introduces additional sources of noise, including (1) textual noise and (2) noise stemming from the open-domain knowledge embedded in LLMs. These issues have received relatively limited attention in traditional RS.

\begin{itemize}
    \item \textit{1) Textual noise.} LLMs typically utilize user or item textual descriptions either to directly generate recommendation results \cite{bao2025bi} or to encode them as embeddings to support downstream recommendation models \cite{ren2024enhancing}. However, such textual data are often affected by missing attributes, corrupted content, or low-quality descriptions, thereby introducing noise at the text level \cite{ren2024representation}. Such noise can substantially degrade model performance given that LLMs rely heavily on the quality of textual input. Moreover, noise may also arise at the token level. Prior studies have shown that inserting specific tokens into prompts can significantly affect model behavior, even when these tokens do not alter the original semantics \cite{ning2025exploring}. 
    Therefore, denoising in LLM-empowered RS should account for noise at both the textual level and the finer-grained token level.
    \item \textit{2) Noise in open knowledge.} Many studies exploit the open-world knowledge of LLMs to enhance recommendation~\cite{peng2025survey}. 
    However, this process may introduce additional noise sources, such as factually incorrect generated content or irrelevant external semantics that are weakly aligned with users' actual interests~\cite{wang2025towards}. 
    To mitigate this implicit noise, S$^2$LENR~\cite{wang2025towards} employs an element-wise gate to decompose each augmented representation into effective and noisy components, and further optimizes them through contrastive learning so that useful preference signals are retained while LLM-introduced irrelevant semantics are isolated. 
    Nevertheless, future work should further strengthen the modeling and mitigation of noise introduced by the open-domain knowledge inherent in LLMs themselves.
\end{itemize}

\end{itemize}


\subsection{Bias and Fairness}
\label{sec:bias_and_fairness}
Bias refers to systematic deviations in recommendation outcomes that favor certain items and users over others in a way that does not reflect their true relevance or intrinsic value~\cite{chen2023bias}. 
For example, popularity bias~\cite{zhang2021causal} refers to the tendency for popular items to be recommended, regardless of the actual preferences of individual users.
These biases can raise fairness concerns, thereby distorting user experience, reducing content diversity, and perpetuating social stereotypes.  Moreover, such biases may be further amplified through feedback loops, resulting in increasingly severe consequences~\cite{chen2023bias}.
As a result, identifying and mitigating bias and unfairness have become crucial research directions for building trustworthy RS.

\begin{figure*}[t]
    \centering
    \includegraphics[width=0.95\linewidth]{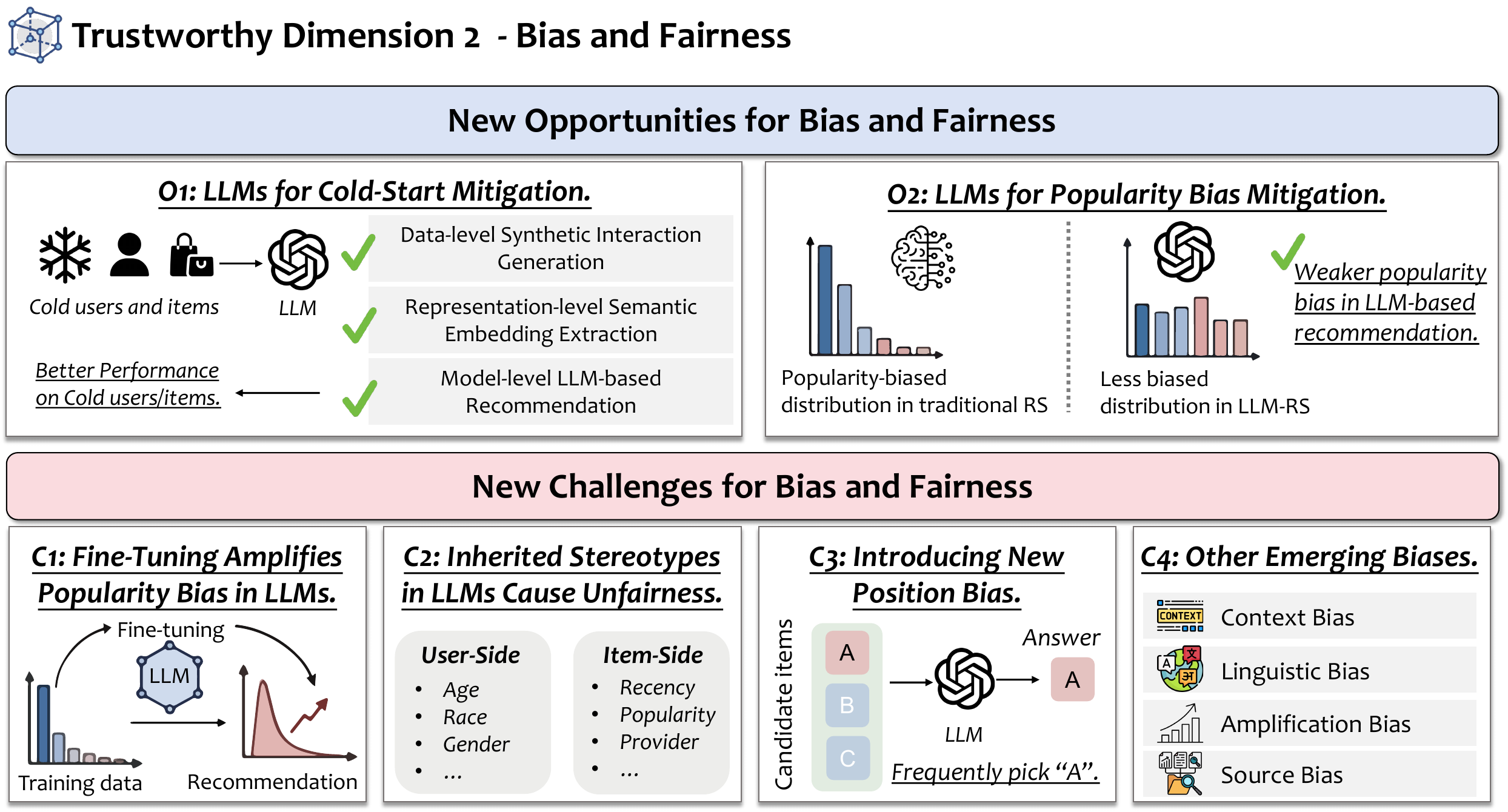} 
    \caption{An illustration of the opportunities and challenges introduced by LLMs for the bias and fairness of recommender systems.}
    \label{fig:bias_and_fairness}
\end{figure*}

The introduction of LLMs into recommendation brings both new opportunities and challenges. 
On one hand, LLMs offer new opportunities to mitigate some biases commonly observed in traditional RS. For example, they can help alleviate the cold start problem, and off the shelf LLMs often exhibit weaker popularity bias than traditional RS, which can help reduce bias in existing recommendation pipelines.
On the other hand, LLMs may amplify existing biases, such as popularity bias and unfairness arising from inherited stereotypes, while also introducing entirely new forms of bias.
This section discusses how LLMs influence bias and fairness in recommender systems, as illustrated in Figure~\ref{fig:bias_and_fairness}.

\subsubsection{New Opportunities for Bias and Fairness}
\label{sec:oppotunity_bias}
LLMs introduce opportunities in bias and fairness primarily in two aspects: enhancing model performance in cold-start scenarios and mitigating popularity bias.
\begin{itemize}
\item \textbf{LLMs for cold-start mitigation.}
LLMs can alleviate the cold-start problem in recommendation systems from three perspectives: data, representation, and model.

\begin{itemize}
    \item \textit{1) Data level mitigation.}
    Some research attempts to leverage LLMs' generative capabilities to create interaction data for users or items with limited historical interactions~\cite{wang2024large, huang2025large, liu2025filterllm, kusano2024data, rungtranont2024using}. The synthetic data can then be used to train traditional recommender systems, thereby improving their ability to handle cold start scenarios.
    As a pioneering work, \citet{wang2024large} directly leverage the powerful generative capabilities of LLMs to synthesize cold-start data for specific cold-start items based on natural language. LLM-search~\cite{rungtranont2024using} enables LLMs to analyze user interaction history, infer the next item a user might like, and then use a search engine (e.g., BM25) to find similar cold-start items for data synthesis. ColdLLM~\cite{huang2025large} efficiently generates high-quality cold-start synthetic data by utilizing LLMs for filtering and training during the synthesis process. In addition to generating individual interactions, \citet{kusano2024data} predicts the attributes of items a user might like based on their history. FilterLLM~\cite{liu2025filterllm} efficiently constructs samples by generating interaction probabilities for multiple potentially liked users for a cold-start item at once, using a Text-to-Distribution approach.
    \item \textit{2) Representation level mitigation.}
    Leveraging the powerful representation extraction capabilities of LLMs, these models can extract semantic representations (e.g., embeddings) from text descriptions to improve cold-start recommendation performance~\cite{yuan2023go, zhang2025llminit}. For example, \citet{gong2023unified} utilize LLMs to extract query and item embeddings in combined search and recommendation scenarios to enhance cold-start recommendation performance. \citet{li2023integrating} use text embeddings to initialize the embeddings of traditional recommendation models, then further enhance cold-start recommendation performance with meta-learning.
    \item \textit{3) Model level mitigation.}
    Directly utilizing LLMs for recommendations can mitigate the cold-start recommendation problem. Early work found that using In-Context Learning (ICL) for LLM-based recommendations achieved good cold-start performance~\cite{sanner2023large,kaur2026efficient}. Subsequent efforts have tried incorporating auxiliary information or modules into LLMs to further enhance cold-start capabilities. \citet{che2024new} further strengthen cold-start ability by training the sample selection process shown to the LLM in ICL via Reinforcement Learning (RL). LLMTreeRec assists LLMs with cold-start recommendations by building an item tree using item categories~\cite{zhang2024llmtreerec}. \citet{yang2025cold} and \citet{sakurai2025llm} leverage knowledge graphs as auxiliary information sources, performing RAG (Retrieval-Augmented Generation) or reasoning on them to enhance cold-start recommendation capabilities. KALM4Rec focuses on scenarios where users provide keywords, using these keywords for RAG to perform cold-start recommendations~\cite{sakurai2025llm}.
\end{itemize}


\item \textbf{LLMs for popularity bias mitigation.}
Although LLM recommendations are inevitably subject to popularity bias~\cite{hou2024large,zhang2024agentcf}, the popularity bias of LLM recommendations is weaker compared to traditional recommendations~\cite{lichtenberg2024large}. Furthermore, this bias can be further mitigated by adjusting the template (e.g., adding "Try to recommend movies that are less popular") or LLM generation hyperparameters (e.g., using a larger temperature) to encourage the LLM to recommend less popular items.
Therefore, some works have explored using LLMs to assist traditional recommendation models in mitigating popularity bias.
Among these, some works have explored using LLMs' generative capabilities for data augmentation to alleviate popularity bias. For instance, \citet{wang2023improving} used LLMs to synthesize conversational recommendation data, and HNLMRec~\cite{zhao2025can} synthesized hard negative samples. Both found that training with this synthesized data could alleviate popularity bias in previous recommendation models. Additionally, LLM-ESR~\cite{liu2024llm} explored introducing semantic information via LLMs to enhance traditional recommendation models, which also served to mitigate popularity bias.

\end{itemize}

\subsubsection{New Challenges for Bias and Fairness}
LLMs can both amplify existing biases in traditional recommenders, such as increased popularity bias caused by fine-tuning LLMs and potential fairness issues, and introduce new types of bias, including position bias and other emerging biases.

\begin{itemize}
    
\item \textbf{Fine-tuning amplifies popularity bias in LLMs.}
Several studies have shown that fine-tuning LLMs on recommendation datasets can exacerbate popularity bias \cite{gao2025process, lu2025dual, jiang2024item, gao2025sprec}. Furthermore, when LLMs are employed as data generators to improve downstream models, the models trained on such biased datasets inherit and potentially intensify the bias present in the original LLMs.
Existing methods predominantly address this issue by adjusting sample weights to reduce bias. For example, IFairLRS \cite{jiang2024item} reweights training samples based on the bias observed between the distribution of target items and that of historical interactions. D2LR \cite{lu2025dual} applies token-wise inverse propensity scoring (IPS) \cite{schnabel2016recommendations} to encourage LLMs to assign greater importance to less popular tokens. Additionally, some approaches integrate the Direct Preference Optimization (DPO) framework and increase the sampling probability of popular items during negative sampling to mitigate bias \cite{liao2024rosepo, gao2025sprec}.

\item \textbf{Inherited stereotypes in LLMs cause unfairness.}

Since LLMs are pre-trained on massive corpora that inevitably encode real-world social biases, they risk reproducing these ingrained stereotypes when deployed in recommendation tasks. This propagation of bias ultimately leads to the unjust treatment of either users or items. Generally, such LLM-induced unfairness manifests across two primary dimensions:

\begin{itemize}
    \item \textit{1) User-side unfairness} \cite{zhang2023chatgpt, deldjoo2024normative, hua2023up5, xu2023study, deldjoo2025cfairllm, sakib2024challenging, tommasel2024fairness, sah2025faireval, hu2025fairwork, das2024unveiling} refers to disparities in how recommenders treat users with different demographic attributes. LLMs can exhibit both explicit and implicit biases when processing demographic cues such as gender, age, ethnicity, or religion \cite{sah2025faireval, zhang2023chatgpt, das2024unveiling, hu2025fairwork, sakib2024challenging}.
    For instance, when a user’s gender information is provided, LLMs may tend to recommend romance movies to female users while suggesting action or science fiction films to male users, resulting in unequal recommendation experiences.
    Furthermore, LLMs can also manifest implicit biases even when sensitive attributes are not explicitly available. As demonstrated by Hua et al. \cite{hua2023up5} and Xu et al. \cite{xu2023study}, LLMs may infer demographic characteristics—such as gender from names or behavioral histories—and inadvertently introduce unfairness into the recommendation.
    Besides, subsequent research has sought to improve fairness evaluation frameworks for LLM-based recommenders, for example by incorporating diverse user attributes \cite{sah2025faireval} and their intersections \cite{deldjoo2025cfairllm}, and by refining metrics to better distinguish genuine personalization from unfair bias \cite{deldjoo2025cfairllm, deldjoo2024normative}.  
    \item \textit{2) Item-side unfairness} \cite{li2023preliminary, deldjoo2024understanding, jiang2024item, zhang2025bifair,deldjoo2025toward} refers to the tendency of LLMs to exhibit bias toward items belonging to specific categories (e.g., genre, creator identity). 
    For off-the-shelf LLMs, Li et al. \cite{li2023preliminary} found that, in news recommendation scenarios, LLMs exhibit a pronounced bias toward popular providers. Deldjoo et al. \cite{deldjoo2024understanding} showed that LLMs display significant favoritism toward newer items, which may result in older items receiving insufficient exposure.
    For fine-tuned LLMs, IFairLRS \cite{jiang2024item} demonstrates that LLMs tend to recommend items belonging to genres that are overrepresented in the training data. 
\end{itemize}

Existing efforts to mitigate these issues can be broadly categorized into these approaches: (1) Re-weighting \cite{zhang2025bifair, jiang2024item}: Adjusting sample weights during training to ensure balanced learning across different user and item groups; (2) Prompt engineering \cite{deldjoo2025cfairllm, li2023preliminary, li2024your, das2024unveiling}: Incorporating fairness requirements into prompts or designing prompts based on fairness principles, such as instructing the system to "act as a fair recommender"; (3) Post-processing \cite{liu2025fairness}: Reorganizing generated recommendation lists from a fairness perspective. Liu et al. \cite{liu2025fairness} leverage LLMs as fairness identifiers and utilize the inherent fairness awareness of LLMs to construct fairer recommendations; and (4) RAG \cite{das2024unveiling}: Das et al. \cite{das2024unveiling} mitigates bias by retrieving contextually relevant but neutral items from a curated external knowledge base and injecting them into the prompt, grounding the model’s generation on balanced factual content rather than on its own biased parameters.

\item \textbf{Introducing new position bias.}
In traditional RS, position bias refers to the phenomenon where items appearing at the top of a recommendation list are more likely to be interacted by users \cite{chen2023bias}, which can lead to feedback data being skewed toward top-ranked items.
In LLM-based RS, position bias takes on a new definition: the recommendations generated by LLMs are influenced by the order of candidate items \cite{ma2023large, hou2024large, bito2025evaluating, jiang2025beyond, xu2025tapping, dai2024bias}. Jiang et al. \cite{jiang2025beyond} demonstrated that LLMs tend to favor items positioned at the beginning of the candidate list. In contrast, in traditional RS, the order of candidate items typically does not affect the recommendation results.

Several methods have been proposed to mitigate position bias, which can be categorized based on the stage of intervention:
(1) Debias during inference~\cite{hou2024large, ma2023large, luo2025recranker}. These methods primarily involve post-processing the recommendation results generated by the LLM to alleviate the impact of position bias. For instance, STELLA \cite{ma2023large} employs a Bayesian framework to calibrate the model’s predictions. Hou et al. \cite{hou2024large} prompt the model to generate multiple rankings by randomly shuffling the candidate items each time, subsequently aggregating the results to produce the final ranking.
(2) Debias during training~\cite{luo2025recranker, zhang2024agentcf}. These approaches introduce position-independent prior knowledge during training to achieve debiasing. For example, RecRanker \cite{luo2025recranker} constructs random shufflings of candidate items for each sample to preserve those responses from the LLM that demonstrate consistency regardless of item position.

\item \textbf{Other emerging biases.}
LLMs can introduce additional biases in RS. 
Wang et al. \cite{wang2026does} identified the \textit{Context Bias}, whereby LLMs excessively rely on auxiliary tokens (\eg task descriptions and prefix-generated tokens) during the generation, resulting in a tendency to favor items highly correlated with these tokens and raising unfairness concerns. 
Zhang et al. \cite{zhang2021language} identified the \textit{Linguistic Bias}, noting that LLMs prioritize producing fluent and grammatically correct text (language modeling) over collaborative filtering (recommending based on user preferences). 
Bao et al. \cite{bao2024decoding} identified the \textit{Amplification Bias}, induced by length normalization, which makes items with more high-probability tokens more likely to be recommended.
Zhou et al. \cite{zhou2025exploring} observed the \textit{Source Bias} that LLM-based RS tend to preferentially recommend Artificial Intelligence Generated Content (AIGC), and this bias is further amplified through the feedback loop. 
Understanding and mitigating these emerging biases is therefore crucial for building fair and reliable LLM-empowered RS.
 
\end{itemize}


\begin{figure*}[t]
    \centering
    \includegraphics[width=0.95\linewidth]{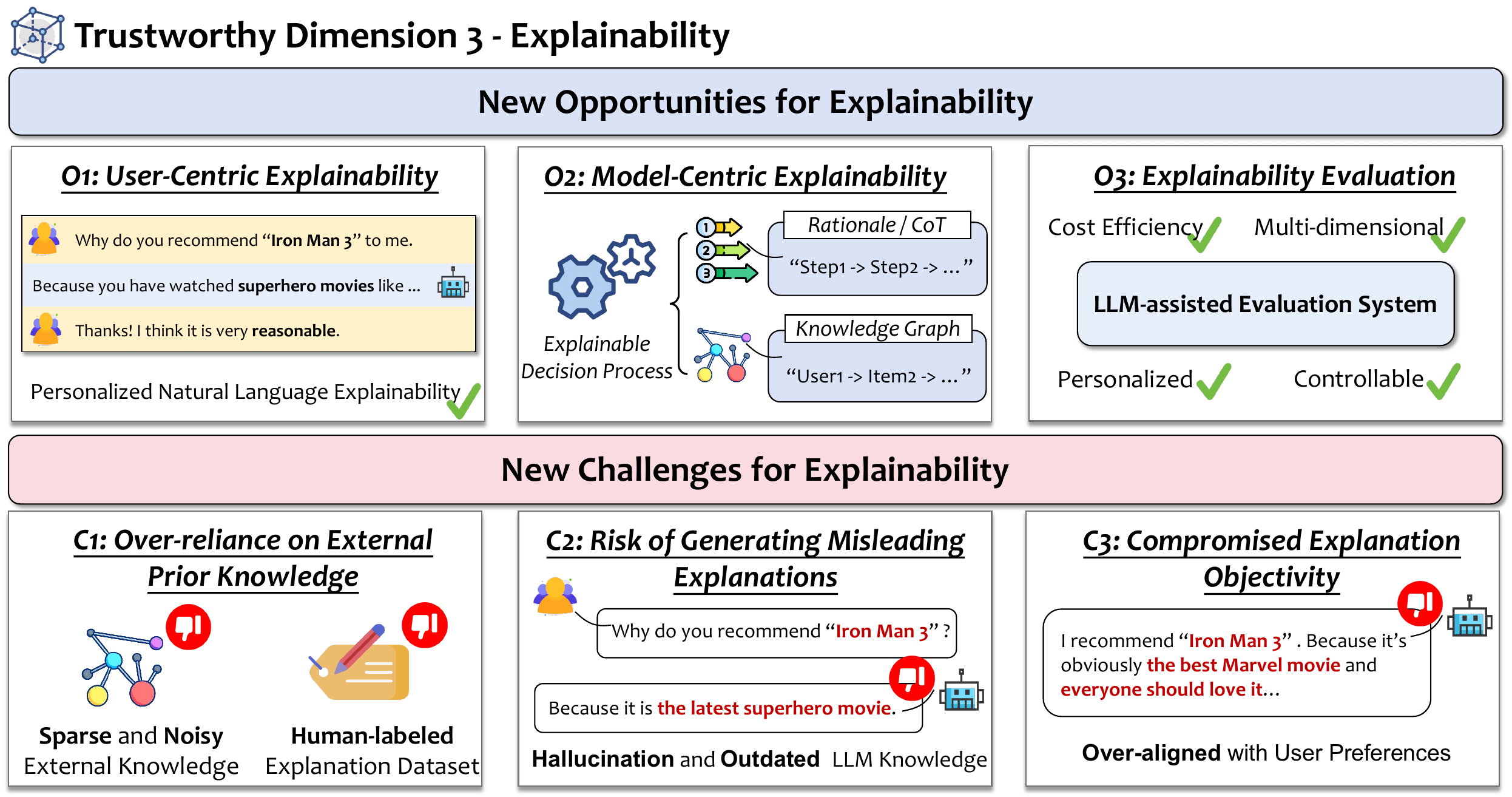
    } 
    \caption{An illustration of the opportunities and challenges introduced by LLMs for the explainability of recommender systems.}
    \label{fig:explainability}
\end{figure*}

\subsection{Explainability}
\label{sec:explainability}

Explainability refers to the ability of the RS to provide understandable insights into its recommendation decisions and underlying mechanisms~\cite{zhang2020explainable}.  
Traditional explainable recommendation methods typically generate explanations based on predefined templates or extract logical rules from recommendation models. However, these approaches often produce low-quality and weakly personalized explanations, and they face substantial challenges in data-sparse and cold-start scenarios~\cite{li2021caesar, chen2021neural, shi2020neural, zhu2021faithfully}.

LLMs have created new opportunities for explainable recommendation~\cite{zhang2024navigating, gao2024dre, feng2024move}. Specifically, they can support: more expressive and personalized user-centric explanations through natural language generation; enhanced model-centric explainability by revealing intermediate reasoning processes or incorporating structured knowledge, such as knowledge graphs; and scalable and flexible evaluation by serving as automatic assessors of explanation quality.  
Despite these advantages, applying LLMs to explainable recommendation also raises new challenges, including over-reliance on external prior knowledge and costly human supervision, the risk of persuasive but misleading explanations caused by hallucination or bias, and the potential loss of objectivity due to excessive personalization. 
In the following, we discuss the opportunities and challenges introduced by LLMs for explainability in trustworthy recommendation, as illustrated in Figure~\ref{fig:explainability}.

\subsubsection{New Opportunities for Explainability}

LLMs introduce opportunities in explainable recommendation primarily in three aspects: more personalized user-centric explanations, more expressive intermediate reasoning processes for model-centric explainability and more flexible explainability evaluation strategies.

\begin{itemize}
\item \textbf{User-centric explainability.}
User-centric explainability concerns explaining to users why a model recommends specific items. LLMs can directly generate recommendation explanations in natural language, thereby enhancing the explainability of the recommendations \cite{zhang2023user,yu2025explainable,yang2024fine,hada2021rexplug}. Many studies have explored using prompt engineering to directly guide pre-trained LLMs (e.g., ChatGPT) to generate explanatory text for recommendation results \cite{silva2024leveraging,ghosh2023jobrecogpt,wang2023llm4vis}; however, these prompt-based methods often produce explanations of poor quality and lack personalization. To enable LLMs to generate natural language explanations that better align with user preferences \cite{luo2024unlocking,peng2024uncertainty,liu2023llmrec}, some studies have explored methods such as reinforcement learning or instruction fine-tuning, using prior reward functions or human-annotated recommendation explanation datasets to train LLMs to generate high-quality personalized recommendation explanations \cite{ petruzzelli2024instructing,lei2024recexplainer}.

\item \textbf{Model-centric explainability.}
Model-Centric explainability focuses on improving the interpretability of the recommendation model itself, including its internal reasoning logic, knowledge utilization, and decision process. Unlike user-centric explanations, which are intended to communicate recommendations to end users, model-centric explainability aims to make the model's underlying mechanisms more transparent and analyzable, thereby supporting model diagnosis, debugging, and trustworthiness assessment.  
Existing LLM-based studies on model-centric explainability in RS can be broadly categorized into two categories:  

\begin{itemize}
    \item \textit{1) Language-based interpretability.}
    These approaches improve model-centric explainability by exposing the intermediate reasoning process of LLM-based recommenders in natural language. Instead of treating explanations solely as post hoc justifications for users, they use language as an interpretable representation of the model’s internal decision-making process. 
    Specifically, LLMs can be employed to infer user preferences from reviews or to identify salient item attributes from textual descriptions, producing intermediate reasoning signals such as human-readable rationales \cite{lyu2023llm,jia2025improving,park2023user} or Chain-of-Thought (CoT) trajectories \cite{yu2025thinkrec,zhang2025reinforced,zhao2025reason}. These textual representations make it easier to inspect how the model connects user-provided information to preference judgments, thereby improving the transparency and analyzability of the recommendation process.
    They can further be encoded into embeddings via text encoders and integrated into downstream recommender models, improving both recommendation performance and the interpretability of the model’s decision process \cite{wang2025blessing, rahdari2024logic}.  
    
    \item \textit{2) KG-based interpretability.} Knowledge Graphs (KGs) capture graph-structured relationships between user and item entities, enriching LLMs with domain-specific recommendation knowledge. Some studies convert KG-structured information into natural language descriptions or soft prompts, then feed them into prompts or fine-tune LLMs with instructions, enabling LLMs to generate higher-quality user preferences and item insights \cite{li2024learning, qiu2024unveiling, li2025g, abu2024knowledge, ma2024xrec}. Additionally, reasoning paths from users to recommended items in KGs can serve as explanations, simulating user decision-making processes. Some studies sample such reasoning paths from KGs to train LLMs with the capability of generating reasoning paths \cite{geng2022path, balloccu2023faithful}. Other studies employ LLMs to extract user interests/intents to augment KGs \cite{wang2024llmrg, zheng2025explain, shi2024llm, wang2024enabling}, and then use graph neural networks to integrate these insights for explainable recommendation.
\end{itemize}

\item \textbf{Explainability evaluation.}
Evaluating the explainability of recommendation methods remains a significant challenge. Traditional evaluation approaches for recommendation explainability can be divided into two main types: human evaluation \cite{hernandez2020effects, musto2019justifying} and automated evaluation \cite{li2017neural, geng2022recommendation}. However, human evaluation is often time-consuming, expensive, and difficult to scale, while automated evaluation methods are typically limited to specific rules (e.g., comparing generated explanations with real user reviews) and fail to effectively assess the personalization, diversity, and other aspects of the generated explanations. The emergence of LLMs has opened up a promising new path for evaluating explainable recommendation methods. LLMasEvaluator \cite{zhang2024large} compared LLM-based evaluations with collected real user evaluations to investigate whether LLMs can serve as evaluators of recommendation explanations, demonstrating that LLMs can provide an accurate, reproducible, and low-cost solution for assessing recommendation explanations. ALERT \cite{li2025alert} designed two LLM-driven evaluation frameworks for recommendation explanations: one leveraging generative LLMs to assess the quality of different explanations in natural language, and another employing discriminative LLMs to score explanations. Shimizu et al. \cite{shimizu2025disentangling} used LLMs to extract positive and negative opinions from users' post-purchase reviews to construct a dataset, evaluating the quality of recommendation explanation models based on whether the generated explanations align well with user sentiment and accurately identify their positive and negative opinions about the target items.
\end{itemize}

\subsubsection{New Challenges for Explainability}
Although leveraging LLMs to improve the explainability of recommendation models has demonstrated significant potential, current approaches still face several challenges, including over-reliance on external knowledge, the risk of generating misleading or unfaithful explanations, and the trade-off between personalization and explanation objectivity.

\begin{itemize}
\item \textbf{Over-reliance on external prior knowledge.}
Most methods rely heavily on external prior knowledge like knowledge graphs, human-designed prompts or annotated explanation datasets \cite{friedman2023leveraging,gao2023chat,zhou2023gpt}. However, knowledge graph-based approaches are often susceptible to data sparsity in recommendation systems and highly sensitive to data noise, and human involvement in prompt design and dataset annotation is costly and difficult to scale. 
Human-designed prompts require substantial expert effort and are difficult to generalize across domains. 
Similarly, constructing annotated explanation datasets is costly, time-consuming, and hard to scale.
These limitations may lead to degraded personalization and diversity in LLM-generated explanations, particularly in scenarios with high data privacy requirements or cold-start situations. 
To mitigate the impact of data sparsity and noise in knowledge graphs, data augmentation methods can be employed to enhance graph-structured data \cite{wei2024llmrec}. To reduce manual costs and resource consumption, some studies have adopted prompt learning to learn soft prompts, thereby reducing reliance on manual prompt design \cite{wang2024rdrec,li2023prompt}; other studies have utilized LLMs to automatically annotate explanation datasets or evaluate the quality of model-generated explanations \cite{li2025alert,shimizu2025disentangling}.

\item \textbf{Risk of generating misleading explanations.}  
LLMs may generate inaccurate yet persuasive recommendation explanations, potentially misleading users \cite{maes2025mitigating}. These misleading behaviors stem from inherent limitations of LLMs, such as generating hallucinated explanations that contradict facts, biases toward long-tail items, and catastrophic forgetting of user contextual knowledge. A feasible approach is to employ RAG (Retrieval-Augmented Generation) \cite{el2024optimizing,hou2024enhancing} to enhance LLMs' knowledge of specific recommendation scenarios (e.g., up-to-date user/item knowledge, collaborative filtering knowledge), thereby alleviating misleadingness in LLM-generated explanations.

\item \textbf{Compromised explanation objectivity.}  
To generate personalized recommendation explanations that better align with user preferences, LLMs are often fine-tuned on user-specific data. However, excessive personalization may compromise the objectivity of the generated explanations \cite{lazovich2023filter}.
For example, the model may emphasize only the positive aspects of an item, since such information is generally more appealing to users, while downplaying or omitting negative aspects. Although these explanations may remain factually correct and non-hallucinatory, they can still present a biased view of the target item, thereby affecting users' ability to form an objective understanding.
To mitigate this issue, additional reward functions or regularization can be designed during LLM training to enforce explanation objectivity, achieving a trade-off between personalization and objectivity.

\end{itemize}


\begin{figure*}[t]
    \centering
    \includegraphics[width=0.95\linewidth]{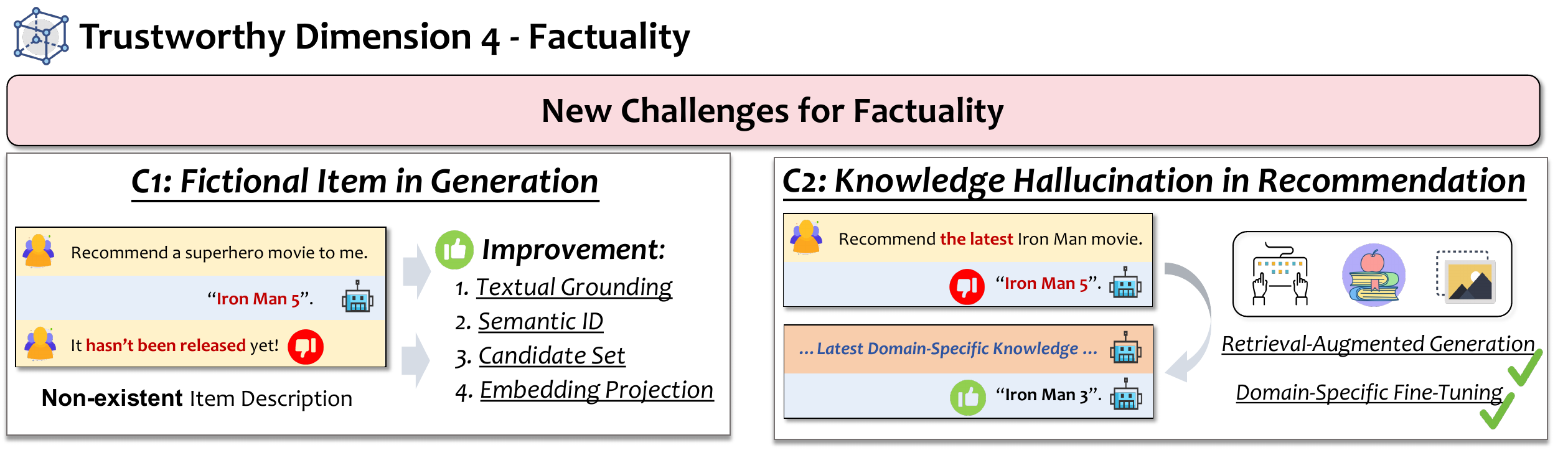} 
    \caption{An illustration of the challenges introduced by LLMs for the factuality of recommender systems.}
    \label{fig:factuality}
\end{figure*}

\subsection{Factuality}
\label{sec:factuality}
Factuality refers to the extent to which the utilized or generated information is factually correct and grounded in reliable evidence. Maintaining strict factuality is fundamental to RS, as it directly ensures user trust, platform credibility, and the overall effectiveness of the system. In the context of LLM-empowered RS, factuality becomes a particularly critical concern. This is primarily because the inherent hallucination issue of LLMs~\cite{huang2025survey}, combined with their lack of domain-specific recommendation knowledge~\cite{bao2023tallrec}, frequently hinders their ability to provide factually accurate recommendations.
Next, we will introduce the two main factuality challenges --- fictional item in generation and knowledge hallucination in recommendation, as illustrated in Figure~\ref{fig:factuality}.

\subsubsection{New Challenges for Factuality}
The factuality challenges introduced by LLMs mainly manifest in two aspects.

\begin{itemize}
\item \textbf{Fictional item in generation.} 
Unlike traditional recommendation models, which rely on a discriminative approach to rank items from a predefined pool thereby inherently ensuring perfect item grounding, LLMs operate on an unconstrained generative paradigm. Consequently, they often fabricate item titles or descriptions that do not map to any actual items, severely compromising recommendation accuracy and system credibility. To mitigate the fictional item challenge, current studies propose solutions across four main paradigms:

\begin{itemize}
    \item \textit{1) Textual based methods.} 
    These methods align the open-ended generation with the real item space at the text level. For instance, BIGRec \cite{bao2025bi} acts as a post-hoc filter, mapping hallucinated predictions to the most semantically similar real items. Other studies intervene during the inference stage via constrained beam search \cite{yang2026bear, bao2024decoding}, which restricts the generation path strictly to a valid token trie tree constructed based on the authentic item set. 
    Additionally, some works focus on the impact of non-existent items during the model training phase. For example, MSL \cite{wang2025msl} masks invalid tokens that do not belong to real items when calculating the token-level loss, preventing the LLM from incorrectly treating non-existent items as negative samples.
    \item \textit{2) Semantic ID based methods.}
    Rather than representing items as language tokens, these approaches leverage structured semantic identifiers to mitigate the generation of non-existent items. Semantic IDs (SIDs) provide a unique mapping from each item to a discrete, fixed-length token sequence \cite{rajput2023recommender, hong2025eager, hua2023index}. Specifically, this paradigm typically begins by encoding item metadata into continuous dense embeddings. These embeddings are then discretized into structured SIDs using quantization techniques (e.g., RQ-VAE). By operating over SID representations, LLMs are constrained to generate recommendations within a controlled and semantically grounded identifier space, thereby substantially reducing factual hallucinations compared to purely text-based generation.
    \item \textit{3) Candidate set based methods.} 
    These methods proactively restrict the LLM's choices by providing a predefined pool of valid items within the prompt \cite{liao2024llara,kim2024large,yue2023llamarec,jiang2025beyond}. While this strategy can partially alleviate hallucinations, it cannot completely eliminate the risk of out-of-set hallucinations. Furthermore, due to the limited context window of LLMs, this approach struggles to scale to real-world recommendation scenarios where the candidate item set is overwhelmingly large.
    \item \textit{4) Embedding based methods.} 
    To bypass the risky auto-regressive decoding process entirely, a representative approach is to append a learnable prediction head to the LLM's output layer \cite{zhu2024collaborative,li2023e4srec,xu2025slmrec,wang2024rethinking,qu2024tokenrec}. The LLM acts as an encoder to output user or context embeddings, which are then directly matched with real item embeddings. This enables efficient and hallucination-free predictions within the exact item space. However, effectively bridging the semantic gap between the LLM's pretrained generation space and the collaborative recommendation space remains an open research question.
\end{itemize}

\item \textbf{Knowledge hallucination in recommendation.}
Another critical factuality challenge arises from LLMs' insufficient or outdated knowledge regarding specific recommendation domains. Pretrained on general and static corpora, LLMs often struggle to fully adapt to specific recommendation tasks and lack accurate information about newly released items or evolving user preferences, leading to hallucinated explanations and unrealistic reasoning.

To bridge this knowledge gap, researchers primarily adopt two strategies. The first is Retrieval-Augmented Generation (RAG), which injects high-quality external knowledge (e.g., recent item metadata, collaborative filtering signals, or Knowledge Graphs) into the LLM context \cite{jiao2025retrieval}. For example, Di et al. \cite{di2023retrieval} extract entity types and reasoning paths from Knowledge Graphs (KGs) to guide LLMs, while K-RagRec \cite{wang2025knowledge} constructs personalized knowledge subgraphs from user historical interactions and integrates them via GNNs, thereby grounding the LLM's reasoning process in factual evidence and generating highly accurate, domain-aware recommendations.
The second strategy is Domain-Specific Fine-Tuning, which adapts LLMs to recommendation tasks by training them on domain-relevant data, such as user-item interactions, item descriptions, and task-specific instruction datasets \cite{liao2024rosepo}. For example, TALLRec \cite{bao2023tallrec} utilizes recommendation domain-specific instruction tuning on LLMs to enrich them with the recommendation dataset knowledge, while LLMSeR \cite{sun2025llmser} constructs high-quality recommendation instruction datasets to improve the faithfulness and reasoning consistency of LLMs, thereby enhancing their understanding of user behavior patterns and mitigating factuality challenges.

\end{itemize}



\begin{figure*}[t]
    \centering
    \includegraphics[width=0.95\linewidth]{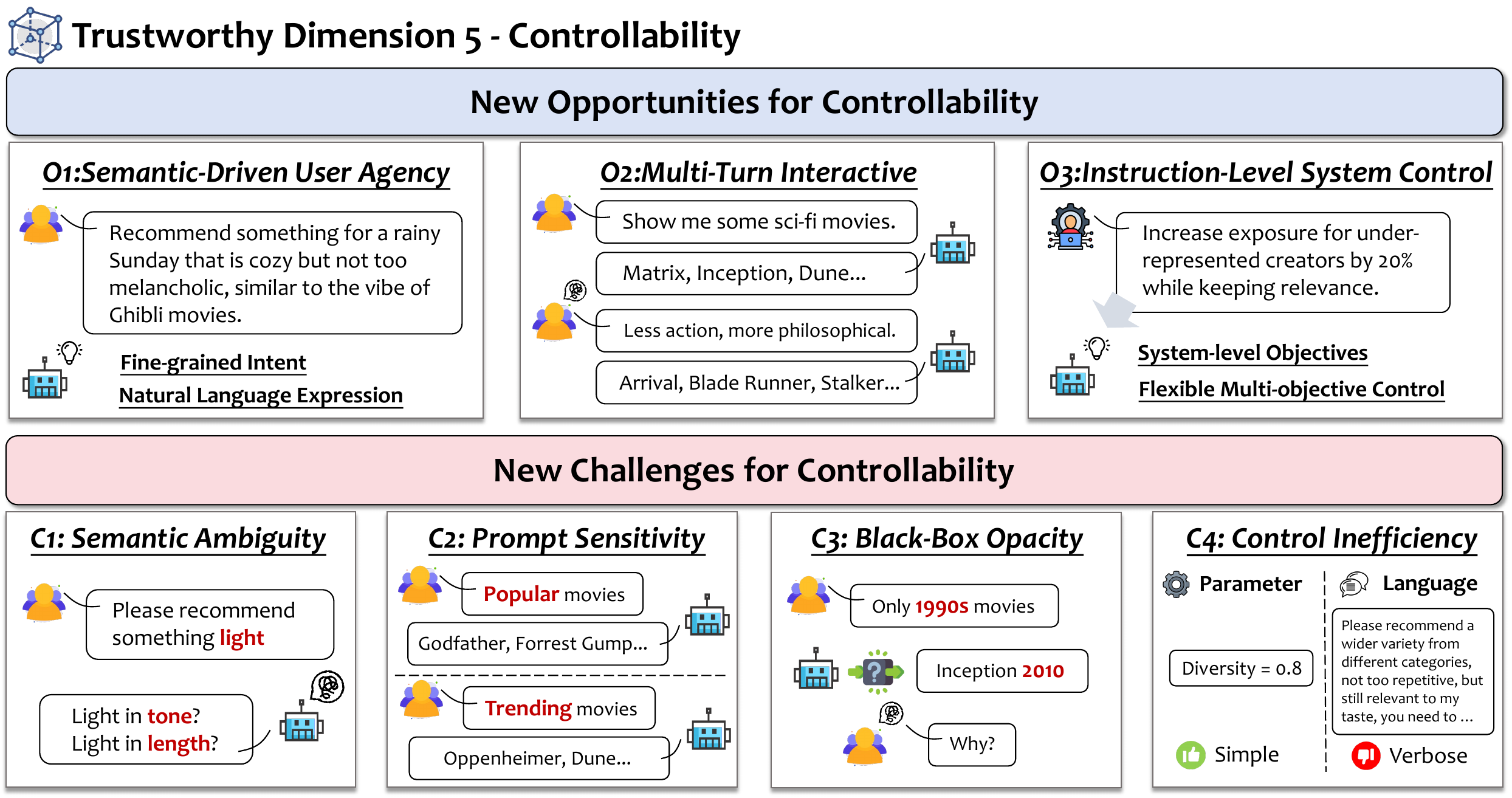} 
    \caption{An illustration of the opportunities and challenges introduced by LLMs for the controllability of recommender systems.}
    \label{fig:controllability}
\end{figure*}

\subsection{Controllability}
\label{sec:controllability}

Controllability in RS is the ability to steer model outputs to satisfy user intents or platform-level objectives~\cite{parra2015user}. It has two complementary forms: user-initiated controllability, where users explicitly specify or revise preferences and constraints to guide recommendations; and system-initiated controllability, where the platform regulates recommendation behavior to meet multi-objective requirements (e.g., accuracy, diversity, fairness, and safety)~\cite{abdollahpouri2020multistakeholder}. As a core ingredient of trustworthy recommendation, controllability supports user agency and transparency while enabling accountable governance of system behavior under changing contexts~\cite{milano2020recommender, wang2025beyond, wang2026two}.

LLMs shift controllability from rigid parameter/loss/constraint-based mechanisms to semantic, instruction-level control~\cite{wei2021finetuned,bao2023tallrec,dai2025token}. Users can express fine-grained intents in natural language, and platforms can modulate objectives via prompting, instruction tuning, or reward shaping—often with minimal or no retraining~\cite{liu2023pre}. This shift broadens the design space for controllable recommendation, but also exposes new failure modes, including ambiguous or unreliable control signals, prompt sensitivity and misuse, and instability under iterative user–system adaptation. This section discusses how LLMs influence controllability in recommender systems, as illustrated in Figure~\ref{fig:controllability}.

\subsubsection{New Opportunities for Controllability}

The incorporation of LLMs into RS has fundamentally transformed the landscape of controllability, presenting new opportunities for both user- and system-initiated control over recommendation outputs.

\begin{itemize}
    \item \textbf{User-initiated control.} 
    Traditionally, user control in recommendation systems has relied on the adjustment of structural parameters, such as diversity factors or weight coefficients~\cite{knijnenburg2012explaining}. These mechanisms are often unintuitive, limited in expressiveness, and fail to capture the nuanced or evolving preferences of users. The interaction is typically static and lacks real-time adaptability, making it difficult for users to actively steer recommendation outcomes according to their high-level intents.

    With the advent of LLMs, controllability is shifting from parameter-level manipulation to semantic-driven interaction. Users can now articulate their preferences, constraints, or objectives through natural language prompts, moving beyond the restrictions of fixed feature spaces~\cite{gao2021advances}. This paradigm enables the expression of complex and personalized requirements—such as desired pace, style, or contextual relevance—directly via dialogue with the system. Furthermore, LLM-based agent architectures facilitate multi-turn interactions, allowing the system to iteratively refine its understanding of user intent and progressively converge towards the user’s goals through semantic negotiation~\cite{lei2020estimation,yao2022react}.

    Recent works exemplify these advancements. For instance, \cite{lu2024aligning} explores controllability at both the item-wise and list-wise level by enabling users to specify category proportions directly in prompts, with reinforcement learning (RL) rewards designed to align outputs with user intentions. Similarly, \cite{wozniak2025improving} empowers users to edit their user profiles in free-form text, enabling explicit, interpretable, and fine-grained control over recommendations, evaluated through offline controllability metrics. Other approaches, such as CTRL-Rec~\cite{carroll2025ctrl}, incorporate user-provided control embeddings—derived from natural language instructions—into the scoring of candidate items, seamlessly integrating explicit user guidance into the ranking process. Interactive agent-based frameworks, like RAH~\cite{shu2024rah} and RecAI~\cite{lian2024recai}, further enhance user controllability by enabling iterative feedback and constraint specification, supporting real-time preference adjustment and collaborative filtering through multi-agent cooperation or prompt-based condition articulation~\cite{shu2024rah}.

    These advances collectively demonstrate an evolution from static, parameter-based configuration to dynamic, semantically rich, and user-driven interaction, substantially enhancing both the expressiveness and effectiveness of user control in recommendation systems.

    \item \textbf{System-initiated control.}
    On the system side, balancing multiple objectives—such as accuracy, diversity, and fairness—has long been a central challenge in RS~\cite{zheng2021disentangling,jannach2023survey}. Traditionally, such trade-offs are embedded within model architectures, loss functions, or through hyperparameter adjustment~\cite{cen2020controllable}. In the era of LLM-empowered recommendation, many works continue to leverage these established mechanisms for system-level controllability. For example, FLAME~\cite{fan2025fine}, PG-Ret~\cite{sharma2024optimizing} achieve flexible trade-offs between system objectives (such as accuracy, safety, novelty) by tuning reward weights or introducing additional regularization terms within their optimization frameworks~\cite{lin2025can}. 
    Agentic models like TriRec~\cite{gong2026breaking} extend this coordination to a platform agent, yet their control mechanism still relies on utility-weight tuning rather than semantic instructions.

    However, these parameter- or reward-based approaches, though effective, still inherit the rigidity of conventional methods: dynamic adjustment or rapid switching of system targets often requires model retraining or fine-tuning, and lacks the semantic expressiveness needed for more nuanced or context-aware control.

    Recently, two main paradigms have emerged for achieving system-initiated controllability in LLM-powered recommendation. The first leverages the unique ability of LLMs to modulate system objectives at the instruction level: for example, DLCRec~\cite{chen2025dlcrec} enables platforms to flexibly control recommendation style—such as category coverage—by simply altering prompt content, allowing real-time adjustment without modifying the underlying model structure. The second paradigm treats LLMs as enhancement modules within traditional frameworks, where controllability is still primarily achieved through parameter or reward tuning. For instance, PG-Ret~\cite{sharma2024optimizing} incorporates LLMs as reward evaluators in a reinforcement learning setup, so that system-level trade-offs like novelty and accuracy can be balanced by adjusting reward weights. Similarly, FELLAS~\cite{yuan2024fellas} utilizes LLMs as external services to enhance item and sequence representations, with system control realized by tuning parameters such as privacy protection and contrastive learning strength in the loss function. These approaches show how LLMs can either directly enable semantic-level system control via prompts, or augment traditional parameter-based methods to achieve more flexible and multi-dimensional system objectives.

    These developments mark a shift from static, structurally-driven system control towards a more flexible, semantic, and instruction-driven paradigm. By leveraging the unique capabilities of LLMs, platforms can achieve multi-objective consistency and policy agility, enhancing the overall adaptability and responsiveness of recommendation strategies.

\end{itemize}

\subsubsection{New Challenges for Controllability}

While LLM-powered RS offer unprecedented flexibility in controllability, they also introduce a range of new challenges that hinder the practical realization of reliable and robust control. These challenges can be broadly categorized into four interrelated issues:

\begin{itemize}

    \item \textbf{Semantic ambiguity.} At the foundation lies the inherent ambiguity and semantic uncertainty of natural language, which makes it difficult to consistently and accurately map user intentions to concrete recommendation actions. Unlike structured parameter configurations, prompts can be vague, context-dependent, or open to multiple interpretations~\cite{webson2022prompt}. This ambiguity becomes even more pronounced in multi-turn dialogues or complex intent formulations, where subtle changes in expression or context can yield divergent outputs~\cite{ribeiro2020beyond,lu2024aligning}.

    \item \textbf{Prompt sensitivity.} This foundational ambiguity directly contributes to a second major issue: prompt sensitivity. Because LLMs operate through stochastic generation conditioned on subtle linguistic cues, even minor variations in wording can result in significantly different recommendation results~\cite{zhao2021calibrate}—altering both the content and order of presented items~\cite{lu2021fantastically,razavi2025benchmarking}. This sensitivity undermines the reproducibility and stability of user control, making it challenging for users and system designers to craft prompts that yield consistent and predictable behaviors. Moreover, when combined with semantic ambiguity, the space of potential failure cases expands dramatically, compounding the difficulty of achieving reliable control.

    \item \textbf{Black-box opacity.} These challenges are further exacerbated by the black-box nature of LLMs. When unexpected outputs occur—whether due to ambiguous intent or prompt sensitivity—the lack of transparent reasoning pathways between input prompts and generated recommendations makes it difficult to diagnose the source of failure~\cite{lipton2018mythos}. This opacity not only obstructs debugging but also impedes the effective calibration of control mechanisms~\cite{fan2026uncertainty}. In other words, when control fails, users and developers are often unable to determine whether the failure lies in prompt interpretation, intent modeling, or output ranking—making it exceedingly hard to isolate root causes and apply targeted adjustments~\cite{polo2024efficient}.

    \item \textbf{Control inefficiency.} Finally, the inefficiency of natural language control introduces practical limitations that feed back into the above issues. Expressing precise or technical constraints through language often requires lengthy, carefully worded instructions, increasing the cognitive load on users and raising the risk of miscommunication. For example, a control signal that might be succinctly expressed through a structured parameter—such as “diversity=0.8”—may require a verbose and potentially ambiguous prompt like “please recommend a wider variety of items from different categories without too much repetition.” This verbosity slows down interaction and makes the system less responsive, particularly in contexts requiring rapid or fine-grained adjustments~\cite{wu2022ai,dohan2022language}. In turn, it increases the likelihood of ambiguous or error-prone prompts, which reintroduce semantic uncertainty and prompt instability—closing the loop of compounded control challenges~\cite{kusano2024longer}.

\end{itemize}

Together, these interlinked challenges reveal that while natural language enables expressive and flexible controllability, it also opens a complex failure surface that demands new methodological innovations. Bridging the gap between the expressive power of language and the reliability, efficiency, and interpretability required for controllable recommendation remains a core obstacle for LLM-driven systems.

\subsection{Privacy}
\label{sec:privacy}

\begin{figure*}[t]
    \centering
    \includegraphics[width=0.95\linewidth]{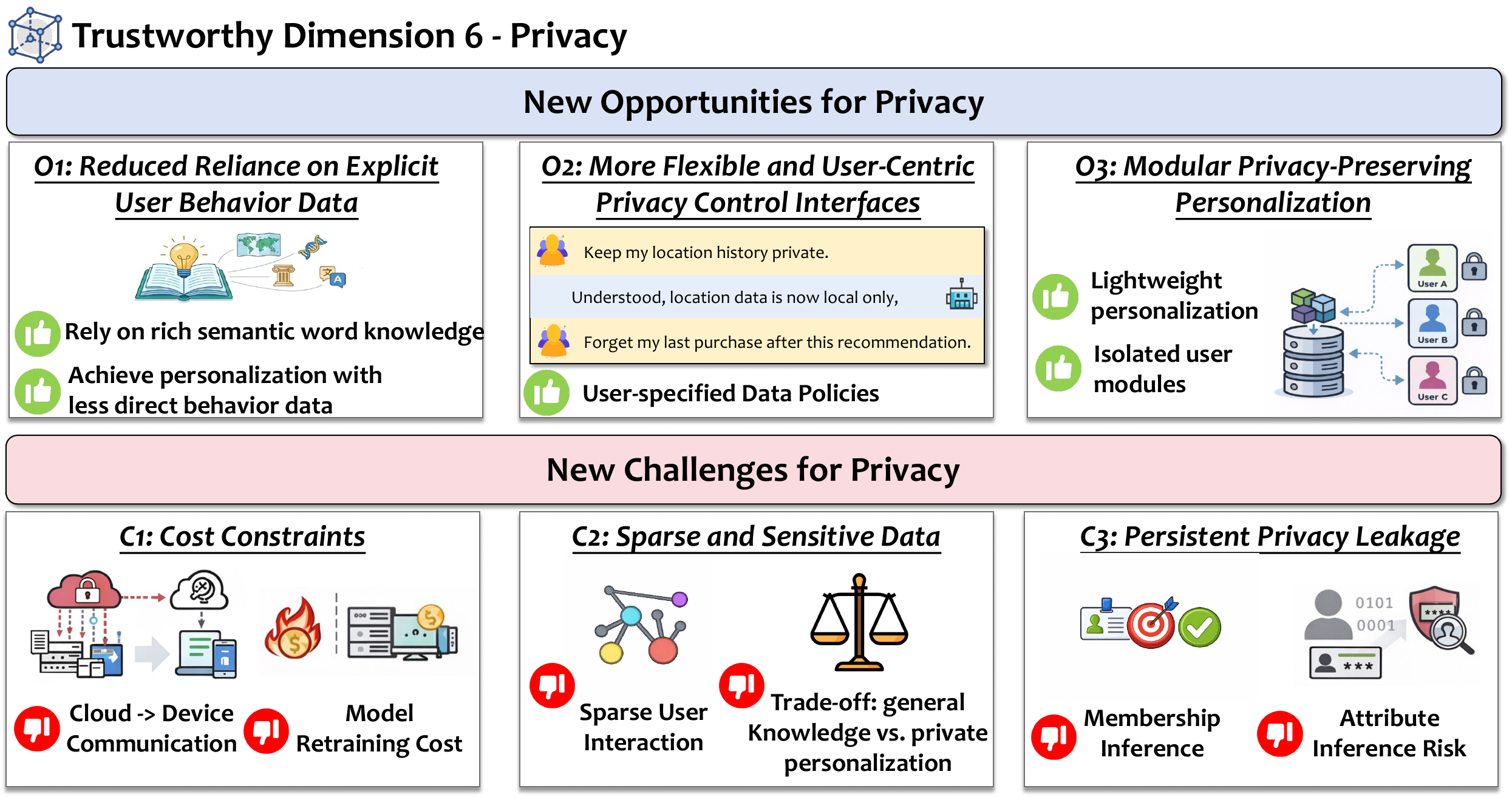} 
    \caption{An illustration of the opportunities and challenges introduced by LLMs for the privacy of recommender systems.}
    \label{fig:privacy}
\end{figure*}

Privacy refers to the protection of users' sensitive information throughout the entire lifecycle of the system, including data collection, model training, deployment, and data deletion~\cite{yang2020federated,sun2024survey}. 
This dimension is particularly critical because recommendation models rely heavily on behavioral traces such as clicks, purchases, browsing histories, and conversational feedback, which can expose sensitive user information and lead to privacy leakage and user profiling~\cite{lin2020fedrec}.

In LLM-empowered RS, privacy considerations become even more prominent. 
On the one hand, LLMs introduce new opportunities: their strong semantic understanding, broad open-world knowledge, and flexible natural language interfaces may create the possibility of reducing the reliance on explicit user data and enable more user-centric privacy control mechanisms. 
On the other hand, their large parameter scale, long-context modeling capabilities, and reliance on cloud-hosted infrastructures also introduce new risks, including increased computational costs for privacy-preserving techniques and more avenues for potential privacy leakage.
In this subsection, we summarize both the emerging opportunities and the new challenges for privacy in trustworthy LLM-empowered RS.

\subsubsection{New Opportunities for Privacy}
Compared with traditional RS, LLMs do not merely introduce stronger models; they also open up new possibilities for privacy-aware recommendation design.
These opportunities mainly arise from the semantic capabilities, interaction flexibility, and modular adaptation mechanisms of LLMs.

\begin{itemize}
    \item \textbf{Reduced reliance on explicit user behavior data.}
    Traditional RS often depend heavily on large amounts of user-item interaction data to infer preferences, which increases privacy risks because personalization quality is tightly coupled with the amount of collected behavioral data.
    In contrast, LLMs possess rich semantic knowledge and strong reasoning abilities, enabling recommendation models to leverage item content, metadata, and world knowledge more effectively~\cite{zhao2024recommender}.
    This creates the possibility of achieving useful personalization with less direct dependence on fine-grained user behavior logs.
    Such a property is particularly valuable in sparse-data or cold-start settings, where recommendation quality may be improved through semantic understanding rather than aggressive collection of personal interaction histories~\cite{zhang2025cold}.

    \item \textbf{More flexible and user-centric privacy control interfaces.}
    LLM-based RS are naturally built on language interfaces, which creates new opportunities for exposing privacy preferences in a more explicit and user-friendly manner~\cite{li2026privacy}.
    Instead of relying only on predefined privacy settings or rigid consent templates, users may specify privacy requirements through natural language, such as what kinds of data can be used, what should remain local, or what should be forgotten after recommendation~\cite{belosevic2025user}.
    This makes privacy control potentially more transparent, interpretable, and adaptable to individual user needs.

    \item \textbf{Modular privacy-preserving personalization.}
    LLM-based RS increasingly adopt parameter-efficient fine-tuning techniques, such as adapters and LoRA, which provide a natural modular structure for privacy-preserving personalization~\cite{sun2024improving,wang2025towards1}.
    This modularity creates opportunities to separate general knowledge from user-specific adaptation, allowing personalized components to be trained, stored, synchronized, or removed more selectively~\cite{wu2024fedlora}.
    As a result, privacy-preserving strategies such as federated personalization and efficient unlearning can be implemented on lightweight user-specific modules rather than on the full LLM, improving controllability and reducing the cost of privacy operations.
\end{itemize}

\subsubsection{New Challenges for Privacy}
Although LLMs create new opportunities for privacy-aware recommendation, these opportunities are far from being fully realized in practice.
Below, we summarize the major challenges that remain in LLM-based recommendation.

\begin{itemize}
    \item \textbf{Communication and resource constraints.}
    Privacy-preserving training methods such as federated learning and machine unlearning are attractive in recommendation because they reduce direct exposure of user data.
    However, when applied to LLM-based recommendation, they face severe efficiency bottlenecks due to the large parameter scale of LLMs.
    Transferring model updates, synchronizing personalized modules, or re-training affected components can still incur substantial computation and communication costs.

    \begin{itemize}
        \item \textit{1) Federated learning.}
        Federated learning (FL) protects privacy by decentralizing data and training models locally on user devices without transferring raw user information to a central server~\cite{zhang2023lightfr, yin2024device}.
        In this paradigm, clients perform local training and share only intermediate model updates, such as gradients or adapter parameters.
        Recent studies adapt FL to LLM-based recommendation by improving efficiency.
        For example, FELLAS leverages external LLM services to enhance item and sequence representations while reducing client-side burden~\cite{yuan2024fellas}.
        FELLRec~\cite{zhao2025federated} introduces client-specific LoRA modules together with dynamic aggregation strategies to better balance privacy, personalization, and training efficiency.
        Despite these advances, the cost of deploying privacy-preserving LLM-based recommendation at scale remains high.
        \item \textit{2) Machine unlearning.}
        Machine unlearning enhances privacy by enabling models to remove the influence of specific user data, which is especially important for supporting deletion requests and the right to be forgotten~\cite{chen2022recommendation, zhang2024recommendation}.
        In recommendation, unlearning is typically achieved through gradient ascent on removed samples, retraining on retained data, or distillation-based forgetting~\cite{li2024survey,sachdeva2024machine}.
        For LLM-based recommendation, the main challenge is that re-training the full model is prohibitively expensive.
        Therefore, current methods mainly focus on updating selective modules instead of the full LLM.
        For example, E2URec employs a teacher-student framework and selectively adjusts LoRA parameters to perform targeted unlearning efficiently while preserving recommendation utility~\cite{wang2025towards1}.
        Nevertheless, efficient and verifiable unlearning for LLM-based recommendation remains underexplored.
    \end{itemize}
    Future research should further reduce the cost of privacy-preserving LLM recommendation through model compression, efficient distillation, lightweight adaptation, and selective synchronization. Techniques such as quantization, pruning, and sparse update transmission may help reduce communication payloads, while server-to-client distillation may enable heavy reasoning to remain in the cloud and only lightweight personalized modules to be maintained locally~\cite{kim2021pqk,zhu2024survey}.

    \item \textbf{Data sparsity and personalized fine-tuning trade-offs.}
    In federated and privacy-constrained recommendation settings, each client usually has access to only a small amount of local data.
    This challenge already exists in traditional recommendation, but becomes more severe in LLM-based recommendation due to the scale and flexibility of LLMs.

    \begin{itemize}
        \item \textit{1) Fine-tuning sensitivity under sparse local data.}
        LLMs contain billions of parameters, making them highly sensitive to fine-tuning on extremely sparse client-specific interaction data.
        Under privacy constraints, the amount of available user data may be further restricted, which increases the risk of overfitting and memorization of individual user behaviors rather than learning generalized preference patterns~\cite{yeom2018privacy}.
        \item \textit{2) Semantic generalization versus private personalization.}
        LLMs are rich in semantic and world knowledge, which can compensate for sparse interactions to some extent.
        However, translating such general knowledge into accurate and privacy-preserving personalization remains difficult~\cite{zhang2024bridging}.
        Over-reliance on general knowledge may produce recommendations that are semantically plausible but insufficiently personalized, whereas excessive adaptation to limited local data may overfit and amplify privacy risks.
    \end{itemize}
    
    A key research question is how to adapt the general knowledge of LLMs to sparse user-specific data while preserving privacy.
    Promising directions include meta-learning, transfer learning, adaptive regularization, and hybrid designs that combine global semantic priors with lightweight private personalization modules~\cite{wei2023personalized}.

    \item \textbf{Persistent privacy leakage in training and inference.} 
    Beyond efficiency and sparsity, LLM-based recommendation also inherits broader privacy risks from LLM systems.
    Even if raw user data are not directly shared, privacy may still be compromised through model memorization, membership inference, attribute inference, or leakage from long-context prompts and cloud-hosted inference services~\cite{carlini2021extracting,zhang2021membership,zhang2023comprehensive}.
    In recommendation scenarios, these risks are especially concerning because the underlying user data are highly behavioral and often highly personal.
    Therefore, privacy in LLM-based recommendation should not be viewed only as a training-time issue. It must also be addressed at inference time, in deployment pipelines, and in privacy auditing and governance mechanisms~\cite{das2025security}.
\end{itemize}


\section{Evaluation Protocol}
\label{sec:evaluation}
Evaluation protocols serve as the cornerstone of trustworthy recommendation research, providing the empirical foundation for rigorously measuring and comparing system reliability. However, as LLMs introduce fundamentally new capabilities—such as natural language interaction, reasoning, free-form explanation—evaluation protocols originally developed for traditional RS are no longer adequate for assessing trustworthiness in this new paradigm. Consequently, the emergence of LLM-empowered RS not only expands the functionality of RS, but also reshapes the very criteria by which trustworthiness is evaluated, making a dedicated evaluation protocol essential.

To provide a structured overview, we organize the evaluation protocol from two complementary perspectives: \textit{datasets} and \textit{metrics}. First, we summarize the benchmark datasets currently used in trustworthy LLM-empowered RS, including both traditional datasets and those specifically designed for LLMs. 
Second, we review the metrics used to quantify the dimensions of trustworthy RS, highlighting both the continued use of traditional RS metrics and the recent emergence of LLM-oriented evaluation metrics, while also discussing the limitations of existing metrics in the LLM era. 

\subsection{Datasets}
The evaluation of RS has long relied on static interaction logs. However, in the era of LLMs, ensuring trustworthy recommendation requires a comprehensive evaluation paradigm that not only leverages existing textual data but also extends to more complex, dynamic, and human-centered behaviors. To facilitate a systematic assessment of LLM-empowered RS, we categorize the relevant datasets into two overarching groups: \textit{Traditional Datasets} and \textit{Emerging Datasets for LLMs}.

\begin{itemize}
    \item \textbf{Traditional Datasets.} Although originally developed for traditional RS, many datasets containing rich textual metadata—such as user reviews, item titles, descriptions, and user profiles—have been extensively repurposed for LLM-based recommendation. These datasets play an important role in evaluating multiple dimensions of trustworthiness. For example, user reviews can be used to assess \textit{Explainability}, while demographic attributes may facilitate the evaluation of \textit{Bias and Fairness}. Representative benchmarks in this category include \textit{Amazon}\footnote{\url{https://jmcauley.ucsd.edu/data/amazon/}}, \textit{MovieLens}~\cite{harper2015movielens}, and Yelp\footnote{\url{https://www.yelp.com/dataset}}.
    
    \item \textbf{Emerging Datasets for LLMs.} To fully evaluate the capabilities of LLMs—such as natural language understanding, autonomous action, and subjective alignment—researchers have introduced new datasets and evaluation environments. These emerging datasets break away from static historical logs and can be further divided into three subcategories:
    
    \begin{itemize}
        \item \textit{Conversational recommendation datasets.} These datasets capture multi-turn natural language dialogues between users and systems/agents. They serve as indispensable testbeds for evaluating specific trustworthy dimensions in long-context linguistic interactions. Notable examples include \textit{ReDial}~\cite{li2018towards}, a renowned large-scale public dataset integrating goal-directed recommendations, open-domain chit-chat, and interactive question-answering, which is used to evaluate Fairness and Explainability. Similarly, \textit{INSPIRED}~\cite{hayati2020inspired} closely mirrors realistic human-human dialogue structures and allows for the assessment of system Explainability during multi-stage natural communication. Furthermore, \textit{TG-ReDial}~\cite{zhou2020towards}, a localized Chinese dataset emphasizing explicit topic guidance and conversational flow transitions, is also applied to evaluate Bias and Fairness and Explainability of conversational recommenders.

        \item \textit{Interactive agent datasets.} Serving as a focal point for evaluating LLMs in dynamic and autonomous recommendation scenarios, these datasets provide critical environments and instructions for agent-oriented tasks while shedding light on key trustworthiness dimensions. For instance, \textit{INSTRUCTREC}~\cite{xu2025iagent} is a comprehensive dataset providing user-instruction formats tailored specifically for agent-oriented recommendation tasks, which plays a crucial role in assessing the Controllability of LLMs under explicit human instructions.

        \item \textit{Human-evaluation datasets.} This category encompasses datasets grounded in manual annotation and subjective human perception. For instance, \textit{multi-dimensional user ratings for recommendations}~\cite{zhang2024navigating} and \textit{user evaluations of recommendation explanation texts}~\cite{lu2023user} are provided to assess the Explainability dimension from user perspectives. Furthermore, \textit{UCP-IMDb}~\cite{wozniak2025improving} incorporates manually modified user profiles with human-annotated ground truth to evaluate the Controllability dimension of recommenders. It also accounts for negative subjective perceptions, such as user discomfort caused by recommended content on \textit{Zhihu}~\cite{liu2025filtering}, which is similarly employed to assess system Controllability.
    \end{itemize}

\end{itemize}

\subsection{Metrics}
To systematically evaluate trustworthy LLM-empowered RS, recent studies have adopted a broad spectrum of metrics spanning both traditional recommendation objectives and emerging LLM-specific concerns. While traditional metrics remain essential, they are no longer sufficient to capture the distinctive behaviors introduced by generative, interactive, and language-driven recommendation pipelines. As a result, the evaluation landscape has undergone substantial changes with the introduction of LLMs, while still facing important limitations that existing metrics fail to fully address. 
In the following, we review representative metrics under each trustworthiness dimension, highlighting both inherited practices from traditional RS and newly proposed metrics tailored to LLM-empowered RS. We further discuss the limitations of current metrics in the LLM era and outline potential directions for future research.

\subsubsection{Robustness.}
Robustness evaluation in LLM-empowered RS concerns the extent to which a model can maintain stable and reliable recommendation behavior under various perturbations. Existing evaluation protocols, however, remain largely centered on traditional recommendation metrics.

\begin{itemize}
    \item \textbf{Traditional Metrics.}
    Most existing studies evaluate robustness by examining the degradation of standard utility metrics, such as \textit{HR}, \textit{NDCG}, and \textit{AUC}, under attacks~\cite{ning2024cheatagent,ning2025retrieval}, noise~\cite{wang2025llm4dsr}, or distribution shifts~\cite{wang2023drdt}.  
    For attack robustness in particular, some metrics focus on the impact of attacks on attacker-specified target items rather than on the entire item set, such as whether the ranking positions of target items are promoted or suppressed. Representative metrics in this line include \textit{T-HR}~\cite{zhang2024lorec}, \textit{T-NDCG}~\cite{zhang2024lorec}, \textit{Prediction Shift (PS)}~\cite{gu2025llm,chiang2023shilling}, and \textit{Attack Success Rate (ASR)}~\cite{ning2025exploring}.
    
    \item \textbf{Limitations.}
    A major limitation of current robustness evaluation is that it largely inherits the traditional RS paradigm. Although these metrics remain informative, they are insufficient for LLM-based RS, where failures may emerge from multiple components of the recommendation pipeline, including prompt sensitivity, reasoning instability, retrieval corruption, and errors in multi-turn interaction. As a result, existing metrics mainly capture the \emph{outcomes} of robustness failures, while offering limited insight into whether the model is resilient to linguistic perturbations, adversarial prompts, or process-level manipulation. Future work should therefore move beyond static performance degradation and develop LLM-native robustness metrics that explicitly assess the stability of prompting, reasoning, retrieval, and interactive recommendation behaviors.

\end{itemize}

\subsubsection{Bias and Fairness.}
Bias and fairness metrics evaluate inequalities in resource allocation, systemic bias, and presentation disparities. Existing evaluation metrics remain largely grounded in traditional fairness metrics, while recent studies have begun to introduce LLM-oriented measures that account for \textit{position bias} and \textit{counterfactual user-side fairness}.

\begin{itemize}
    \item \textbf{Traditional Metrics.}
    On the item side, representative inherited metrics include \textit{Catalog Coverage}~\cite{deldjoo2024understanding}, \textit{Gini Coefficient}~\cite{jiang2025beyond,lichtenberg2024large,deldjoo2024understanding}, \textit{HHI}~\cite{deldjoo2024understanding}, and \textit{ARP}~\cite{jiang2025beyond}, which are used to assess exposure concentration and popularity bias. On the user side, classical group-fairness metrics such as \textit{SPD}~\cite{sakib2024challenging}, \textit{DPD}~\cite{jiang2025beyond}, and \textit{JSD}~\cite{das2024unveiling} are still widely used to quantify disparities in recommendation outcomes across demographic groups. More generally, most current studies continue to evaluate fairness through distributional imbalance, exposure inequality, or utility gaps, indicating that fairness assessment in LLMs remains largely rooted in traditional RS fairness frameworks~\cite{deldjoo2024understanding,jiang2025beyond,deldjoo2024normative}.
    
    \item \textbf{Emerging Metrics for LLMs.} Evaluation along this dimension has increasingly incorporated LLM-oriented metrics, primarily in the following two aspects.
    \begin{itemize}
        \item \textit{Position bias evaluation.} This bias is caused by LLMs’ sensitivity to the ordering of candidate items in prompts. Metrics such as \textit{Positional Consistency}~\cite{bito2025evaluating}, \textit{Output Similarity}~\cite{bito2025evaluating}, and \textit{Input Sensitivity}~\cite{bito2025evaluating} have been specifically proposed to assess whether recommendation results remain stable when the same candidate set is reordered or shuffled—an issue that is nonexistent in traditional RS.
        \item \textit{Counterfactual user-side fairness evaluation.} Some metrics are based on perturbation and counterfactual analysis, measuring unfairness by comparing recommendation outcomes before and after altering sensitive user attributes in the input (e.g., LLM prompts). Metrics such as \textit{PAFS}~\cite{sah2025faireval}, \textit{SNSR}~\cite{deldjoo2025cfairllm,zhang2023chatgpt}, \textit{SNSV}~\cite{deldjoo2025cfairllm,zhang2023chatgpt}, \textit{NSD}, \textit{NCSD}~\cite{deldjoo2024normative}, and \textit{True Preference Alignment}~\cite{deldjoo2025cfairllm} are introduced to measure how recommendation outputs change when demographic attributes are perturbed in prompts. Moreover, ranking-aware list comparison metrics such as \textit{SERP}~\cite{sah2025faireval} and \textit{PRAG}~\cite{sah2025faireval} have been adopted or designed to better evaluate consistency between recommendation lists generated under different sensitive-attribute conditions. 
    \end{itemize}

    \item \textbf{Limitations.}
    Current fairness evaluation for LLM-based RS still faces two major limitations. First, counterfactual metrics mainly measure output changes after modifying sensitive user attributes, but such changes do not necessarily imply unfairness. In LLM-RS, recommendation differences may result from either biased treatment or legitimate personalization, making these metrics difficult to interpret. Second, the field still lacks unified benchmarks and standardized evaluation protocols. Different studies often use different prompt templates and metric definitions, making results difficult to compare and reproduce across works.
    Future work should develop fairness metrics that can better distinguish unjust bias from legitimate personalization, especially in rich contextual and conversational settings. In addition, unified benchmarks and standardized evaluation protocols are needed to support reliable and reproducible fairness assessment.
\end{itemize}

\subsubsection{Explainability.}
Explainability evaluates the transparency and persuasiveness of recommendations.
In the LLM era, explainability evaluation has evolved from feature-level justification toward semantic reasoning, personalization, and discourse-level assessment.

\begin{itemize}
    \item \textbf{Traditional Metrics.}
    Early metrics primarily relied on structured features and template-based explanation generation. Metrics such as \textit{FMR}~\cite{shimizu2025disentangling} and \textit{FCR}~\cite{peng2024uncertainty,cui2022m6} quantify whether generated explanations explicitly mention ground-truth item features derived from reviews or metadata. To avoid rigid or repetitive explanations, metrics such as \textit{Distinct}~\cite{yang2024fine,hada2021rexplug} and \textit{DIV}~\cite{shimizu2025disentangling,balloccu2023faithful} measure the diversity of mentioned features across explanations, typically by calculating overlap statistics. These metrics remain applicable, especially when explanations are still grounded in identifiable item attributes. However, they mainly evaluate surface-level alignment and linguistic variation.
    
    \item \textbf{Emerging Metrics for LLMs.} LLMs can generate personalized explanations without relying on predefined templates or explicit feature constraints. This shift has led to new explainability evaluation criteria. 
    \begin{itemize}
        \item \textit{Semantic consistency evaluation.} Unlike earlier lexical matching metrics, \textit{Aspect Score}~\cite{gao2024dre} leverages LLMs to extract item aspects and evaluates semantic overlap between generated explanations and ground-truth aspects, moving toward semantic-level faithfulness. Similarly, \textit{Concept Matching Ratio (CMR)}~\cite{yang2024fine} measures higher-level conceptual consistency embedded within reasoning chains rather than isolated features.
        
        \item \textit{Personalization evaluation.} LLMs enable adaptive explanations tailored to individual users. \textit{Acceptance Ratio}~\cite{maes2025mitigating} measures changes in user acceptance likelihood after exposure to explanations. \textit{Concept Overlapping Ratio (COR)}~\cite{yang2024fine} evaluates whether the same item is justified using distinct reasoning concepts across users, while \textit{User Preference Alignment (UPA)}~\cite{qiu2024unveiling} assesses whether explanations align with user intents under the LLM-as-a-judge paradigm. These metrics reflect a shift from item-centric correctness to user-centric justification.
        
        \item \textit{Persuasiveness evaluation.} LLM-powered explanations often involve multi-step reasoning and rhetorical structuring. Metrics such as \textit{Reasoning Proficiency (RP)}~\cite{qiu2024unveiling} and \textit{Explainability (EX)}~\cite{qiu2024unveiling} employ evaluator LLMs to assess logical coherence, inference validity, and overall persuasive quality at the discourse level.
    \end{itemize}
    
    \item \textbf{Limitations.}
    A major limitation of LLM-powered explainability evaluation is that it largely measures textual plausibility rather than causal faithfulness. Existing metrics assess whether explanations appear coherent, persuasive, and aligned with user preferences. However, fluent and semantically consistent explanations can still be post-hoc rationalizations that do not reflect the true decision process. Moreover, heavy reliance on LLM-powered evaluators introduces potential circularity, as generator and judge models may share similar biases and reward stylistic fluency over genuine grounding. As a result, current metrics capture explanatory plausibility but provide limited evidence of causal responsibility. Future evaluation should therefore move beyond overlap-based and single-model judging schemes toward counterfactual, and externally grounded protocols that better assess whether explanations faithfully represent the underlying recommendation mechanism.
\end{itemize}

\subsubsection{Factuality.} Factuality measures the truthfulness of generated content. In traditional RS, this was rarely evaluated explicitly, since items were selected from a closed candidate set. In contrast, LLMs can generate incorrect or hallucinated outputs, making factuality an independent evaluation dimension.

\begin{itemize}
    \item \textbf{Emerging Metrics for LLMs.}
    Recent studies introduce dedicated factuality metrics to evaluate whether generated outputs remain grounded in real entities and factual knowledge.
    \begin{itemize}
        \item \textit{Hallucination detection evaluation.}   Metrics such as \textit{Hallucination Rate}~\cite{xu2025iagent,jiang2025beyond,lu2024aligning} and \textit{Unmatched Ratio (UR)}~\cite{spurlock2024chatgpt} quantify the proportion of generated items that cannot be matched to a valid database entry or catalog entity. Conversely, \textit{Valid Ratio}~\cite{liao2024llara} measures the proportion of outputs that successfully correspond to real items. These metrics directly address a new failure mode introduced by generative recommenders: fabricated recommendations. 

        \item \textit{Structural faithfulness evaluation.}   Beyond entity existence, LLM-powered systems may generate reasoning paths or relational explanations grounded in knowledge graphs. \textit{Path Faithfulness Rate (PFR)}~\cite{balloccu2023faithful} evaluates whether generated relational paths are free from nonexistent triplets, ensuring that intermediate reasoning steps correspond to valid knowledge graph relations. This extends factuality from entity-level correctness to structural consistency. 
        
        \item \textit{Factual consistency evaluation.}   Under the LLM-as-a-judge paradigm, \textit{Factual Consistency (FC)}~\cite{qiu2024unveiling} prompts evaluator LLMs to assess whether textual claims about item attributes align with known facts. Unlike entity matching metrics, FC focuses on attribute-level correctness and semantic factual alignment within explanations or recommendation justifications.         
    \end{itemize}

    \item \textbf{Limitations.} Current evaluation metrics primarily focus on observable factual consistency against available knowledge sources, but they do not fully capture the dynamic real-world recommendation environments. In practice, item information, user preferences, and external knowledge evolve continuously, while LLMs rely on static pre-trained representations that may lag behind current reality. Conversely, a product that once existed may be discontinued but still be confidently recommended by an LLM drawing on stale knowledge. Future work should incorporate temporal grounding, fine-grained attribute validation, and stronger external knowledge alignment to more rigorously assess factual consistency in recommendation.
\end{itemize}

\subsubsection{Controllability.}
Controllability assesses the precision in steering system outputs to align with predefined user expectations. Unlike traditional ID-based pipelines, the autoregressive paradigm of LLMs introduces inherent uncontrollability. Consequently, existing evaluations adapt traditional rule-based metrics while introducing new measures tailored for \textit{generative redundancy}.
\begin{itemize}
    \item \textbf{Traditional Metrics.}
Traditional controllability metrics assess whether the system satisfies explicit rules and user constraints. Specifically, one category of metrics measures constraint satisfaction, which includes \textit{Category Proportion Accuracy~(CPA)} and \textit{Top-$k$ Target Category Proportion~(TCP)}~\cite{lu2024aligning}, along with \textit{Mean Average Error of Coverage~(MAE\_Cov)}~\cite{chen2025dlcrec}. Another category measures constraint violations, such as \textit{Average Filtering Burden}~\cite{liu2025filtering} and \textit{Filtered Rate~(FR)}~\cite{xu2025iagent}. In LLM settings they evaluate how well generated outputs follow filtering instructions.
    
\item \textbf{Emerging Metrics for LLMs.} 
LLM-based RS are susceptible to \textit{generative redundancy}. To address this, recent studies introduce dedicated metrics (e.g., \textit{RepeatItem}, \textit{CorrectCount}, and \textit{InHistory} \cite{lu2024aligning}) to quantify LLM-specific flaws, capturing issues like intra-list repetition, output-length deviations, and the resurfacing of historical items.

\item \textbf{Limitations.}
While current metrics effectively capture rigid, list-level constraints, they inherit a static, item-centric paradigm that falls short of evaluating true conversational controllability. Specifically, existing metrics rely heavily on superficial rule-matching for isolated, single-turn outputs. They fail to capture the interactive and semantic complexity of LLMs, completely overlooking whether a system can accurately comprehend nuanced natural language instructions and persistently maintain, or dynamically update, these constraints across lengthy multi-turn sessions.
\end{itemize}

\subsubsection{Privacy.}
Privacy metrics evaluate the extent to which the RS safeguards sensitive user data. With the advent of LLM-empowered RS, privacy evaluation has expanded beyond traditional data-level leakage to encompass model- and pipeline-level risks.

\begin{itemize}
\item \textbf{Traditional Metrics.} Privacy evaluation largely inherits metrics from traditional machine learning. These metrics primarily fall into two categories. 
One category measures \textit{privacy leakage risks}, encompassing both the exposure of sensitive user data and the explicit memorization of the model. This includes metrics such as \textit{Binary Privacy Leakage~(BPL)} and \textit{Semantic Privacy Leakage~(SPL)}~\cite{khezresmaeilzadeh2025preserving}, as well as \textit{Attack AUC}~\cite{zhang2021membership}. The other category evaluates \textit{utility preservation} (i.e., the privacy-utility trade-off), which measures the degradation in recommendation quality (e.g., drops in Recall or NDCG) caused by applying privacy-preserving methods~\cite{li2024survey}. 

\item \textbf{Emerging Metrics for LLMs.} LLMs introduce new privacy risks, which has motivated the development of LLM-oriented privacy metrics. However, dedicated privacy metrics specifically designed for LLM-based RS remain limited. As a result, many existing metrics are adapted from the broader LLM privacy literature and can be viewed as potential directions for LLM-RS evaluation, while their application in recommendation settings is still largely underexplored.
\begin{itemize}
    \item \textit{Prompt and embedding reconstruction risk.} 
    Adversaries may attempt to recover the original textual inputs from intermediate representations or model outputs. Metrics such as \textit{Reconstruction Similarity} (e.g., ROUGE or BLEU between reconstructed and original prompts)~\cite{morris2023text} and \textit{Attribute Inference Accuracy}~\cite{staab2023beyond}, originally developed for general LLM privacy research, can be naturally extended to quantify how much sensitive textual information can be recovered in LLM-RS, capturing a leakage channel that does not exist in traditional RS.
    \item \textit{Memorization and extraction risk.} Due to the large parameter scale of LLMs, fine-tuning on user interaction data may cause models to memorize rare or individual-specific behaviors. Metrics adapted from the general LLM privacy literature, such as \textit{Extraction Rate}~\cite{carlini2021extracting} and \textit{Memorization Score}~\cite{carlini2022quantifying}, provide a natural basis for measuring how often training-specific user data can be elicited from LLM-RS outputs under crafted queries, although their systematic application in recommendation settings remains underexplored.

    \item \textit{Unlearning effectiveness.} As retraining the full LLM is prohibitively expensive, the evaluation of unlearning has become a distinct dimension. Metrics such as \textit{Forgetting Rate}~\cite{wang2025towards1}, which measures the extent to which the influence of removed samples has been eliminated, are commonly adopted to assess unlearning effectiveness in LLM-RS.
\end{itemize}

\item \textbf{Limitations.} 
Despite recent progress, privacy evaluation in LLM-based RS still faces several key limitations. First, existing metrics primarily focus on training-time leakage, leaving crucial inference-time risks (e.g., prompt leakage via cloud APIs or conversational histories) largely unmeasured. Second, unlearning evaluation lacks verifiability; current metrics mainly observe output behavior and cannot guarantee that the removed data's influence is truly erased from model parameters. Third, the lack of unified privacy benchmarks and standardized attack protocols tailored to LLM-RS makes cross-study comparisons difficult.
Future work should prioritize inference-time assessments, develop verifiable unlearning metrics, and establish benchmarks that jointly evaluate privacy, utility, and cost.
\end{itemize}

\section{Open Problems and Future Directions}
\label{sec:future}

In this section, we identify five important open problems that warrant further investigation: (i) the trade-offs between multiple dimensions of trustworthiness and recommendation utility, (ii) emerging trustworthiness challenges introduced by new agentic recommendation paradigms, (iii) the limited theoretical understanding of trustworthiness in LLM-empowered RS, (iv) the inefficiency introduced by LLM integration, and (v) the urgent need for comprehensive benchmarks and evaluation methods that take into account trustworthiness. 
To this end, we highlight several promising directions to address these challenges and advance the field toward more trustworthy LLM-empowered RS.

\subsection{Toward Multi-objective Trustworthiness in LLM-empowered Recommenders}

Trustworthiness in RS is inherently a multi-objective problem, involving fundamental trade-offs among competing goals~\cite{abdollahpouri2020multistakeholder,zheng2021disentangling,jannach2023survey}.
However, existing research on trustworthy LLM-empowered RS largely focuses on individual trustworthiness dimensions in isolation, with limited efforts toward their joint optimization.
Consequently, future research should move beyond single-objective mitigation strategies and explicitly formulate trustworthy LLM-empowered RS as a multi-objective optimization problem.
Moreover, not all trustworthiness objectives should be treated equally. Some requirements should be enforced as strict constraints rather than flexible objectives, particularly when violations may result in serious user harm or regulatory issues, such as privacy leakage or factually incorrect outputs.

\begin{itemize}
	\item \textbf{Accuracy-trustworthiness trade-off.}
	Trustworthiness-oriented objectives are not always aligned with standard recommendation utility metrics. For example, some debiasing strategies may fail to improve recommendation performance and can even degrade ranking accuracy~\cite{jiang2024item}. A promising direction is to develop adaptive optimization strategies that dynamically balance recommendation quality and trustworthiness objectives. For instance, future work could explore Pareto-aware optimization~\cite{lampinen2000multiobjective} and constraint-based formulations, where trustworthy recommendation is modeled as a constrained multi-objective optimization problem. Under such formulations, system designers can explicitly determine which objectives should be jointly optimized and which dimensions should be imposed as hard constraints.

    \item \textbf{Explainability-robustness trade-off.}
	Enhancing the explainability of recommender systems inherently conflicts with robustness against malicious attacks. 
	In LLM-empowered RS, generating detailed reasoning traces or item-feature attributions may expose the model's underlying decision boundaries. 
	Although such transparency can improve explainability, it may also reduce attack robustness.
	For example, malicious actors can exploit this transparency to reverse-engineer the recommendation logic and craft precise adversarial attacks, such as prompt injection or profile poisoning~\cite{shokri2021privacy,kumbam2025exploiting}. 
	To defend against these threats, models are often equipped with strict safety guardrails and defensive alignment mechanisms. 
	However, such protections may cause the LLM to hide its internal reasoning process and produce overly conservative responses, thereby weakening explainability~\cite{wang2023decodingtrust}. 
	Future research should therefore explore explainability methods that provide meaningful transparency to users while avoiding the exposure of sensitive model behaviors and maintaining robustness against adversarial attacks.

    \item \textbf{Fairness-privacy trade-off.} 
    Mitigating biases and unfairness in RS typically relies on collecting and utilizing sensitive user information (e.g., demographic attributes) to detect biases and ensure equitable outcomes~\cite{chen2023bias}.
	However, privacy-preserving recommendation restricts the collection and utilization of such information~\cite{yeom2018privacy}. 
	This trade-off becomes more pronounced in LLM-empowered RS, where improving fairness often requires richer user contexts, long-text interactions, and fine-grained semantic profiling~\cite{gallegos2024bias}, all of which may increase the risk of privacy leakage. 
	Future research should investigate how to improve fairness under strict privacy constraints, for example, by developing debiasing methods that minimize the reliance on sensitive user information while limiting privacy leakage risks.
\end{itemize}

\subsection{Trustworthiness in Agentic Recommendation}
Agentic RS represent an emerging paradigm in which recommendation is modeled as an autonomous and goal-driven process.
Compared with one-shot recommendation, agentic LLM-empowered RS can maintain memory, invoke external tools, and iteratively interact with users to accomplish complex recommendation tasks~\cite{yao2022react,peng2025survey,shu2024rah}.
However, these capabilities also introduce new trustworthiness risks. Long-term memory, multi-step reasoning, and external tool use may all become sources of unreliable or untrustworthy behaviors. Consequently, future research should pay greater attention to trustworthiness issues in agentic RS.

\begin{itemize}
	\item \textbf{Robustness against memory attacks.}
    Agentic recommenders often maintain memory to store user preferences and feedback~\cite{zhu2025recommender}. While memory improves personalization, it also expands the attack surface: poisoned memories may continuously influence future recommendations~\cite{dong2025memory}.
    A promising direction is to develop robust memory governance mechanisms, such as memory provenance tracking, suspicious content sanitization, and efficient memory correction or unlearning methods that can prevent the long-term propagation of malicious or corrupted information~\cite{yang2025drunkagent}.

	\item \textbf{Factuality via tool use.}
	Hallucination remains a major challenge for trustworthy LLM-empowered recommendation.
	Beyond post-hoc detection, agentic recommendation provides an opportunity to enforce factual grounding throughout the recommendation process.
    Specifically, agents can retrieve supporting evidence, invoke structured tools to verify item attributes, and iteratively validate generated claims before producing final recommendations.
    In addition, process-level supervision that rewards faithful intermediate reasoning steps, rather than only the final output, may help reduce plausible but factually incorrect recommendation rationales~\cite{wei2022chain,bang2025hallulens}.

    \item \textbf{Fairness through multi-agent verification.} Relying on a single agent to simultaneously generate and evaluate recommendations may limit the reliability of fairness assessment. Existing studies show that LLMs often exhibit confirmation bias and overconfidence, while effective self-correction remains difficult without external feedback~\cite{huang2023large,du2024improving}. 
    To address this issue, future work may explore multi-agent recommendation frameworks in which dedicated verifier agents independently examine the outputs and reasoning processes of the main recommendation agent for potential unfairness or biased behaviors~\cite{gong2026breaking,yao2022react}. 
    Such collaborative verification mechanisms may provide more reliable fairness guarantees for complex recommendation workflows.
\end{itemize}

\subsection{Theoretical Foundations for Trustworthiness in LLM-empowered Recommenders}

Current research on trustworthy LLM-empowered RS is predominantly empirical, relying heavily on heuristic prompting~\cite{rotar2026can} and empirical observations~\cite{zhang2021language}. However, a rigorous theoretical foundation is essential to guarantee the reliability of these systems, especially in high-stakes domains. Bridging the gap between empirical success and theoretical guarantees represents a critical frontier. Future work should focus on developing a theoretical understanding of trustworthiness in LLM-empowered RS, with several promising directions outlined below.

\begin{itemize}
	\item \textbf{Theoretical guarantees for robustness.} Although recent studies have explored improving the robustness of LLM-empowered RS against attacks~\cite{wang2025id},  noise~\cite{wang2025llm4dsr}, and out-of-distribution (OOD) scenarios~\cite{wang2023drdt}, most existing approaches remain largely empirical. 
	A promising direction is to move beyond heuristic defenses and develop rigorous theoretical foundations for robustness. Traditional machine learning theories and optimization frameworks, such as Distributionally Robust Optimization (DRO)~\cite{rahimian2019distributionally}, Invariant Risk Minimization (IRM)~\cite{arjovsky2019invariant}, and Domain Adaptation (DA)~\cite{ben2006analysis}, could be introduced into LLM-empowered RS to provide principled robustness guarantees under dynamic recommendation environments. Such formulations may enable the derivation of formal generalization and robustness bounds, thereby laying the foundation for recommendation algorithms with \emph{provable robustness} guarantees.

    \item \textbf{Theoretical guarantees for trustworthy agentic recommendation.} 
    As LLM-empowered RS evolve from single-shot predictors into autonomous agentic RS, trustworthiness becomes a long-term sequential decision-making problem rather than a static prediction problem~\cite{yao2022react,peng2025survey,shu2024rah}. Existing agentic RS rely on multi-turn interactions, memory updates, reasoning, and tool usage, yet their long-term trustworthiness properties remain poorly understood. 
    A promising direction is to introduce sequential decision-making frameworks, such as Markov Decision Processes (MDPs)~\cite{puterman1990markov}, Partially Observable MDPs (POMDPs)~\cite{hausknecht2015deep}, and reinforcement learning theory~\cite{li2017deep}, to formally model long-term trustworthiness objectives. 
    Furthermore, recent multi-agent RS involve interactions and collaborations among multiple autonomous agents~\cite{wang2024macrec}. This motivates the incorporation of game-theoretic~\cite{mesterton2019introduction} frameworks and mechanism design to analyze strategic agent behaviors and ensure that agent interactions lead to trustworthy outcomes.
\end{itemize}

\subsection{Efficiency in LLM Integration for Trustworthy Recommendation}
Efficiency is a fundamental requirement for RS, as high inference latency degrades user experience while excessive training overhead increases deployment and maintenance costs. This challenge becomes even more critical in trustworthy LLM-empowered RS. On the one hand, integrating LLMs introduces substantial computational overhead due to their large parameter scale and costly auto-regressive generation~\cite{liu2024deepseek}. On the other hand, many trustworthiness-enhancing techniques, such as fairness intervention~\cite{gao2025sprec}, explanation generation~\cite{yu2025explainable}, and privacy protection~\cite{li2026privacy}, further increase system complexity through additional modules or multi-stage pipelines. Therefore, trustworthy LLM-empowered RS face dual efficiency challenges: the high cost of LLM integration and the additional overhead introduced by trustworthiness mechanisms. Future work may address these challenges from two perspectives: training efficiency and inference efficiency.

\begin{itemize}
\item \textbf{Training efficiency.}  
Trustworthy RS require frequent updates to respond to newly identified risks, evolving user needs, and user feedback on problematic recommendations. 
However, repeatedly fine-tuning large LLMs is often prohibitively expensive, limiting the timely correction of trust-related failures. 
Future work should explore \emph{trust-aware parameter-efficient learning}, where trustworthiness constraints can be incorporated through lightweight adaptation (\eg LoRA~\cite{hu2022lora}) rather than full-model retraining. 
Another important direction is \emph{incremental and continual trust learning}~\cite{wang2024comprehensive}, which allows the system to update its behavior based on newly observed risks, user complaints, or regulatory requirements without catastrophic forgetting or repeated large-scale training. 

\item \textbf{Inference efficiency.}  
Inference efficiency is closely tied to the perceived reliability and usability of RS, especially in high-concurrency online settings. Meanwhile, trustworthy LLM-empowered RS often require extra inference-time modules, such as factual verification~\cite{lin2026verifiable}, and explanation generation~\cite{yu2025explainable}, which may substantially increase latency. 
Future work should develop \emph{cache-augmented trust-aware inference}. Systems can precompute reusable trust-related information offline, such as evidence snippets for factual verification and cached rationales for explanation generation. During online inference, the recommender can retrieve this information and combine it with lightweight ranking models, thereby reducing repeated LLM calls and lowering latency. A key challenge is to keep this information up to date and reliable as users, items, and external requirements change.
\end{itemize}

\subsection{Evaluation and Benchmarks for Trustworthy LLM-empowered Recommendation}
Existing benchmarks usually focus on specific tasks or isolated aspects of trustworthiness rather than providing a comprehensive evaluation framework. Moreover, some important properties, such as whether users genuinely perceive the system as trustworthy, are difficult to assess using only offline metrics~\cite{zhang2024large}. 
In addition, LLM-based RS can be highly sensitive to prompts and deployment-time distribution shifts, meaning that results on a single static benchmark may lead to overly optimistic conclusions~\cite{ebrat2024lusifer}. 
Future research should therefore focus on building evaluation frameworks that are comprehensive, human-centered, and aligned with real-world deployment.

\begin{itemize}
    \item \textbf{Comprehensive benchmarks.}
    Current benchmarks often focus on only a single trust-related metric, making it difficult to assess the overall trustworthiness of LLM-empowered RS.
    Future benchmarks should instead evaluate recommendation quality together with multiple trustworthiness dimensions in a unified framework.
    Such benchmarks should cover diverse domains, user groups, and interaction scenarios, while also considering prompt sensitivity, adversarial inputs, and distribution shifts.
    Rather than reporting a single leaderboard score, they should present multi-dimensional results and trade-offs, helping researchers understand whether improvements in recommendation performance come at the expense of trustworthiness.

	\item \textbf{Human-centered protocols.}
	Building human-centered evaluation protocols is particularly important for trustworthy LLM-empowered RS. 
	For example, explainability and controllability are inherently user-centered properties, and their quality cannot be fully measured by quantitative metrics alone. 
	Future evaluation protocols should therefore combine user studies with interactive tasks to assess whether users can understand the recommendations, feel in control of the system, and perceive the system as reliable, while also considering risks such as persuasion and over-reliance.

	\item \textbf{Online evaluation.}
	Offline benchmarks alone are often insufficient to capture the trustworthiness of LLM-empowered RS, since many trust-related issues only become visible during real user interactions and long-term deployment~\cite{ebrat2024lusifer}. 
	For example, users may gradually change their behaviors in response to the system, while the recommendation environment itself can evolve over time. 
	Future work should therefore place greater emphasis on \emph{online evaluation} for trustworthy LLM-empowered RS, including safer A/B testing protocols, interleaving-based comparisons, and safety-aware experimentation capable of detecting regressions in trust-related metrics before they lead to harmful outcomes. 
	Such online evaluation frameworks should also incorporate clear stop conditions, rollback mechanisms, and incident-response procedures to ensure that harmful behaviors can be mitigated rapidly once detected.
\end{itemize}


\section{Conclusion}
\label{sec:conclusion}
Given the pervasive societal impact of recommender systems, trustworthiness has become a central concern in their design and deployment. In recent years, the integration of LLMs has substantially reshaped RS, improving their performance while fundamentally transforming the landscape of trustworthy recommendation. In this survey, we provide a systematic review of trustworthy LLM-empowered recommendation from six key dimensions: robustness, bias and fairness, explainability, factuality, controllability, and privacy.
Across these dimensions, we examine the dual role of LLMs in trustworthy recommendation: they offer new opportunities for improving trustworthiness, while also introducing challenges such as hallucination, bias amplification, expanded attack surfaces, controllability failures, and privacy leakage.
In addition, we review existing evaluation protocols, including commonly used datasets and metrics, and discuss the limitations of current evaluation frameworks. 
Finally, we highlight several important directions for future research.
Overall, we hope this survey can provide researchers and practitioners with a structured understanding of trustworthy LLM-empowered recommendation, and encourage further research toward more trustworthy recommender systems.

\begin{acks}
This work is supported by the National Natural Science Foundation of China (62372399, 62476244), and the advanced computing resources provided by the Super Computing Center of Hangzhou City University.
\end{acks}

\bibliographystyle{ACM-Reference-Format}
\bibliography{sample-base}

@String{Computing = "Computing" }

@String{Computer = "{IEEE} Computer" }

@String{Springer = "Springer-Verlag" }

@article{chung2013betap,
  title={$\beta$P: A novel approach to filter out malicious rating profiles from recommender systems},
  author={Chung, Chen-Yao and Hsu, Ping-Yu and Huang, Shih-Hsiang},
  journal={Decision Support Systems},
  volume={55},
  number={1},
  pages={314--325},
  year={2013},
  publisher={Elsevier}
}

@article{nawara2024shilling,
  title={Shilling attacks and fake reviews injection: Principles, models, and datasets},
  author={Nawara, Dina and Aly, Ahmed and Kashef, Rasha},
  journal={IEEE Transactions on Computational Social Systems},
  year={2024},
  publisher={IEEE}
}

@article{chen2025llm,
  title={LLM-as-Critic: Contrastive and Adversarial Strategies for Authentic Text Verification},
  author={Chen, Wei and Chen, Dexin},
  year={2025}
}

@article{peng2025survey,
  title={A survey on llm-powered agents for recommender systems},
  author={Peng, Qiyao and Liu, Hongtao and Huang, Hua and Yang, Qing and Shao, Minglai},
  journal={arXiv preprint arXiv:2502.10050},
  year={2025}
}

@inproceedings{wang2025towards,
  title={Towards S$^2$-Challenges Underlying LLM-Based Augmentation for Personalized News Recommendation},
  author={Wang, Shicheng and Tang, Hengzhu and Gao, Li and Guo, Shu and Cheng, Suqi and Wang, Junfeng and Yin, Dawei and Liu, Tingwen and Wang, Lihong},
  booktitle={Proceedings of the AAAI Conference on Artificial Intelligence},
  volume={39},
  number={12},
  pages={12739--12747},
  year={2025}
}

@inproceedings{truong2021data,
  title={Data-free model extraction},
  author={Truong, Jean-Baptiste and Maini, Pratyush and Walls, Robert J and Papernot, Nicolas},
  booktitle={Proceedings of the IEEE/CVF conference on computer vision and pattern recognition},
  pages={4771--4780},
  year={2021}
}

@inproceedings{chiang2023shilling,
  title={Shilling black-box review-based recommender systems through fake review generation},
  author={Chiang, Hung-Yun and Chen, Yi-Syuan and Song, Yun-Zhu and Shuai, Hong-Han and Chang, Jason S},
  booktitle={Proceedings of the 29th ACM SIGKDD Conference on Knowledge Discovery and Data Mining},
  pages={286--297},
  year={2023}
}

@inproceedings{wu2023attacking,
  title={Attacking pre-trained recommendation},
  author={Wu, Yiqing and Xie, Ruobing and Zhang, Zhao and Zhu, Yongchun and Zhuang, Fuzhen and Zhou, Jie and Xu, Yongjun and He, Qing},
  booktitle={Proceedings of the 46th International ACM SIGIR Conference on Research and Development in Information Retrieval},
  pages={1811--1815},
  year={2023}
}

@article{zhao2025llm4mea,
  title={LLM4MEA: Data-free Model Extraction Attacks on Sequential Recommenders via Large Language Models},
  author={Zhao, Shilong and Sun, Fei and Zhang, Kaike and Jing, Shaoling and Su, Du and Shi, Zhichao and Yin, Zhiyi and Shen, Huawei and Cheng, Xueqi},
  journal={arXiv preprint arXiv:2507.16969},
  year={2025}
}

@article{gallegos2024bias,
  title={Bias and fairness in large language models: A survey},
  author={Gallegos, Isabel O and Rossi, Ryan A and Barrow, Joe and Tanjim, Md Mehrab and Kim, Sungchul and Dernoncourt, Franck and Yu, Tong and Zhang, Ruiyi and Ahmed, Nesreen K},
  journal={Computational Linguistics},
  volume={50},
  number={3},
  pages={1097--1179},
  year={2024},
  publisher={MIT Press 255 Main Street, 9th Floor, Cambridge, Massachusetts 02142, USA~…}
}

@article{li2024your,
  title={Your large language model is secretly a fairness proponent and you should prompt it like one},
  author={Li, Tianlin and Zhang, Xiaoyu and Du, Chao and Pang, Tianyu and Liu, Qian and Guo, Qing and Shen, Chao and Liu, Yang},
  journal={arXiv preprint arXiv:2402.12150},
  year={2024}
}

@inproceedings{schnabel2016recommendations,
  title={Recommendations as treatments: Debiasing learning and evaluation},
  author={Schnabel, Tobias and Swaminathan, Adith and Singh, Ashudeep and Chandak, Navin and Joachims, Thorsten},
  booktitle={international conference on machine learning},
  pages={1670--1679},
  year={2016},
  organization={PMLR}
}

@inproceedings{wang2021clicks,
  title={Clicks can be cheating: Counterfactual recommendation for mitigating clickbait issue},
  author={Wang, Wenjie and Feng, Fuli and He, Xiangnan and Zhang, Hanwang and Chua, Tat-Seng},
  booktitle={Proceedings of the 44th international ACM SIGIR conference on research and development in information retrieval},
  pages={1288--1297},
  year={2021}
}

@inproceedings{lin2023self,
  title={A self-correcting sequential recommender},
  author={Lin, Yujie and Wang, Chenyang and Chen, Zhumin and Ren, Zhaochun and Xin, Xin and Yan, Qiang and de Rijke, Maarten and Cheng, Xiuzhen and Ren, Pengjie},
  booktitle={Proceedings of the ACM Web Conference 2023},
  pages={1283--1293},
  year={2023}
}

@article{chen2023bias,
  title={Bias and debias in recommender system: A survey and future directions},
  author={Chen, Jiawei and Dong, Hande and Wang, Xiang and Feng, Fuli and Wang, Meng and He, Xiangnan},
  journal={ACM Transactions on Information Systems},
  volume={41},
  number={3},
  pages={1--39},
  year={2023},
  publisher={ACM New York, NY}
}

@article{morris2020textattack,
  title={Textattack: A framework for adversarial attacks, data augmentation, and adversarial training in nlp},
  author={Morris, John X and Lifland, Eli and Yoo, Jin Yong and Grigsby, Jake and Jin, Di and Qi, Yanjun},
  journal={arXiv preprint arXiv:2005.05909},
  year={2020}
}

@inproceedings{zhang2020gcn,
  title={Gcn-based user representation learning for unifying robust recommendation and fraudster detection},
  author={Zhang, Shijie and Yin, Hongzhi and Chen, Tong and Hung, Quoc Viet Nguyen and Huang, Zi and Cui, Lizhen},
  booktitle={Proceedings of the 43rd international ACM SIGIR conference on research and development in information retrieval},
  pages={689--698},
  year={2020}
}

@article{li2021caesar,
  title={Caesar: context-aware explanation based on supervised attention for service recommendations},
  author={Li, Lei and Chen, Li and Dong, Ruihai},
  journal={Journal of Intelligent Information Systems},
  volume={57},
  number={1},
  pages={147--170},
  year={2021},
  publisher={Springer}
}

@inproceedings{chen2021neural,
  title={Neural collaborative reasoning},
  author={Chen, Hanxiong and Shi, Shaoyun and Li, Yunqi and Zhang, Yongfeng},
  booktitle={Proceedings of the web conference 2021},
  pages={1516--1527},
  year={2021}
}

@inproceedings{shi2020neural,
  title={Neural logic reasoning},
  author={Shi, Shaoyun and Chen, Hanxiong and Ma, Weizhi and Mao, Jiaxin and Zhang, Min and Zhang, Yongfeng},
  booktitle={Proceedings of the 29th ACM International Conference on Information \& Knowledge Management},
  pages={1365--1374},
  year={2020}
}

@article{zhu2021faithfully,
  title={Faithfully explainable recommendation via neural logic reasoning},
  author={Zhu, Yaxin and Xian, Yikun and Fu, Zuohui and De Melo, Gerard and Zhang, Yongfeng},
  journal={arXiv preprint arXiv:2104.07869},
  year={2021}
}

@inproceedings{hernandez2020effects,
  title={Effects of argumentative explanation types on the perception of review-based recommendations},
  author={Hernandez-Bocanegra, Diana C and Donkers, Tim and Ziegler, J{\"u}rgen},
  booktitle={Adjunct Publication of the 28th ACM Conference on User Modeling, Adaptation and Personalization},
  pages={219--225},
  year={2020}
}

@inproceedings{musto2019justifying,
  title={Justifying recommendations through aspect-based sentiment analysis of users reviews},
  author={Musto, Cataldo and Lops, Pasquale and de Gemmis, Marco and Semeraro, Giovanni},
  booktitle={Proceedings of the 27th ACM conference on user modeling, adaptation and personalization},
  pages={4--12},
  year={2019}
}

@inproceedings{li2017neural,
  title={Neural attentive session-based recommendation},
  author={Li, Jing and Ren, Pengjie and Chen, Zhumin and Ren, Zhaochun and Lian, Tao and Ma, Jun},
  booktitle={Proceedings of the 2017 ACM on Conference on Information and Knowledge Management},
  pages={1419--1428},
  year={2017}
}

@inproceedings{geng2022recommendation,
  title={Recommendation as language processing (rlp): A unified pretrain, personalized prompt \& predict paradigm (p5)},
  author={Geng, Shijie and Liu, Shuchang and Fu, Zuohui and Ge, Yingqiang and Zhang, Yongfeng},
  booktitle={Proceedings of the 16th ACM conference on recommender systems},
  pages={299--315},
  year={2022}
}

@inproceedings{wang2024can,
  title={Can small language models be good reasoners for sequential recommendation?},
  author={Wang, Yuling and Tian, Changxin and Hu, Binbin and Yu, Yanhua and Liu, Ziqi and Zhang, Zhiqiang and Zhou, Jun and Pang, Liang and Wang, Xiao},
  booktitle={Proceedings of the ACM Web Conference 2024},
  pages={3876--3887},
  year={2024}
}

@inproceedings{park2023user,
  title={A user preference and intent extraction framework for explainable conversational recommender systems},
  author={Park, Jieun and Kim, Sangyeon and Lee, Sangwon},
  booktitle={Companion Proceedings of the 2023 ACM SIGCHI Symposium on Engineering Interactive Computing Systems},
  pages={16--23},
  year={2023}
}

@article{wang2025blessing,
  title={The Blessing of Reasoning: LLM-Based Contrastive Explanations in Black-Box Recommender Systems},
  author={Wang, Yuyan and Li, Pan and Chen, Minmin},
  journal={arXiv preprint arXiv:2502.16759},
  year={2025}
}

@article{lyu2023llm,
  title={Llm-rec: Personalized recommendation via prompting large language models},
  author={Lyu, Hanjia and Jiang, Song and Zeng, Hanqing and Xia, Yinglong and Wang, Qifan and Zhang, Si and Chen, Ren and Leung, Christopher and Tang, Jiajie and Luo, Jiebo},
  journal={arXiv preprint arXiv:2307.15780},
  year={2023}
}

@article{jia2025improving,
  title={Improving LLM Interpretability and Performance via Guided Embedding Refinement for Sequential Recommendation},
  author={Jia, Nanshan and Yuan, Chenfei and Wu, Yuhang and Zheng, Zeyu},
  journal={arXiv preprint arXiv:2504.11658},
  year={2025}
}

@article{yu2025thinkrec,
  title={ThinkRec: Thinking-based recommendation via LLM},
  author={Yu, Qihang and Fu, Kairui and Zhang, Shengyu and Lv, Zheqi and Wu, Fan and Wu, Fei},
  journal={arXiv preprint arXiv:2505.15091},
  year={2025}
}

@article{zhang2025reinforced,
  title={Reinforced Latent Reasoning for LLM-based Recommendation},
  author={Zhang, Yang and Xu, Wenxin and Zhao, Xiaoyan and Wang, Wenjie and Feng, Fuli and He, Xiangnan and Chua, Tat-Seng},
  journal={arXiv preprint arXiv:2505.19092},
  year={2025}
}

@article{zhao2025reason,
  title={Reason-to-Recommend: Using Interaction-of-Thought Reasoning to Enhance LLM Recommendation},
  author={Zhao, Keyu and Xu, Fengli and Li, Yong},
  journal={arXiv preprint arXiv:2506.05069},
  year={2025}
}

@article{li2024learning,
  title={Learning Structure and Knowledge Aware Representation with Large Language Models for Concept Recommendation},
  author={Li, Qingyao and Xia, Wei and Du, Kounianhua and Zhang, Qiji and Zhang, Weinan and Tang, Ruiming and Yu, Yong},
  journal={arXiv preprint arXiv:2405.12442},
  year={2024}
}

@article{qiu2024unveiling,
  title={Unveiling User Preferences: A Knowledge Graph and LLM-Driven Approach for Conversational Recommendation},
  author={Qiu, Zhangchi and Luo, Linhao and Pan, Shirui and Liew, Alan Wee-Chung},
  journal={arXiv preprint arXiv:2411.14459},
  year={2024}
}

@inproceedings{geng2022path,
  title={Path language modeling over knowledge graphsfor explainable recommendation},
  author={Geng, Shijie and Fu, Zuohui and Tan, Juntao and Ge, Yingqiang and De Melo, Gerard and Zhang, Yongfeng},
  booktitle={Proceedings of the ACM web conference 2022},
  pages={946--955},
  year={2022}
}

@article{balloccu2023faithful,
  title={Faithful Path Language Modeling for Explainable Recommendation over Knowledge Graph},
  author={Balloccu, Giacomo and Boratto, Ludovico and Cancedda, Christian and Fenu, Gianni and Marras, Mirko},
  journal={arXiv preprint arXiv:2310.16452},
  year={2023}
}

@inproceedings{zhang2023user,
  title={User-centric conversational recommendation: Adapting the need of user with large language models},
  author={Zhang, Gangyi},
  booktitle={Proceedings of the 17th ACM Conference on Recommender Systems},
  pages={1349--1354},
  year={2023}
}

@article{zheng2025explain,
  title={Explain What You Mean: Intent Augmented Knowledge Graph Recommender Built With LLM},
  author={Zheng, Wenqing and Fatsi, Noah and Barcklow, Daniel and Kalaev, Dmitri and Yao, Steven and Reinert, Owen and Bruss, C Bayan and Rosa, Daniele},
  journal={arXiv preprint arXiv:2505.10900},
  year={2025}
}

@article{wang2024enabling,
  title={Enabling Explainable Recommendation in E-commerce with LLM-powered Product Knowledge Graph},
  author={Wang, Menghan and Guo, Yuchen and Zhang, Duanfeng and Jin, Jianian and Li, Minnie and Schonfeld, Dan and Zhou, Shawn},
  journal={arXiv preprint arXiv:2412.01837},
  year={2024}
}

@inproceedings{wang2024llmrg,
  title={Llmrg: Improving recommendations through large language model reasoning graphs},
  author={Wang, Yan and Chu, Zhixuan and Ouyang, Xin and Wang, Simeng and Hao, Hongyan and Shen, Yue and Gu, Jinjie and Xue, Siqiao and Zhang, James and Cui, Qing and others},
  booktitle={Proceedings of the AAAI Conference on Artificial Intelligence},
  volume={38},
  number={17},
  pages={19189--19196},
  year={2024}
}

@article{shi2024llm,
  title={Llm-powered explanations: Unraveling recommendations through subgraph reasoning},
  author={Shi, Guangsi and Deng, Xiaofeng and Luo, Linhao and Xia, Lijuan and Bao, Lei and Ye, Bei and Du, Fei and Pan, Shirui and Li, Yuxiao},
  journal={arXiv preprint arXiv:2406.15859},
  year={2024}
}

@article{gao2023chat,
  title={Chat-rec: Towards interactive and explainable llms-augmented recommender system},
  author={Gao, Yunfan and Sheng, Tao and Xiang, Youlin and Xiong, Yun and Wang, Haofen and Zhang, Jiawei},
  journal={arXiv preprint arXiv:2303.14524},
  year={2023}
}

@inproceedings{li2025g,
  title={G-Refer: Graph Retrieval-Augmented Large Language Model for Explainable Recommendation},
  author={Li, Yuhan and Zhang, Xinni and Luo, Linhao and Chang, Heng and Ren, Yuxiang and King, Irwin and Li, Jia},
  booktitle={Proceedings of the ACM on Web Conference 2025},
  pages={240--251},
  year={2025}
}

@inproceedings{yu2025explainable,
  title={Explainable ctr prediction via llm reasoning},
  author={Yu, Xiaohan and Zhang, Li and Chen, Chong},
  booktitle={Proceedings of the Eighteenth ACM International Conference on Web Search and Data Mining},
  pages={707--716},
  year={2025}
}

@article{zhang2024navigating,
  title={Navigating user experience of chatgpt-based conversational recommender systems: The effects of prompt guidance and recommendation domain},
  author={Zhang, Yizhe and Jin, Yucheng and Chen, Li and Yang, Ting},
  journal={arXiv preprint arXiv:2405.13560},
  year={2024}
}

@inproceedings{luo2024unlocking,
  title={Unlocking the potential of large language models for explainable recommendations},
  author={Luo, Yucong and Cheng, Mingyue and Zhang, Hao and Lu, Junyu and Chen, Enhong},
  booktitle={International Conference on Database Systems for Advanced Applications},
  pages={286--303},
  year={2024},
  organization={Springer}
}

@inproceedings{yang2024fine,
  title={Fine-tuning large language model based explainable recommendation with explainable quality reward},
  author={Yang, Mengyuan and Zhu, Mengying and Wang, Yan and Chen, Linxun and Zhao, Yilei and Wang, Xiuyuan and Han, Bing and Zheng, Xiaolin and Yin, Jianwei},
  booktitle={Proceedings of the AAAI Conference on Artificial Intelligence},
  volume={38},
  number={8},
  pages={9250--9259},
  year={2024}
}

@inproceedings{hada2021rexplug,
  title={Rexplug: Explainable recommendation using plug-and-play language model},
  author={Hada, Deepesh V and Shevade, Shirish K},
  booktitle={Proceedings of the 44th international ACM SIGIR conference on research and development in information retrieval},
  pages={81--91},
  year={2021}
}

@inproceedings{wang2023llm4vis,
  title={LLM4Vis: Explainable Visualization Recommendation using ChatGPT},
  author={Wang, Lei and Zhang, Songheng and Wang, Yun and Lim, Ee-Peng and Wang, Yong},
  booktitle={Proceedings of the 2023 Conference on Empirical Methods in Natural Language Processing: Industry Track},
  pages={675--692},
  year={2023}
}

@article{peng2024uncertainty,
  title={Uncertainty-Aware Explainable Recommendation with Large Language Models},
  author={Peng, Yicui and Chen, Hao and Lin, Chingsheng and Huang, Guo and Hu, Jinrong and Guo, Hui and Kong, Bin and Hu, Shu and Wu, Xi and Wang, Xin},
  journal={arXiv preprint arXiv:2402.03366},
  year={2024}
}

@article{liu2023llmrec,
  title={Llmrec: Benchmarking large language models on recommendation task},
  author={Liu, Junling and Liu, Chao and Zhou, Peilin and Ye, Qichen and Chong, Dading and Zhou, Kang and Xie, Yueqi and Cao, Yuwei and Wang, Shoujin and You, Chenyu and others},
  journal={arXiv preprint arXiv:2308.12241},
  year={2023}
}

@article{ma2024xrec,
  title={Xrec: Large language models for explainable recommendation},
  author={Ma, Qiyao and Ren, Xubin and Huang, Chao},
  journal={arXiv preprint arXiv:2406.02377},
  year={2024}
}

@inproceedings{petruzzelli2024instructing,
  title={Instructing and prompting large language models for explainable cross-domain recommendations},
  author={Petruzzelli, Alessandro and Musto, Cataldo and Laraspata, Lucrezia and Rinaldi, Ivan and de Gemmis, Marco and Lops, Pasquale and Semeraro, Giovanni},
  booktitle={Proceedings of the 18th ACM Conference on Recommender Systems},
  pages={298--308},
  year={2024}
}

@inproceedings{lei2024recexplainer,
  title={Recexplainer: Aligning large language models for explaining recommendation models},
  author={Lei, Yuxuan and Lian, Jianxun and Yao, Jing and Huang, Xu and Lian, Defu and Xie, Xing},
  booktitle={Proceedings of the 30th ACM SIGKDD Conference on Knowledge Discovery and Data Mining},
  pages={1530--1541},
  year={2024}
}

@article{gao2024dre,
  title={DRE: Generating Recommendation Explanations by Aligning Large Language Models at Data-level},
  author={Gao, Shen and Wang, Yifan and Fang, Jiabao and Chen, Lisi and Han, Peng and Shang, Shuo},
  journal={arXiv preprint arXiv:2404.06311},
  year={2024}
}

@inproceedings{feng2024move,
  title={Where to move next: Zero-shot generalization of llms for next poi recommendation},
  author={Feng, Shanshan and Lyu, Haoming and Li, Fan and Sun, Zhu and Chen, Caishun},
  booktitle={2024 IEEE Conference on Artificial Intelligence (CAI)},
  pages={1530--1535},
  year={2024},
  organization={IEEE}
}

@inproceedings{silva2024leveraging,
  title={Leveraging chatgpt for automated human-centered explanations in recommender systems},
  author={Silva, {\'I}tallo and Marinho, Leandro and Said, Alan and Willemsen, Martijn C},
  booktitle={Proceedings of the 29th International Conference on Intelligent User Interfaces},
  pages={597--608},
  year={2024}
}

@inproceedings{rahdari2024logic,
  title={Logic-scaffolding: Personalized aspect-instructed recommendation explanation generation using llms},
  author={Rahdari, Behnam and Ding, Hao and Fan, Ziwei and Ma, Yifei and Chen, Zhuotong and Deoras, Anoop and Kveton, Branislav},
  booktitle={Proceedings of the 17th ACM International Conference on Web Search and Data Mining},
  pages={1078--1081},
  year={2024}
}

@article{friedman2023leveraging,
  title={Leveraging large language models in conversational recommender systems},
  author={Friedman, Luke and Ahuja, Sameer and Allen, David and Tan, Zhenning and Sidahmed, Hakim and Long, Changbo and Xie, Jun and Schubiner, Gabriel and Patel, Ajay and Lara, Harsh and others},
  journal={arXiv preprint arXiv:2305.07961},
  year={2023}
}

@article{zhou2023gpt,
  title={Gpt as a baseline for recommendation explanation texts},
  author={Zhou, Joyce and Joachims, Thorsten},
  journal={arXiv preprint arXiv:2309.08817},
  year={2023}
}

@article{ghosh2023jobrecogpt,
  title={JobRecoGPT--Explainable job recommendations using LLMs},
  author={Ghosh, Preetam and Sadaphal, Vaishali},
  journal={arXiv preprint arXiv:2309.11805},
  year={2023}
}

@inproceedings{abu2024knowledge,
  title={Knowledge graphs as context sources for llm-based explanations of learning recommendations},
  author={Abu-Rasheed, Hasan and Weber, Christian and Fathi, Madjid},
  booktitle={2024 IEEE Global Engineering Education Conference (EDUCON)},
  pages={1--5},
  year={2024},
  organization={IEEE}
}

@inproceedings{li2025alert,
  title={ALERT: An LLM-powered Benchmark for Automatic Evaluation of Recommendation Explanations},
  author={Li, Yichuan and Zhang, Xinyang and Zhang, Chenwei and Li, Mao and Liu, Tianyi and Chen, Pei and Gao, Yifan and Lee, Kyumin and Ding, Kaize and Wang, Zhengyang and others},
  booktitle={Proceedings of the 2025 Conference of the Nations of the Americas Chapter of the Association for Computational Linguistics: Human Language Technologies (Volume 1: Long Papers)},
  pages={2704--2719},
  year={2025}
}

@inproceedings{zhang2024large,
  title={Large language models as evaluators for recommendation explanations},
  author={Zhang, Xiaoyu and Li, Yishan and Wang, Jiayin and Sun, Bowen and Ma, Weizhi and Sun, Peijie and Zhang, Min},
  booktitle={Proceedings of the 18th ACM Conference on Recommender Systems},
  pages={33--42},
  year={2024}
}

@inproceedings{shimizu2025disentangling,
  title={Disentangling likes and dislikes in personalized generative explainable recommendation},
  author={Shimizu, Ryotaro and Wada, Takashi and Wang, Yu and Kruse, Johannes and O'Brien, Sean and HtaungKham, Sai and Song, Linxin and Yoshikawa, Yuya and Saito, Yuki and Tsung, Fugee and others},
  booktitle={Proceedings of the ACM on Web Conference 2025},
  pages={4793--4809},
  year={2025}
}

@inproceedings{bao2024decoding,
  title={Decoding matters: Addressing amplification bias and homogeneity issue in recommendations for large language models},
  author={Bao, Keqin and Zhang, Jizhi and Zhang, Yang and Huo, Xinyue and Chen, Chong and Feng, Fuli},
  booktitle={Proceedings of the 2024 Conference on Empirical Methods in Natural Language Processing},
  pages={10540--10552},
  year={2024}
}

@article{yang2026bear,
  title={BEAR: Towards Beam-Search-Aware Optimization for Recommendation with Large Language Models},
  author={Yang, Weiqin and Wang, Bohao and Xu, Zhenxiang and Chen, Jiawei and Zhang, Shengjia and Chen, Jingbang and Jin, Canghong and Wang, Can},
  journal={arXiv preprint arXiv:2601.22925},
  year={2026}
}

@inproceedings{lazovich2023filter,
  title={Filter bubbles and affective polarization in user-personalized large language model outputs},
  author={Lazovich, Tomo},
  booktitle={Proceedings on},
  pages={29--37},
  year={2023},
  organization={PMLR}
}

@inproceedings{maes2025mitigating,
  title={Mitigating Misleadingness in LLM-Generated Natural Language Explanations for Recommender Systems: Ensuring Broad Truthfulness Through Factuality and Faithfulness},
  author={Maes, Ulysse and Michiels, Lien and Smets, Annelien},
  booktitle={Proceedings of the 29th Annual ACM Conference on Intelligent User Interfaces},
  year={2025}
}

@inproceedings{wei2024llmrec,
  title={Llmrec: Large language models with graph augmentation for recommendation},
  author={Wei, Wei and Ren, Xubin and Tang, Jiabin and Wang, Qinyong and Su, Lixin and Cheng, Suqi and Wang, Junfeng and Yin, Dawei and Huang, Chao},
  booktitle={Proceedings of the 17th ACM International Conference on Web Search and Data Mining},
  pages={806--815},
  year={2024}
}

@article{huang2025survey,
  title={A survey on hallucination in large language models: Principles, taxonomy, challenges, and open questions},
  author={Huang, Lei and Yu, Weijiang and Ma, Weitao and Zhong, Weihong and Feng, Zhangyin and Wang, Haotian and Chen, Qianglong and Peng, Weihua and Feng, Xiaocheng and Qin, Bing and others},
  journal={ACM Transactions on Information Systems},
  volume={43},
  number={2},
  pages={1--55},
  year={2025},
  publisher={ACM New York, NY}
}

@inproceedings{hua2023index,
  title={How to index item ids for recommendation foundation models},
  author={Hua, Wenyue and Xu, Shuyuan and Ge, Yingqiang and Zhang, Yongfeng},
  booktitle={Proceedings of the Annual International ACM SIGIR Conference on Research and Development in Information Retrieval in the Asia Pacific Region},
  pages={195--204},
  year={2023}
}

@inproceedings{hong2025eager,
  title={EAGER-LLM: Enhancing Large Language Models as Recommenders through Exogenous Behavior-Semantic Integration},
  author={Hong, Minjie and Xia, Yan and Wang, Zehan and Zhu, Jieming and Wang, Ye and Cai, Sihang and Yang, Xiaoda and Dai, Quanyu and Dong, Zhenhua and Zhang, Zhimeng and others},
  booktitle={Proceedings of the ACM on Web Conference 2025},
  pages={2754--2762},
  year={2025}
}

@inproceedings{zhu2024collaborative,
  title={Collaborative large language model for recommender systems},
  author={Zhu, Yaochen and Wu, Liang and Guo, Qi and Hong, Liangjie and Li, Jundong},
  booktitle={Proceedings of the ACM Web Conference 2024},
  pages={3162--3172},
  year={2024}
}

@article{li2023e4srec,
  title={E4srec: An elegant effective efficient extensible solution of large language models for sequential recommendation},
  author={Li, Xinhang and Chen, Chong and Zhao, Xiangyu and Zhang, Yong and Xing, Chunxiao},
  journal={arXiv preprint arXiv:2312.02443},
  year={2023}
}

@inproceedings{xu2025slmrec,
  title={SLMRec: Distilling large language models into small for sequential recommendation},
  author={Xu, Wujiang and Wu, Qitian and Liang, Zujie and Han, Jiaojiao and Ning, Xuying and Shi, Yunxiao and Lin, Wenfang and Zhang, Yongfeng},
  booktitle={The Thirteenth International Conference on Learning Representations},
  year={2025}
}

@article{wang2024rethinking,
  title={Rethinking large language model architectures for sequential recommendations},
  author={Wang, Hanbing and Liu, Xiaorui and Fan, Wenqi and Zhao, Xiangyu and Kini, Venkataramana and Yadav, Devendra and Wang, Fei and Wen, Zhen and Tang, Jiliang and Liu, Hui},
  journal={arXiv preprint arXiv:2402.09543},
  year={2024}
}

@article{qu2024tokenrec,
  title={Tokenrec: learning to tokenize id for llm-based generative recommendation},
  author={Qu, Haohao and Fan, Wenqi and Zhao, Zihuai and Li, Qing},
  journal={arXiv preprint arXiv:2406.10450},
  year={2024}
}

@article{bao2025bi,
  title={A bi-step grounding paradigm for large language models in recommendation systems},
  author={Bao, Keqin and Zhang, Jizhi and Wang, Wenjie and Zhang, Yang and Yang, Zhengyi and Luo, Yanchen and Chen, Chong and Feng, Fuli and Tian, Qi},
  journal={ACM Transactions on Recommender Systems},
  volume={3},
  number={4},
  pages={1--27},
  year={2025},
  publisher={ACM New York, NY}
}

@inproceedings{wang2025msl,
  title={Msl: Not all tokens are what you need for tuning llm as a recommender},
  author={Wang, Bohao and Liu, Feng and Chen, Jiawei and Lou, Xingyu and Zhang, Changwang and Wang, Jun and Sun, Yuegang and Feng, Yan and Chen, Chun and Wang, Can},
  booktitle={Proceedings of the 48th international ACM SIGIR conference on research and development in information retrieval},
  pages={1912--1922},
  year={2025}
}

@inproceedings{liao2024llara,
  title={Llara: Large language-recommendation assistant},
  author={Liao, Jiayi and Li, Sihang and Yang, Zhengyi and Wu, Jiancan and Yuan, Yancheng and Wang, Xiang and He, Xiangnan},
  booktitle={Proceedings of the 47th International ACM SIGIR Conference on Research and Development in Information Retrieval},
  pages={1785--1795},
  year={2024}
}

@inproceedings{kim2024large,
  title={Large language models meet collaborative filtering: An efficient all-round llm-based recommender system},
  author={Kim, Sein and Kang, Hongseok and Choi, Seungyoon and Kim, Donghyun and Yang, Minchul and Park, Chanyoung},
  booktitle={Proceedings of the 30th ACM SIGKDD Conference on Knowledge Discovery and Data Mining},
  pages={1395--1406},
  year={2024}
}

@article{yue2023llamarec,
  title={Llamarec: Two-stage recommendation using large language models for ranking},
  author={Yue, Zhenrui and Rabhi, Sara and Moreira, Gabriel de Souza Pereira and Wang, Dong and Oldridge, Even},
  journal={arXiv preprint arXiv:2311.02089},
  year={2023}
}

@inproceedings{jiang2025beyond,
  title={Beyond Utility: Evaluating LLM as Recommender},
  author={Jiang, Chumeng and Wang, Jiayin and Ma, Weizhi and Clarke, Charles LA and Wang, Shuai and Wu, Chuhan and Zhang, Min},
  booktitle={Proceedings of the ACM on Web Conference 2025},
  pages={3850--3862},
  year={2025}
}

@article{wang2025llm4dsr,
  title={Llm4dsr: Leveraging large language model for denoising sequential recommendation},
  author={Wang, Bohao and Liu, Feng and Zhang, Changwang and Chen, Jiawei and Wu, Yudi and Zhou, Sheng and Lou, Xingyu and Wang, Jun and Feng, Yan and Chen, Chun and others},
  journal={ACM Transactions on Information Systems},
  volume={44},
  number={1},
  pages={1--32},
  year={2025},
  publisher={ACM New York, NY}
}

@article{sun2025llmser,
  title={LLMSeR: Enhancing Sequential Recommendation via LLM-based Data Augmentation},
  author={Sun, Yuqi and Liu, Qidong and Zhu, Haiping and Tian, Feng},
  journal={arXiv preprint arXiv:2503.12547},
  year={2025}
}

@article{liao2024rosepo,
  title={RosePO: Aligning LLM-based Recommenders with Human Values},
  author={Liao, Jiayi and He, Xiangnan and Xie, Ruobing and Wu, Jiancan and Yuan, Yancheng and Sun, Xingwu and Kang, Zhanhui and Wang, Xiang},
  journal={arXiv preprint arXiv:2410.12519},
  year={2024}
}

@article{ning2025retrieval,
  title={Retrieval-Augmented Purifier for Robust LLM-Empowered Recommendation},
  author={Ning, Liangbo and Fan, Wenqi and Li, Qing},
  journal={arXiv preprint arXiv:2504.02458},
  year={2025}
}

@article{wang2025knowledge,
  title={Knowledge Graph Retrieval-Augmented Generation for LLM-based Recommendation},
  author={Wang, Shijie and Fan, Wenqi and Feng, Yue and Ma, Xinyu and Wang, Shuaiqiang and Yin, Dawei},
  journal={arXiv preprint arXiv:2501.02226},
  year={2025}
}

@inproceedings{jiao2025retrieval,
  title={Retrieval and Structuring Augmented Generation with LLMs for Web Applications},
  author={Jiao, Yizhu and Ouyang, Siru and Zhong, Ming and Zhang, Yunyi and Ding, Linyi and Zhou, Sizhe and Han, Jiawei},
  booktitle={Companion Proceedings of the ACM on Web Conference 2025},
  pages={25--28},
  year={2025}
}

@article{hou2024enhancing,
  title={Enhancing Dietary Supplement Question Answer via Retrieval-Augmented Generation (RAG) with LLM},
  author={Hou, Yu and Zhang, Rui},
  journal={medRxiv},
  pages={2024--09},
  year={2024},
  publisher={Cold Spring Harbor Laboratory Press}
}

@inproceedings{el2024optimizing,
  title={Optimizing Recommendation Systems in E-Learning: Synergistic Integration of Lang Chain, GPT Models, and Retrieval Augmented Generation (RAG)},
  author={EL Maazouzi, Qamar and Retbi, Asma{\^a} and Bennani, Samir},
  booktitle={International Conference on Smart Applications and Data Analysis},
  pages={106--118},
  year={2024},
  organization={Springer}
}

@inproceedings{di2023retrieval,
  title={Retrieval-augmented recommender system: Enhancing recommender systems with large language models},
  author={Di Palma, Dario},
  booktitle={Proceedings of the 17th ACM Conference on Recommender Systems},
  pages={1369--1373},
  year={2023}
}

@article{kaur2026efficient,
  title={Efficient and responsible adaptation of large language models for robust top-k recommendations},
  author={Kaur, Kirandeep and Gupta, Vinayak and Chadha, Manya and Shah, Chirag},
  journal={ACM Transactions on Recommender Systems},
  volume={4},
  number={3},
  pages={1--31},
  year={2026},
  publisher={ACM New York, NY}
}

@inproceedings{zhao2025federated,
  title={A Federated Framework for LLM-based Recommendation},
  author={Zhao, Jujia and Wang, Wenjie and Xu, Chen and Ng, See Kiong and Chua, Tat-Seng},
  booktitle={Findings of the Association for Computational Linguistics: NAACL 2025},
  pages={2852--2865},
  year={2025}
}

@inproceedings{bao2023tallrec,
  title={Tallrec: An effective and efficient tuning framework to align large language model with recommendation},
  author={Bao, Keqin and Zhang, Jizhi and Zhang, Yang and Wang, Wenjie and Feng, Fuli and He, Xiangnan},
  booktitle={Proceedings of the 17th ACM Conference on Recommender Systems},
  pages={1007--1014},
  year={2023}
}

@article{cui2022m6,
  title={M6-rec: Generative pretrained language models are open-ended recommender systems},
  author={Cui, Zeyu and Ma, Jianxin and Zhou, Chang and Zhou, Jingren and Yang, Hongxia},
  journal={arXiv preprint arXiv:2205.08084},
  year={2022}
}

@inproceedings{li2023prompt,
  title={Prompt distillation for efficient llm-based recommendation},
  author={Li, Lei and Zhang, Yongfeng and Chen, Li},
  booktitle={Proceedings of the 32nd ACM International Conference on Information and Knowledge Management},
  pages={1348--1357},
  year={2023}
}

@article{wang2025towards1,
  title={Towards efficient and effective unlearning of large language models for recommendation},
  author={Wang, Hangyu and Lin, Jianghao and Chen, Bo and Yang, Yang and Tang, Ruiming and Zhang, Weinan and Yu, Yong},
  journal={Frontiers of Computer Science},
  volume={19},
  number={3},
  pages={193327},
  year={2025},
  publisher={Springer}
}

@inproceedings{cui2024distillation,
  title={Distillation matters: empowering sequential recommenders to match the performance of large language models},
  author={Cui, Yu and Liu, Feng and Wang, Pengbo and Wang, Bohao and Tang, Heng and Wan, Yi and Wang, Jun and Chen, Jiawei},
  booktitle={Proceedings of the 18th ACM Conference on Recommender Systems},
  pages={507--517},
  year={2024}
}

@inproceedings{sun2024large,
  title={Large language models enhanced collaborative filtering},
  author={Sun, Zhongxiang and Si, Zihua and Zang, Xiaoxue and Zheng, Kai and Song, Yang and Zhang, Xiao and Xu, Jun},
  booktitle={Proceedings of the 33rd ACM International Conference on Information and Knowledge Management},
  pages={2178--2188},
  year={2024}
}

@inproceedings{wang2024rdrec,
  title={RDRec: Rationale Distillation for LLM-based Recommendation},
  author={Wang, Xinfeng and Cui, Jin and Suzuki, Yoshimi and Fukumoto, Fumiyo},
  booktitle={Proceedings of the 62nd Annual Meeting of the Association for Computational Linguistics (Volume 2: Short Papers)},
  pages={65--74},
  year={2024}
}

@inproceedings{sanner2023large,
  title={Large language models are competitive near cold-start recommenders for language-and item-based preferences},
  author={Sanner, Scott and Balog, Krisztian and Radlinski, Filip and Wedin, Ben and Dixon, Lucas},
  booktitle={Proceedings of the 17th ACM conference on recommender systems},
  pages={890--896},
  year={2023}
}

@inproceedings{wang2024large,
  title={Large language models as data augmenters for cold-start item recommendation},
  author={Wang, Jianling and Lu, Haokai and Caverlee, James and Chi, Ed H and Chen, Minmin},
  booktitle={Companion Proceedings of the ACM Web Conference 2024},
  pages={726--729},
  year={2024}
}

@inproceedings{huang2025large,
  title={Large Language Model Simulator for Cold-Start Recommendation},
  author={Huang, Feiran and Bei, Yuanchen and Yang, Zhenghang and Jiang, Junyi and Chen, Hao and Shen, Qijie and Wang, Senzhang and Karray, Fakhri and Yu, Philip S},
  booktitle={Proceedings of the Eighteenth ACM International Conference on Web Search and Data Mining},
  pages={261--270},
  year={2025}
}

@article{liu2025filterllm,
  title={FilterLLM: Text-To-Distribution LLM for Billion-Scale Cold-Start Recommendation},
  author={Liu, Ruochen and Chen, Hao and Bei, Yuanchen and Zhou, Zheyu and Chen, Lijia and Shen, Qijie and Huang, Feiran and Karray, Fakhri and Wang, Senzhang},
  journal={arXiv preprint arXiv:2502.16924},
year={2025}
}

@inproceedings{kusano2024data,
  title={Data Augmentation using Reverse Prompt for Cost-Efficient Cold-Start Recommendation},
  author={Kusano, Genki},
  booktitle={Proceedings of the 18th ACM Conference on Recommender Systems},
  pages={861--865},
  year={2024}
}

@inproceedings{rungtranont2024using,
  title={Using Large Language Models as user Interests Interpretation for Solving Cold-Start Item Recommendation},
  author={Rungtranont, Panjawat and Mongkolnavin, Janjao},
  booktitle={2024 23rd International Symposium on Communications and Information Technologies (ISCIT)},
  pages={36--40},
  year={2024},
  organization={IEEE}
}

@inproceedings{yuan2023go,
  title={Where to go next for recommender systems? id-vs. modality-based recommender models revisited},
  author={Yuan, Zheng and Yuan, Fajie and Song, Yu and Li, Youhua and Fu, Junchen and Yang, Fei and Pan, Yunzhu and Ni, Yongxin},
  booktitle={Proceedings of the 46th International ACM SIGIR Conference on Research and Development in Information Retrieval},
  pages={2639--2649},
  year={2023}
}

@article{zhang2025llminit,
  title={Llminit: A free lunch from large language models for selective initialization of recommendation},
  author={Zhang, Weizhi and Yang, Liangwei and Yang, Wooseong and Zou, Henry Peng and Liu, Yuqing and Xu, Ke and Medya, Sourav and Yu, Philip S},
  journal={arXiv preprint arXiv:2503.01814},
  year={2025}
}

@inproceedings{gong2023unified,
  title={An unified search and recommendation foundation model for cold-start scenario},
  author={Gong, Yuqi and Ding, Xichen and Su, Yehui and Shen, Kaiming and Liu, Zhongyi and Zhang, Guannan},
  booktitle={Proceedings of the 32nd ACM International Conference on Information and Knowledge Management},
  pages={4595--4601},
  year={2023}
}

@article{li2023integrating,
  title={Integrating Prior Knowledge from Meta-Learning and Large Language Models for Cold-Start Recommendation},
  author={Li, Yu and Liu, Yixiao and Furukawa, Tetsuya},
  year={2023},
  publisher={Interdisciplinary Graduate School of Engineering Sciences, Kyushu University}
}

@inproceedings{lu2024aligning,
  title={Aligning large language models for controllable recommendations},
  author={Lu, Wensheng and Lian, Jianxun and Zhang, Wei and Li, Guanghua and Zhou, Mingyang and Liao, Hao and Xie, Xing},
  booktitle={Proceedings of the 62nd Annual Meeting of the Association for Computational Linguistics (Volume 1: Long Papers)},
  pages={8159--8172},
  year={2024}
}

@inproceedings{wozniak2025improving,
  title={Improving LLM-Based Recommender Systems with User-Controllable Profiles},
  author={Wo{\'z}niak, Stanis{\l}aw and Duszenko, Jacek and Koco{\'n}, Jan and Kazienko, Przemysaw},
  booktitle={Companion Proceedings of the ACM on Web Conference 2025},
  pages={2102--2111},
  year={2025}
}

@inproceedings{chen2025dlcrec,
  title={DLCRec: A Novel Approach for Managing Diversity in LLM-Based Recommender Systems},
  author={Chen, Jiaju and Gao, Chongming and Yuan, Shuai and Liu, Shuchang and Cai, Qingpeng and Jiang, Peng},
  booktitle={Proceedings of the Eighteenth ACM International Conference on Web Search and Data Mining},
  pages={857--865},
  year={2025}
}

@article{shu2024rah,
  title={RAH! RecSys--Assistant--Human: A Human-Centered Recommendation Framework With LLM Agents},
  author={Shu, Yubo and Zhang, Haonan and Gu, Hansu and Zhang, Peng and Lu, Tun and Li, Dongsheng and Gu, Ning},
  journal={IEEE Transactions on Computational Social Systems},
  year={2024},
  publisher={IEEE}
}

@inproceedings{carroll2025ctrl,
  title={CTRL-Rec: Controlling Recommender Systems With Natural Language},
  author={Carroll, Micah and Foote, Adeline and Williams, Marcus and Dragan, Anca and Knox, W Bradley and Milli, Smitha},
  booktitle={ICLR 2025 Workshop on Bidirectional Human-AI Alignment}
}

@article{fan2025fine,
  title={Fine-grained List-wise Alignment for Generative Medication Recommendation},
  author={Fan, Chenxiao and Gao, Chongming and Shi, Wentao and Gong, Yaxin and Zhao, Zihao and Feng, Fuli},
  journal={arXiv preprint arXiv:2505.20218},
  year={2025}
}

@inproceedings{sharma2024optimizing,
  title={Optimizing Novelty of Top-k Recommendations using Large Language Models and Reinforcement Learning},
  author={Sharma, Amit and Li, Hua and Li, Xue and Jiao, Jian},
  booktitle={Proceedings of the 30th ACM SIGKDD Conference on Knowledge Discovery and Data Mining},
  pages={5669--5679},
  year={2024}
}

@article{yuan2024fellas,
  title={FELLAS: Enhancing Federated Sequential Recommendation with LLM as External Services},
  author={Yuan, Wei and Yang, Chaoqun and Ye, Guanhua and Chen, Tong and Hung, Nguyen Quoc Viet and Yin, Hongzhi},
  journal={ACM Transactions on Information Systems},
  year={2024},
  publisher={ACM New York, NY}
}

@inproceedings{lian2024recai,
  title={Recai: Leveraging large language models for next-generation recommender systems},
  author={Lian, Jianxun and Lei, Yuxuan and Huang, Xu and Yao, Jing and Xu, Wei and Xie, Xing},
  booktitle={Companion Proceedings of the ACM Web Conference 2024},
  pages={1031--1034},
  year={2024}
}

@inproceedings{liu2025filtering,
  title={Filtering Discomforting Recommendations with Large Language Models},
  author={Liu, Jiahao and Shao, Yiyang and Zhang, Peng and Li, Dongsheng and Gu, Hansu and Chen, Chao and Du, Longzhi and Lu, Tun and Gu, Ning},
  booktitle={Proceedings of the ACM on Web Conference 2025},
  pages={3639--3650},
  year={2025}
}

@article{xu2025iagent,
  title={iAgent: LLM Agent as a Shield between User and Recommender Systems},
  author={Xu, Wujiang and Shi, Yunxiao and Liang, Zujie and Ning, Xuying and Mei, Kai and Wang, Kun and Zhu, Xi and Xu, Min and Zhang, Yongfeng},
  journal={arXiv preprint arXiv:2502.14662},
  year={2025}
}

@article{ribeiro2020beyond,
  title={Beyond accuracy: Behavioral testing of NLP models with CheckList},
  author={Ribeiro, Marco Tulio and Wu, Tongshuang and Guestrin, Carlos and Singh, Sameer},
  journal={arXiv preprint arXiv:2005.04118},
  year={2020}
}

@inproceedings{lu2021fantastically,
  title={Fantastically ordered prompts and where to find them: Overcoming few-shot prompt order sensitivity},
  author={Lu, Yao and Bartolo, Max and Moore, Alastair and Riedel, Sebastian and Stenetorp, Pontus},
  booktitle={Proceedings of the 60th Annual Meeting of the Association for Computational Linguistics (Volume 1: Long Papers)},
  pages={8086--8098},
  year={2022}
}

@inproceedings{razavi2025benchmarking,
  title={Benchmarking prompt sensitivity in large language models},
  author={Razavi, Amirhossein and Soltangheis, Mina and Arabzadeh, Negar and Salamat, Sara and Zihayat, Morteza and Bagheri, Ebrahim},
  booktitle={European Conference on Information Retrieval},
  pages={303--313},
  year={2025},
  organization={Springer}
}

@article{polo2024efficient,
  title={Efficient multi-prompt evaluation of llms},
  author={Polo, Felipe M and Xu, Ronald and Weber, Lucas and Silva, M{\'\i}rian and Bhardwaj, Onkar and Choshen, Leshem and de Oliveira, Allysson F and Sun, Yuekai and Yurochkin, Mikhail},
  journal={Advances in Neural Information Processing Systems},
  volume={37},
  pages={22483--22512},
  year={2024}
}

@article{kusano2024longer,
  title={Are Longer Prompts Always Better? Prompt Selection in Large Language Models for Recommendation Systems},
  author={Kusano, Genki and Akimoto, Kosuke and Takeoka, Kunihiro},
  journal={arXiv preprint arXiv:2412.14454},
  year={2024}
}

@article{zhang2024llmtreerec,
  title={LLMTreeRec: Unleashing the Power of Large Language Models for Cold-Start Recommendations},
  author={Zhang, Wenlin and Wu, Chuhan and Li, Xiangyang and Wang, Yuhao and Dong, Kuicai and Wang, Yichao and Dai, Xinyi and Zhao, Xiangyu and Guo, Huifeng and Tang, Ruiming},
  journal={arXiv preprint arXiv:2404.00702},
  year={2024}
}

@article{che2024new,
  title={New Community Cold-Start Recommendation: A Novel Large Language Model-based Method},
  author={Che, Shangkun and Mao, Minjia and Liu, Hongyan},
  year={2024}
}

@article{yang2025cold,
  title={Cold-Start Recommendation with Knowledge-Guided Retrieval-Augmented Generation},
  author={Yang, Wooseong and Zhang, Weizhi and Liu, Yuqing and Han, Yuwei and Wang, Yu and Lee, Junhyun and Yu, Philip S},
  journal={arXiv preprint arXiv:2505.20773},
  year={2025}
}

@inproceedings{sakurai2025llm,
  title={Llm is knowledge graph reasoner: Llm’s intuition-aware knowledge graph reasoning for cold-start sequential recommendation},
  author={Sakurai, Keigo and Togo, Ren and Ogawa, Takahiro and Haseyama, Miki},
  booktitle={European Conference on Information Retrieval},
  pages={263--278},
  year={2025},
  organization={Springer}
}

@article{zhang2025cold,
  title={Cold-Start Recommendation towards the Era of Large Language Models (LLMs): A Comprehensive Survey and Roadmap},
  author={Zhang, Weizhi and Bei, Yuanchen and Yang, Liangwei and Zou, Henry Peng and Zhou, Peilin and Liu, Aiwei and Li, Yinghui and Chen, Hao and Wang, Jianling and Wang, Yu and others},
  journal={arXiv preprint arXiv:2501.01945},
  year={2025}
}

@article{lichtenberg2024large,
  title={Large language models as recommender systems: A study of popularity bias},
  author={Lichtenberg, Jan Malte and Buchholz, Alexander and Schw{\"o}bel, Pola},
  journal={arXiv preprint arXiv:2406.01285},
  year={2024}
}

@article{spurlock2024chatgpt,
  title={Chatgpt for conversational recommendation: Refining recommendations by reprompting with feedback},
  author={Spurlock, Kyle Dylan and Acun, Cagla and Saka, Esin and Nasraoui, Olfa},
  journal={arXiv preprint arXiv:2401.03605},
  year={2024}
}

@inproceedings{zhang2024agentcf,
  title={Agentcf: Collaborative learning with autonomous language agents for recommender systems},
  author={Zhang, Junjie and Hou, Yupeng and Xie, Ruobing and Sun, Wenqi and McAuley, Julian and Zhao, Wayne Xin and Lin, Leyu and Wen, Ji-Rong},
  booktitle={Proceedings of the ACM Web Conference 2024},
  pages={3679--3689},
  year={2024}
}

@article{wang2023improving,
  title={Improving conversational recommendation systems via bias analysis and language-model-enhanced data augmentation},
  author={Wang, Xi and Rahmani, Hossein A and Liu, Jiqun and Yilmaz, Emine},
  journal={arXiv preprint arXiv:2310.16738},
  year={2023}
}

@article{zhao2025can,
  title={Can LLM-Driven Hard Negative Sampling Empower Collaborative Filtering? Findings and Potentials},
  author={Zhao, Chu and Yang, Enneng and Liu, Yuting and Zhao, Jianzhe and Guo, Guibing and Wang, Xingwei},
  journal={arXiv preprint arXiv:2504.04726},
  year={2025}
}

@article{liu2024llm,
  title={Llm-esr: Large language models enhancement for long-tailed sequential recommendation},
  author={Liu, Qidong and Wu, Xian and Wang, Yejing and Zhang, Zijian and Tian, Feng and Zheng, Yefeng and Zhao, Xiangyu},
  journal={Advances in Neural Information Processing Systems},
  volume={37},
  pages={26701--26727},
  year={2024}
}

@article{huang2021data,
  title={Data poisoning attacks to deep learning based recommender systems},
  author={Huang, Hai and Mu, Jiaming and Gong, Neil Zhenqiang and Li, Qi and Liu, Bin and Xu, Mingwei},
  journal={arXiv preprint arXiv:2101.02644},
  year={2021}
}

@inproceedings{zhang2024lorec,
  title={Lorec: Combating poisons with large language model for robust sequential recommendation},
  author={Zhang, Kaike and Cao, Qi and Wu, Yunfan and Sun, Fei and Shen, Huawei and Cheng, Xueqi},
  booktitle={Proceedings of the 47th International ACM SIGIR Conference on Research and Development in Information Retrieval},
  pages={1733--1742},
  year={2024}
}

@inproceedings{wang2025id,
  title={Id-free not risk-free: Llm-powered agents unveil risks in id-free recommender systems},
  author={Wang, Zongwei and Gao, Min and Yu, Junliang and Gao, Xinyi and Nguyen, Quoc Viet Hung and Sadiq, Shazia and Yin, Hongzhi},
  booktitle={Proceedings of the 48th International ACM SIGIR Conference on Research and Development in Information Retrieval},
  pages={1902--1911},
  year={2025}
}

@article{ning2025exploring,
  title={Exploring Backdoor Attack and Defense for LLM-empowered Recommendations},
  author={Ning, Liangbo and Fan, Wenqi and Li, Qing},
  journal={arXiv preprint arXiv:2504.11182},
  year={2025}
}

@article{zhang2024stealthy,
  title={Stealthy attack on large language model based recommendation},
  author={Zhang, Jinghao and Liu, Yuting and Liu, Qiang and Wu, Shu and Guo, Guibing and Wang, Liang},
  journal={arXiv preprint arXiv:2402.14836},
  year={2024}
}

@article{filandrianos2025bias,
  title={Bias Beware: The Impact of Cognitive Biases on LLM-Driven Product Recommendations},
  author={Filandrianos, Giorgos and Dimitriou, Angeliki and Lymperaiou, Maria and Thomas, Konstantinos and Stamou, Giorgos},
  journal={arXiv preprint arXiv:2502.01349},
  year={2025}
}

@inproceedings{nazary2025stealthy,
  title={Stealthy LLM-Driven Data Poisoning Attacks Against Embedding-Based Retrieval-Augmented Recommender Systems},
  author={Nazary, Fatemeh and Deldjoo, Yashar and Di Noia, Tommaso and Di Sciascio, Eugenio},
  booktitle={Adjunct Proceedings of the 33rd ACM Conference on User Modeling, Adaptation and Personalization},
  pages={98--102},
  year={2025}
}

@inproceedings{ning2024cheatagent,
  title={Cheatagent: Attacking llm-empowered recommender systems via llm agent},
  author={Ning, Liang-bo and Wang, Shijie and Fan, Wenqi and Li, Qing and Xu, Xin and Chen, Hao and Huang, Feiran},
  booktitle={Proceedings of the 30th ACM SIGKDD Conference on Knowledge Discovery and Data Mining},
  pages={2284--2295},
  year={2024}
}

@article{gu2025llm,
  title={LLM-Based User Simulation for Low-Knowledge Shilling Attacks on Recommender Systems},
  author={Gu, Shengkang and Liu, Jiahao and Li, Dongsheng and Zhang, Guangping and Han, Mingzhe and Gu, Hansu and Zhang, Peng and Gu, Ning and Shang, Li and Lu, Tun},
  journal={arXiv preprint arXiv:2505.13528},
  year={2025}
}

@article{song2024large,
  title={Large language model enhanced hard sample identification for denoising recommendation},
  author={Song, Tianrui and Chao, Wenshuo and Liu, Hao},
  journal={arXiv preprint arXiv:2409.10343},
  year={2024}
}

@inproceedings{wang2025unleashing,
  title={Unleashing the Power of Large Language Model for Denoising Recommendation},
  author={Wang, Shuyao and Zheng, Zhi and Sui, Yongduo and Xiong, Hui},
  booktitle={Proceedings of the ACM on Web Conference 2025},
  pages={252--263},
  year={2025}
}

@article{wang2025ruleagent,
  title={RuleAgent: Discovering Rules for Recommendation Denoising with Autonomous Language Agents},
  author={Wang, Zongwei and Gao, Min and Yu, Junliang and Hou, Yupeng and Sadiq, Shazia and Yin, Hongzhi},
  journal={arXiv preprint arXiv:2503.23374},
  year={2025}
}

@inproceedings{sun2021does,
  title={Does Every Data Instance Matter? Enhancing Sequential Recommendation by Eliminating Unreliable Data.},
  author={Sun, Yatong and Wang, Bin and Sun, Zhu and Yang, Xiaochun},
  booktitle={IJCAI},
  pages={1579--1585},
  year={2021}
}

@inproceedings{zhang2022hierarchical,
  title={Hierarchical item inconsistency signal learning for sequence denoising in sequential recommendation},
  author={Zhang, Chi and Du, Yantong and Zhao, Xiangyu and Han, Qilong and Chen, Rui and Li, Li},
  booktitle={Proceedings of the 31st ACM international conference on information \& knowledge management},
  pages={2508--2518},
  year={2022}
}

@inproceedings{zhang2024ssdrec,
  title={Ssdrec: Self-augmented sequence denoising for sequential recommendation},
  author={Zhang, Chi and Han, Qilong and Chen, Rui and Zhao, Xiangyu and Tang, Peng and Song, Hongtao},
  booktitle={2024 IEEE 40th International Conference on Data Engineering (ICDE)},
  pages={803--815},
  year={2024},
  organization={IEEE}
}

@inproceedings{wang2021denoising,
  title={Denoising implicit feedback for recommendation},
  author={Wang, Wenjie and Feng, Fuli and He, Xiangnan and Nie, Liqiang and Chua, Tat-Seng},
  booktitle={Proceedings of the 14th ACM international conference on web search and data mining},
  pages={373--381},
  year={2021}
}

@article{peng2025denoising,
  title={Denoising alignment with large language model for recommendation},
  author={Peng, Yingtao and Gao, Chen and Zhang, Yu and Dan, Tangpeng and Du, Xiaoyi and Luo, Hengliang and Li, Yong and Meng, Xiaofeng},
  journal={ACM Transactions on Information Systems},
  volume={43},
  number={2},
  pages={1--35},
  year={2025},
  publisher={ACM New York, NY}
}

@inproceedings{ren2024representation,
  title={Representation learning with large language models for recommendation},
  author={Ren, Xubin and Wei, Wei and Xia, Lianghao and Su, Lixin and Cheng, Suqi and Wang, Junfeng and Yin, Dawei and Huang, Chao},
  booktitle={Proceedings of the ACM web conference 2024},
  pages={3464--3475},
  year={2024}
}

@inproceedings{wang2024distributionally,
  title={Distributionally robust graph-based recommendation system},
  author={Wang, Bohao and Chen, Jiawei and Li, Changdong and Zhou, Sheng and Shi, Qihao and Gao, Yang and Feng, Yan and Chen, Chun and Wang, Can},
  booktitle={Proceedings of the ACM web conference 2024},
  pages={3777--3788},
  year={2024}
}

@inproceedings{yang2023generic,
  title={A generic learning framework for sequential recommendation with distribution shifts},
  author={Yang, Zhengyi and He, Xiangnan and Zhang, Jizhi and Wu, Jiancan and Xin, Xin and Chen, Jiawei and Wang, Xiang},
  booktitle={Proceedings of the 46th International ACM SIGIR Conference on Research and Development in Information Retrieval},
  pages={331--340},
  year={2023}
}

@article{wang2023drdt,
  title={Drdt: Dynamic reflection with divergent thinking for llm-based sequential recommendation},
  author={Wang, Yu and Liu, Zhiwei and Zhang, Jianguo and Yao, Weiran and Heinecke, Shelby and Yu, Philip S},
  journal={arXiv preprint arXiv:2312.11336},
  year={2023}
}

@article{ebrat2024lusifer,
  title={Lusifer: LLM-based user simulated feedback environment for online recommender systems},
  author={Ebrat, Danial and Paradalis, Eli and Rueda, Luis},
  journal={arXiv preprint arXiv:2405.13362},
  year={2024}
}

@inproceedings{gao2025process,
  title={Process-supervised llm recommenders via flow-guided tuning},
  author={Gao, Chongming and Gao, Mengyao and Fan, Chenxiao and Yuan, Shuai and Shi, Wentao and He, Xiangnan},
  booktitle={Proceedings of the 48th International ACM SIGIR Conference on Research and Development in Information Retrieval},
  pages={1934--1943},
  year={2025}
}

@inproceedings{lu2025dual,
  title={Dual Debiasing in LLM-based Recommendation},
  author={Lu, Sijin and Man, Zhibo and Luo, Fangyuan and Wu, Jun},
  booktitle={Proceedings of the 48th International ACM SIGIR Conference on Research and Development in Information Retrieval},
  pages={2685--2689},
  year={2025}
}

@inproceedings{gao2025sprec,
  title={Sprec: Self-play to debias llm-based recommendation},
  author={Gao, Chongming and Chen, Ruijun and Yuan, Shuai and Huang, Kexin and Yu, Yuanqing and He, Xiangnan},
  booktitle={Proceedings of the ACM on Web Conference 2025},
  pages={5075--5084},
  year={2025}
}

@inproceedings{deldjoo2025toward,
  title={Toward Holistic Evaluation of Recommender Systems Powered by Generative Models},
  author={Deldjoo, Yashar and Mehta, Nikhil and Sathiamoorthy, Maheswaran and Zhang, Shuai and Castells, Pablo and McAuley, Julian},
  booktitle={Proceedings of the 48th International ACM SIGIR Conference on Research and Development in Information Retrieval},
  pages={3932--3942},
  year={2025}
}

@article{ma2023large,
  title={Large language models are not stable recommender systems},
  author={Ma, Tianhui and Cheng, Yuan and Zhu, Hengshu and Xiong, Hui},
  journal={arXiv preprint arXiv:2312.15746},
  year={2023}
}

@inproceedings{hou2024large,
  title={Large language models are zero-shot rankers for recommender systems},
  author={Hou, Yupeng and Zhang, Junjie and Lin, Zihan and Lu, Hongyu and Xie, Ruobing and McAuley, Julian and Zhao, Wayne Xin},
  booktitle={European Conference on Information Retrieval},
  pages={364--381},
  year={2024},
  organization={Springer}
}

@article{bito2025evaluating,
  title={Evaluating Position Bias in Large Language Model Recommendations},
  author={Bito, Ethan and Ren, Yongli and He, Estrid},
  journal={arXiv preprint arXiv:2508.02020},
  year={2025}
}

@article{luo2025recranker,
  title={Recranker: Instruction tuning large language model as ranker for top-k recommendation},
  author={Luo, Sichun and He, Bowei and Zhao, Haohan and Shao, Wei and Qi, Yanlin and Huang, Yinya and Zhou, Aojun and Yao, Yuxuan and Li, Zongpeng and Xiao, Yuanzhang and others},
  journal={ACM Transactions on Information Systems},
  volume={43},
  number={5},
  pages={1--31},
  year={2025},
  publisher={ACM New York, NY}
}

@article{zhang2021language,
  title={Language models as recommender systems: Evaluations and limitations},
  author={Zhang, Yuhui and Ding, Hao and Shui, Zeren and Ma, Yifei and Zou, James and Deoras, Anoop and Wang, Hao},
  year={2021}
}

@inproceedings{zhang2023chatgpt,
  title={Is chatgpt fair for recommendation? evaluating fairness in large language model recommendation},
  author={Zhang, Jizhi and Bao, Keqin and Zhang, Yang and Wang, Wenjie and Feng, Fuli and He, Xiangnan},
  booktitle={Proceedings of the 17th ACM Conference on Recommender Systems},
  pages={993--999},
  year={2023}
}

@article{deldjoo2024normative,
  title={A Normative Framework for Benchmarking Consumer Fairness in Large Language Model Recommender System},
  author={Deldjoo, Yashar and Nazary, Fatemeh},
  journal={arXiv preprint arXiv:2405.02219},
  year={2024}
}

@article{hua2023up5,
  title={Up5: Unbiased foundation model for fairness-aware recommendation},
  author={Hua, Wenyue and Ge, Yingqiang and Xu, Shuyuan and Ji, Jianchao and Zhang, Yongfeng},
  journal={arXiv preprint arXiv:2305.12090},
  year={2023}
}

@article{xu2025tapping,
  title={Tapping the potential of large language models as recommender systems: A comprehensive framework and empirical analysis},
  author={Xu, Lanling and Zhang, Junjie and Li, Bingqian and Wang, Jinpeng and Chen, Sheng and Zhao, Wayne Xin and Wen, Ji-Rong},
  journal={ACM Transactions on Knowledge Discovery from Data},
  volume={19},
  number={5},
  pages={1--51},
  year={2025},
  publisher={ACM New York, NY}
}

@inproceedings{wang2026does,
  title={Does LLM Focus on the Right Words? Mitigating Context Bias in LLM-based Recommenders},
  author={Wang, Bohao and Chen, Jiawei and Liu, Feng and Zhang, Changwang and Wang, Jun and Jin, Canghong and Chen, Chun and Wang, Can},
  booktitle={Proceedings of the ACM Web Conference 2026},
  pages={6688--6699},
  year={2026}
}

@article{xu2023study,
  title={A study of implicit ranking unfairness in large language models},
  author={Xu, Chen and Wang, Wenjie and Li, Yuxin and Pang, Liang and Xu, Jun and Chua, Tat-Seng},
  journal={arXiv preprint arXiv:2311.07054},
  year={2023}
}

@article{deldjoo2025cfairllm,
  title={Cfairllm: Consumer fairness evaluation in large-language model recommender system},
  author={Deldjoo, Yashar and Di Noia, Tommaso},
  journal={ACM Transactions on Intelligent Systems and Technology},
  year={2025},
  publisher={ACM New York, NY}
}

@inproceedings{sakib2024challenging,
  title={Challenging fairness: A comprehensive exploration of bias in llm-based recommendations},
  author={Sakib, Shahnewaz Karim and Das, Anindya Bijoy},
  booktitle={2024 IEEE International Conference on Big Data (BigData)},
  pages={1585--1592},
  year={2024},
  organization={IEEE}
}

@inproceedings{tommasel2024fairness,
  title={Fairness Matters: A look at LLM-generated group recommendations},
  author={Tommasel, Antonela},
  booktitle={Proceedings of the 18th ACM Conference on Recommender Systems},
  pages={993--998},
  year={2024}
}

@article{sah2025faireval,
  title={FairEval: Evaluating Fairness in LLM-Based Recommendations with Personality Awareness},
  author={Sah, Chandan Kumar and Lian, Xiaoli and Xu, Tony and Zhang, Li},
  journal={arXiv preprint arXiv:2504.07801},
  year={2025}
}

@article{li2023preliminary,
  title={A preliminary study of chatgpt on news recommendation: Personalization, provider fairness, fake news},
  author={Li, Xinyi and Zhang, Yongfeng and Malthouse, Edward C},
  journal={arXiv preprint arXiv:2306.10702},
  year={2023}
}

@article{deldjoo2024understanding,
  title={Understanding biases in ChatGPT-based recommender systems: Provider fairness, temporal stability, and recency},
  author={Deldjoo, Yashar},
  journal={ACM Transactions on Recommender Systems},
  year={2024},
  publisher={ACM New York, NY}
}

@inproceedings{jiang2024item,
  title={Item-side fairness of large language model-based recommendation system},
  author={Jiang, Meng and Bao, Keqin and Zhang, Jizhi and Wang, Wenjie and Yang, Zhengyi and Feng, Fuli and He, Xiangnan},
  booktitle={Proceedings of the ACM Web Conference 2024},
  pages={4717--4726},
  year={2024}
}

@article{zhang2025bifair,
  title={BiFair: A Fairness-aware Training Framework for LLM-enhanced Recommender Systems via Bi-level Optimization},
  author={Zhang, Jiaming and Li, Yuyuan and Xu, Yiqun and Zhang, Li and Feng, Xiaohua and Ren, Zhifei and Chen, Chaochao},
  journal={arXiv preprint arXiv:2507.04294},
  year={2025}
}

@article{das2024unveiling,
  title={Unveiling and Mitigating Bias in Large Language Model Recommendations: A Path to Fairness},
  author={Das, Anindya Bijoy and Sakib, Shahnewaz Karim},
  journal={arXiv preprint arXiv:2409.10825},
  year={2024}
}

@article{liu2025fairness,
  title={Fairness identification of large language models in recommendation},
  author={Liu, Wei and Liu, Baisong and Qin, Jiangcheng and Zhang, Xueyuan and Huang, Weiming and Wang, Yangyang},
  journal={Scientific Reports},
  volume={15},
  number={1},
  pages={5516},
  year={2025},
  publisher={Nature Publishing Group UK London}
}

@inproceedings{dai2024bias,
  title={Bias and unfairness in information retrieval systems: New challenges in the llm era},
  author={Dai, Sunhao and Xu, Chen and Xu, Shicheng and Pang, Liang and Dong, Zhenhua and Xu, Jun},
  booktitle={Proceedings of the 30th ACM SIGKDD Conference on Knowledge Discovery and Data Mining},
  pages={6437--6447},
  year={2024}
}

@inproceedings{hu2025fairwork,
  title={FairWork: A Generic Framework For Evaluating Fairness In LLM-Based Job Recommender System},
  author={Hu, Yuhan and Lyu, Ziyu and Bai, Lu and Cui, Lixin},
  booktitle={Proceedings of the 48th International ACM SIGIR Conference on Research and Development in Information Retrieval},
  pages={3964--3968},
  year={2025}
}

@inproceedings{zhou2025exploring,
  title={Exploring the Escalation of Source Bias in User, Data, and Recommender System Feedback Loop},
  author={Zhou, Yuqi and Dai, Sunhao and Pang, Liang and Wang, Gang and Dong, Zhenhua and Xu, Jun and Wen, Ji-Rong},
  booktitle={Proceedings of the 48th International ACM SIGIR Conference on Research and Development in Information Retrieval},
  pages={1676--1686},
  year={2025}
}

@article{nguyen2025enhancing,
  title={Enhancing Graph-based Recommendations with Majority-Voting LLM-Rerank Augmentation},
  author={Nguyen, Minh-Anh and Nguyen, Bao and Hoang, Tuan Anh and Le, Duc-Trong and Le, Dung D and others},
  journal={arXiv preprint arXiv:2507.21563},
  year={2025}
}

@article{wei2022chain,
  title={Chain-of-thought prompting elicits reasoning in large language models},
  author={Wei, Jason and Wang, Xuezhi and Schuurmans, Dale and Bosma, Maarten and Xia, Fei and Chi, Ed and Le, Quoc V and Zhou, Denny and others},
  journal={Advances in neural information processing systems},
  volume={35},
  pages={24824--24837},
  year={2022}
}

@inproceedings{sun2025llm4rsr,
  title={LLM4RSR: Large Language Models as Data Correctors for Robust Sequential Recommendation},
  author={Sun, Yatong and Yang, Xiaochun and Sun, Zhu and Wang, Yan and Wang, Bin and Qu, Xinghua},
  booktitle={Proceedings of the AAAI Conference on Artificial Intelligence},
  volume={39},
  number={12},
  pages={12604--12612},
  year={2025}
}

@article{yang2025drunkagent,
  title={DrunkAgent: Stealthy Memory Corruption in LLM-Powered Recommender Agents},
  author={Yang, Shiyi and Hu, Zhibo and Li, Xinshu and Wang, Chen and Yu, Tong and Xu, Xiwei and Zhu, Liming and Yao, Lina},
  journal={arXiv preprint arXiv:2503.23804},
  year={2025}
}

@inproceedings{morris2023text,
  title={Text embeddings reveal (almost) as much as text},
  author={Morris, John and Kuleshov, Volodymyr and Shmatikov, Vitaly and Rush, Alexander M},
  booktitle={Proceedings of the 2023 Conference on Empirical Methods in Natural Language Processing},
  pages={12448--12460},
  year={2023}
}

@inproceedings{carlini2022quantifying,
  title={Quantifying memorization across neural language models},
  author={Carlini, Nicholas and Ippolito, Daphne and Jagielski, Matthew and Lee, Katherine and Tramer, Florian and Zhang, Chiyuan},
  booktitle={The Eleventh International Conference on Learning Representations},
  year={2022}
}

@article{staab2023beyond,
  title={Beyond memorization: Violating privacy via inference with large language models},
  author={Staab, Robin and Vero, Mark and Balunovi{\'c}, Mislav and Vechev, Martin},
  journal={arXiv preprint arXiv:2310.07298},
  year={2023}
}

@inproceedings{ren2024enhancing,
  title={Enhancing sequential recommenders with augmented knowledge from aligned large language models},
  author={Ren, Yankun and Chen, Zhongde and Yang, Xinxing and Li, Longfei and Jiang, Cong and Cheng, Lei and Zhang, Bo and Mo, Linjian and Zhou, Jun},
  booktitle={Proceedings of the 47th International ACM SIGIR Conference on Research and Development in Information Retrieval},
  pages={345--354},
  year={2024}
}

@inproceedings{chen2022recommendation,
  title={Recommendation unlearning},
  author={Chen, Chong and Sun, Fei and Zhang, Min and Ding, Bolin},
  booktitle={Proceedings of the ACM web conference 2022},
  pages={2768--2777},
  year={2022}
}

@article{zhang2023lightfr,
  title={LightFR: Lightweight federated recommendation with privacy-preserving matrix factorization},
  author={Zhang, Honglei and Luo, Fangyuan and Wu, Jun and He, Xiangnan and Li, Yidong},
  journal={ACM Transactions on Information Systems},
  volume={41},
  number={4},
  pages={1--28},
  year={2023},
  publisher={ACM New York, NY}
}

@article{yin2024device,
  title={On-device recommender systems: A comprehensive survey},
  author={Yin, Hongzhi and Qu, Liang and Chen, Tong and Yuan, Wei and Zheng, Ruiqi and Long, Jing and Xia, Xin and Shi, Yuhui and Zhang, Chengqi},
  journal={arXiv preprint arXiv:2401.11441},
  year={2024}
}

@article{khezresmaeilzadeh2025preserving,
  title={Preserving privacy and utility in llm-based product recommendations},
  author={Khezresmaeilzadeh, Tina and Zhang, Jiang and Andreadis, Dimitrios and Psounis, Konstantinos},
  journal={arXiv preprint arXiv:2505.00951},
  year={2025}
}

@article{zhang2024recommendation,
  title={Recommendation unlearning via influence function},
  author={Zhang, Yang and Hu, Zhiyu and Bai, Yimeng and Wu, Jiancan and Wang, Qifan and Feng, Fuli},
  journal={ACM Transactions on Recommender Systems},
  volume={3},
  number={2},
  pages={1--23},
  year={2024},
  publisher={ACM New York, NY}
}

@article{li2024survey,
  title={A survey on recommendation unlearning: Fundamentals, taxonomy, evaluation, and open questions},
  author={Li, Yuyuan and Feng, Xiaohua and Chen, Chaochao and Yang, Qiang},
  journal={arXiv preprint arXiv:2412.12836},
  year={2024}
}

@inproceedings{sachdeva2024machine,
  title={Machine unlearning for recommendation systems: An insight},
  author={Sachdeva, Bhavika and Rathee, Harshita and Sristi and Sharma, Arun and Wydma{\'n}ski, Witold},
  booktitle={International Conference On Innovative Computing And Communication},
  pages={415--430},
  year={2024},
  organization={Springer}
}

@article{kim2021pqk,
  title={PQK: model compression via pruning, quantization, and knowledge distillation},
  author={Kim, Jangho and Chang, Simyung and Kwak, Nojun},
  journal={arXiv preprint arXiv:2106.14681},
  year={2021}
}

@article{zhu2024survey,
  title={A survey on model compression for large language models},
  author={Zhu, Xunyu and Li, Jian and Liu, Yong and Ma, Can and Wang, Weiping},
  journal={Transactions of the Association for Computational Linguistics},
  volume={12},
  pages={1556--1577},
  year={2024},
  publisher={MIT Press 255 Main Street, 9th Floor, Cambridge, Massachusetts 02142, USA~…}
}

@inproceedings{yeom2018privacy,
  title={Privacy risk in machine learning: Analyzing the connection to overfitting},
  author={Yeom, Samuel and Giacomelli, Irene and Fredrikson, Matt and Jha, Somesh},
  booktitle={2018 IEEE 31st computer security foundations symposium (CSF)},
  pages={268--282},
  year={2018},
  organization={IEEE}
}

@inproceedings{zhang2024bridging,
  title={Bridging the information gap between domain-specific model and general llm for personalized recommendation},
  author={Zhang, Wenxuan and Liu, Hongzhi and Dong, Zhijin and Du, Yingpeng and Zhu, Chen and Song, Yang and Zhu, Hengshu and Wu, Zhonghai},
  booktitle={Asia-Pacific Web (APWeb) and Web-Age Information Management (WAIM) Joint International Conference on Web and Big Data},
  pages={280--294},
  year={2024},
  organization={Springer}
}

@article{wei2023personalized,
  title={Personalized federated learning with differential privacy and convergence guarantee},
  author={Wei, Kang and Li, Jun and Ma, Chuan and Ding, Ming and Chen, Wen and Wu, Jun and Tao, Meixia and Poor, H Vincent},
  journal={IEEE Transactions on Information Forensics and Security},
  volume={18},
  pages={4488--4503},
  year={2023},
  publisher={IEEE}
}

@article{zhu2025recommender,
  title={Recommender systems meet large language model agents: A survey},
  author={Zhu, Xi and Wang, Yu and Gao, Hang and Xu, Wujiang and Wang, Chen and Liu, Zhiwei and Wang, Kun and Jin, Mingyu and Pang, Linsey and Weng, Qingsong and others},
  journal={Foundations and Trends{\textregistered} in Privacy and Security},
  volume={7},
  number={4},
  pages={247--396},
  year={2025},
  publisher={Emerald Publishing Limited}
}

@article{lin2020fedrec,
  title={Fedrec: Federated recommendation with explicit feedback},
  author={Lin, Guanyu and Liang, Feng and Pan, Weike and Ming, Zhong},
  journal={IEEE Intelligent Systems},
  volume={36},
  number={5},
  pages={21--30},
  year={2020},
  publisher={IEEE}
}

@article{sun2024survey,
  title={A survey on federated recommendation systems},
  author={Sun, Zehua and Xu, Yonghui and Liu, Yong and He, Wei and Kong, Lanju and Wu, Fangzhao and Jiang, Yali and Cui, Lizhen},
  journal={IEEE Transactions on Neural Networks and Learning Systems},
  volume={36},
  number={1},
  pages={6--20},
  year={2024},
  publisher={IEEE}
}

@incollection{yang2020federated,
  title={Federated recommendation systems},
  author={Yang, Liu and Tan, Ben and Zheng, Vincent W and Chen, Kai and Yang, Qiang},
  booktitle={Federated Learning: Privacy and Incentive},
  pages={225--239},
  year={2020},
  publisher={Springer}
}

@inproceedings{carlini2021extracting,
  title={Extracting training data from large language models},
  author={Carlini, Nicholas and Tramer, Florian and Wallace, Eric and Jagielski, Matthew and Herbert-Voss, Ariel and Lee, Katherine and Roberts, Adam and Brown, Tom and Song, Dawn and Erlingsson, Ulfar and others},
  booktitle={30th USENIX security symposium (USENIX Security 21)},
  pages={2633--2650},
  year={2021}
}

@article{das2025security,
  title={Security and privacy challenges of large language models: A survey},
  author={Das, Badhan Chandra and Amini, M Hadi and Wu, Yanzhao},
  journal={ACM Computing Surveys},
  volume={57},
  number={6},
  pages={1--39},
  year={2025},
  publisher={ACM New York, NY}
}

@article{zhang2023comprehensive,
  title={Comprehensive privacy analysis on federated recommender system against attribute inference attacks},
  author={Zhang, Shijie and Yuan, Wei and Yin, Hongzhi},
  journal={IEEE Transactions on Knowledge and Data Engineering},
  volume={36},
  number={3},
  pages={987--999},
  year={2023},
  publisher={IEEE}
}

@inproceedings{zhang2021membership,
  title={Membership inference attacks against recommender systems},
  author={Zhang, Minxing and Ren, Zhaochun and Wang, Zihan and Ren, Pengjie and Chen, Zhunmin and Hu, Pengfei and Zhang, Yang},
  booktitle={Proceedings of the 2021 ACM SIGSAC Conference on Computer and Communications Security},
  pages={864--879},
  year={2021}
}

@article{wu2024fedlora,
  title={FedLoRA: When personalized federated learning meets low-rank adaptation},
  author={Wu, Xinghao and Liu, Xuefeng and Niu, Jianwei and Wang, Haolin and Tang, Shaojie and Zhu, Guogang},
  year={2024}
}

@article{sun2024improving,
  title={Improving lora in privacy-preserving federated learning},
  author={Sun, Youbang and Li, Zitao and Li, Yaliang and Ding, Bolin},
  journal={arXiv preprint arXiv:2403.12313},
  year={2024}
}

@inproceedings{belosevic2025user,
  title={User-Centric Design Paradigms for Trust and Control in Human-LLM-Interactions: A Survey},
  author={Belosevic, Milena},
  booktitle={Proceedings of the Fourth Workshop on Bridging Human-Computer Interaction and Natural Language Processing (HCI+ NLP)},
  pages={17--32},
  year={2025}
}

@article{li2026privacy,
  title={Privacy Control in Conversational LLM Platforms: A Walkthrough Study},
  author={Li, Zhuoyang and Wu, Yanlai and Li, Yao and Gui, Xinning and Luo, Yuhan},
  journal={arXiv preprint arXiv:2602.10684},
  year={2026}
}

@article{wei2021finetuned,
  title={Finetuned language models are zero-shot learners},
  author={Wei, Jason and Bosma, Maarten and Zhao, Vincent Y and Guu, Kelvin and Yu, Adams Wei and Lester, Brian and Du, Nan and Dai, Andrew M and Le, Quoc V},
  journal={arXiv preprint arXiv:2109.01652},
  year={2021}
}

@article{liu2023pre,
  title={Pre-train, prompt, and predict: A systematic survey of prompting methods in natural language processing},
  author={Liu, Pengfei and Yuan, Weizhe and Fu, Jinlan and Jiang, Zhengbao and Hayashi, Hiroaki and Neubig, Graham},
  journal={ACM computing surveys},
  volume={55},
  number={9},
  pages={1--35},
  year={2023},
  publisher={ACM New York, NY}
}

@article{knijnenburg2012explaining,
  title={Explaining the user experience of recommender systems},
  author={Knijnenburg, Bart P and Willemsen, Martijn C and Gantner, Zeno and Soncu, Hakan and Newell, Chris},
  journal={User modeling and user-adapted interaction},
  volume={22},
  number={4},
  pages={441--504},
  year={2012},
  publisher={Springer}
}

@inproceedings{cen2020controllable,
  title={Controllable multi-interest framework for recommendation},
  author={Cen, Yukuo and Zhang, Jianwei and Zou, Xu and Zhou, Chang and Yang, Hongxia and Tang, Jie},
  booktitle={Proceedings of the 26th ACM SIGKDD international conference on knowledge discovery \& data mining},
  pages={2942--2951},
  year={2020}
}

@article{gong2026breaking,
  title={Breaking User-Centric Agency: A Tri-Party Framework for Agent-Based Recommendation},
  author={Gong, Yaxin and Gao, Chongming and Fan, Chenxiao and Wang, Wenjie and Feng, Fuli and He, Xiangnan},
  journal={arXiv preprint arXiv:2603.10673},
  year={2026}
}

@article{lin2025can,
  title={How can recommender systems benefit from large language models: A survey},
  author={Lin, Jianghao and Dai, Xinyi and Xi, Yunjia and Liu, Weiwen and Chen, Bo and Zhang, Hao and Liu, Yong and Wu, Chuhan and Li, Xiangyang and Zhu, Chenxu and others},
  journal={ACM Transactions on Information Systems},
  volume={43},
  number={2},
  pages={1--47},
  year={2025},
  publisher={ACM New York, NY}
}

@inproceedings{webson2022prompt,
  title={Do prompt-based models really understand the meaning of their prompts?},
  author={Webson, Albert and Pavlick, Ellie},
  booktitle={Proceedings of the 2022 conference of the north american chapter of the association for computational linguistics: Human language technologies},
  pages={2300--2344},
  year={2022}
}

@inproceedings{zheng2021disentangling,
  title={Disentangling user interest and conformity for recommendation with causal embedding},
  author={Zheng, Yu and Gao, Chen and Li, Xiang and He, Xiangnan and Li, Yong and Jin, Depeng},
  booktitle={Proceedings of the web conference 2021},
  pages={2980--2991},
  year={2021}
}

@article{jannach2023survey,
  title={A survey on multi-objective recommender systems},
  author={Jannach, Dietmar and Abdollahpouri, Himan},
  journal={Frontiers in big Data},
  volume={6},
  pages={1157899},
  year={2023},
  publisher={Frontiers Media SA}
}

@article{gao2021advances,
  title={Advances and challenges in conversational recommender systems: A survey},
  author={Gao, Chongming and Lei, Wenqiang and He, Xiangnan and De Rijke, Maarten and Chua, Tat-Seng},
  journal={AI open},
  volume={2},
  pages={100--126},
  year={2021},
  publisher={Elsevier}
}

@inproceedings{lei2020estimation,
  title={Estimation-action-reflection: Towards deep interaction between conversational and recommender systems},
  author={Lei, Wenqiang and He, Xiangnan and Miao, Yisong and Wu, Qingyun and Hong, Richang and Kan, Min-Yen and Chua, Tat-Seng},
  booktitle={Proceedings of the 13th international conference on web search and data mining},
  pages={304--312},
  year={2020}
}

@article{parra2015user,
  title={User-controllable personalization: A case study with SetFusion},
  author={Parra, Denis and Brusilovsky, Peter},
  journal={International Journal of Human-Computer Studies},
  volume={78},
  pages={43--67},
  year={2015},
  publisher={Elsevier}
}

@article{abdollahpouri2020multistakeholder,
  title={Multistakeholder recommendation: Survey and research directions: H. Abdollahpouri et al.},
  author={Abdollahpouri, Himan and Adomavicius, Gediminas and Burke, Robin and Guy, Ido and Jannach, Dietmar and Kamishima, Toshihiro and Krasnodebski, Jan and Pizzato, Luiz},
  journal={User Modeling and User-Adapted Interaction},
  volume={30},
  number={1},
  pages={127--158},
  year={2020},
  publisher={Springer}
}

@inproceedings{wu2022ai,
  title={Ai chains: Transparent and controllable human-ai interaction by chaining large language model prompts},
  author={Wu, Tongshuang and Terry, Michael and Cai, Carrie Jun},
  booktitle={Proceedings of the 2022 CHI conference on human factors in computing systems},
  pages={1--22},
  year={2022}
}

@article{dohan2022language,
  title={Language model cascades},
  author={Dohan, David and Xu, Winnie and Lewkowycz, Aitor and Austin, Jacob and Bieber, David and Lopes, Raphael Gontijo and Wu, Yuhuai and Michalewski, Henryk and Saurous, Rif A and Sohl-Dickstein, Jascha and others},
  journal={arXiv preprint arXiv:2207.10342},
  year={2022}
}

@article{fan2026uncertainty,
  title={Uncertainty-aware Generative Recommendation},
  author={Fan, Chenxiao and Gao, Chongming and Gong, Yaxin and Liu, Haoyan and Feng, Fuli and He, Xiangnan},
  journal={arXiv preprint arXiv:2602.11719},
  year={2026}
}

@article{dai2025token,
  title={Token-Controlled Re-ranking for Sequential Recommendation via LLMs},
  author={Dai, Wenxi and Xu, Wujiang and Wang, Pinhuan and Metaxas, Dimitris N},
  journal={arXiv preprint arXiv:2511.17913},
  year={2025}
}

@article{milano2020recommender,
  title={Recommender systems and their ethical challenges},
  author={Milano, Silvia and Taddeo, Mariarosaria and Floridi, Luciano},
  journal={Ai \& Society},
  volume={35},
  number={4},
  pages={957--967},
  year={2020},
  publisher={Springer}
}

@article{lipton2018mythos,
  title={The mythos of model interpretability: In machine learning, the concept of interpretability is both important and slippery.},
  author={Lipton, Zachary C},
  journal={Queue},
  volume={16},
  number={3},
  pages={31--57},
  year={2018},
  publisher={ACM New York, NY, USA}
}

@inproceedings{zhao2021calibrate,
  title={Calibrate before use: Improving few-shot performance of language models},
  author={Zhao, Zihao and Wallace, Eric and Feng, Shi and Klein, Dan and Singh, Sameer},
  booktitle={International conference on machine learning},
  pages={12697--12706},
  year={2021},
  organization={Pmlr}
}

@inproceedings{yao2022react,
  title={React: Synergizing reasoning and acting in language models},
  author={Yao, Shunyu and Zhao, Jeffrey and Yu, Dian and Du, Nan and Shafran, Izhak and Narasimhan, Karthik R and Cao, Yuan},
  booktitle={The eleventh international conference on learning representations},
  year={2022}
}

@inproceedings{kang2018self,
  title={Self-attentive sequential recommendation},
  author={Kang, Wang-Cheng and McAuley, Julian},
  booktitle={2018 IEEE international conference on data mining (ICDM)},
  pages={197--206},
  year={2018},
  organization={IEEE}
}

@inproceedings{he2020lightgcn,
  title={Lightgcn: Simplifying and powering graph convolution network for recommendation},
  author={He, Xiangnan and Deng, Kuan and Wang, Xiang and Li, Yan and Zhang, Yongdong and Wang, Meng},
  booktitle={Proceedings of the 43rd International ACM SIGIR conference on research and development in Information Retrieval},
  pages={639--648},
  year={2020}
}

@inproceedings{ma2008sorec,
  title={Sorec: social recommendation using probabilistic matrix factorization},
  author={Ma, Hao and Yang, Haixuan and Lyu, Michael R and King, Irwin},
  booktitle={Proceedings of the 17th ACM conference on Information and knowledge management},
  pages={931--940},
  year={2008}
}

@article{adcock2026llama,
  title={The Llama 4 Herd: Architecture, Training, Evaluation, and Deployment Notes},
  author={Adcock, Aaron and Srivastava, Aayushi and Dubey, Abhimanyu and Jauhri, Abhinav and Pande, Abhinav and Pandey, Abhinav and Sharma, Abhinav and Kadian, Abhishek and Kumawat, Abhishek and Kelsey, Adam and others},
  journal={arXiv preprint arXiv:2601.11659},
  year={2026}
}

@article{liu2024deepseek,
  title={Deepseek-v3 technical report},
  author={Liu, Aixin and Feng, Bei and Xue, Bing and Wang, Bingxuan and Wu, Bochao and Lu, Chengda and Zhao, Chenggang and Deng, Chengqi and Zhang, Chenyu and Ruan, Chong and others},
  journal={arXiv preprint arXiv:2412.19437},
  year={2024}
}

@article{zhao2024recommender,
  title={Recommender systems in the era of large language models (llms)},
  author={Zhao, Zihuai and Fan, Wenqi and Li, Jiatong and Liu, Yunqing and Mei, Xiaowei and Wang, Yiqi and Wen, Zhen and Wang, Fei and Zhao, Xiangyu and Tang, Jiliang and others},
  journal={IEEE Transactions on Knowledge and Data Engineering},
  volume={36},
  number={11},
  pages={6889--6907},
  year={2024},
  publisher={IEEE}
}

@article{ge2024survey,
  title={A survey on trustworthy recommender systems},
  author={Ge, Yingqiang and Liu, Shuchang and Fu, Zuohui and Tan, Juntao and Li, Zelong and Xu, Shuyuan and Li, Yunqi and Xian, Yikun and Zhang, Yongfeng},
  journal={ACM Transactions on Recommender Systems},
  volume={3},
  number={2},
  pages={1--68},
  year={2024},
  publisher={ACM New York, NY}
}

@article{wang2024trustworthy,
  title={Trustworthy recommender systems},
  author={Wang, Shoujin and Zhang, Xiuzhen and Wang, Yan and Ricci, Francesco},
  journal={ACM Transactions on Intelligent Systems and Technology},
  volume={15},
  number={4},
  pages={1--20},
  year={2024},
  publisher={ACM New York, NY}
}

@article{fan2022comprehensive,
  title={A comprehensive survey on trustworthy recommender systems},
  author={Fan, Wenqi and Zhao, Xiangyu and Chen, Xiao and Su, Jingran and Gao, Jingtong and Wang, Lin and Liu, Qidong and Wang, Yiqi and Xu, Han and Chen, Lei and others},
  journal={arXiv preprint arXiv:2209.10117},
  year={2022}
}

@article{li2023trustworthy,
  title={Trustworthy AI: From principles to practices},
  author={Li, Bo and Qi, Peng and Liu, Bo and Di, Shuai and Liu, Jingen and Pei, Jiquan and Yi, Jinfeng and Zhou, Bowen},
  journal={ACM Computing Surveys},
  volume={55},
  number={9},
  pages={1--46},
  year={2023},
  publisher={ACM New York, NY}
}

@inproceedings{bang2025hallulens,
  title={Hallulens: Llm hallucination benchmark},
  author={Bang, Yejin and Ji, Ziwei and Schelten, Alan and Hartshorn, Anthony and Fowler, Tara and Zhang, Cheng and Cancedda, Nicola and Fung, Pascale},
  booktitle={Proceedings of the 63rd Annual Meeting of the Association for Computational Linguistics (Volume 1: Long Papers)},
  pages={24128--24156},
  year={2025}
}

@inproceedings{tang2018personalized,
  title={Personalized top-n sequential recommendation via convolutional sequence embedding},
  author={Tang, Jiaxi and Wang, Ke},
  booktitle={Proceedings of the eleventh ACM international conference on web search and data mining},
  pages={565--573},
  year={2018}
}

@article{hidasi2015session,
  title={Session-based recommendations with recurrent neural networks},
  author={Hidasi, Bal{\'a}zs and Karatzoglou, Alexandros and Baltrunas, Linas and Tikk, Domonkos},
  journal={arXiv preprint arXiv:1511.06939},
  year={2015}
}

@inproceedings{sun2019bert4rec,
  title={BERT4Rec: Sequential recommendation with bidirectional encoder representations from transformer},
  author={Sun, Fei and Liu, Jun and Wu, Jian and Pei, Changhua and Lin, Xiao and Ou, Wenwu and Jiang, Peng},
  booktitle={Proceedings of the 28th ACM international conference on information and knowledge management},
  pages={1441--1450},
  year={2019}
}

@article{wang2024enhanced,
  title={Enhanced generative recommendation via content and collaboration integration},
  author={Wang, Yidan and Ren, Zhaochun and Sun, Weiwei and Yang, Jiyuan and Liang, Zhixiang and Chen, Xin and Xie, Ruobing and Yan, Su and Zhang, Xu and Ren, Pengjie and others},
  journal={arXiv preprint arXiv:2403.18480},
  year={2024}
}

@inproceedings{chang2024conversational,
  title={Conversational product recommendation using LLM},
  author={Chang, Ting-Jui and Lin, Lydia Hsiao-Mei and Tsai, Richard Tzong-Han},
  booktitle={2024 IEEE 4th International Conference on Electronic Communications, Internet of Things and Big Data (ICEIB)},
  pages={340--343},
  year={2024},
  organization={IEEE}
}

@article{rajput2023recommender,
  title={Recommender systems with generative retrieval},
  author={Rajput, Shashank and Mehta, Nikhil and Singh, Anima and Hulikal Keshavan, Raghunandan and Vu, Trung and Heldt, Lukasz and Hong, Lichan and Tay, Yi and Tran, Vinh and Samost, Jonah and others},
  journal={Advances in Neural Information Processing Systems},
  volume={36},
  pages={10299--10315},
  year={2023}
}

@article{nguyen2024manipulating,
  title={Manipulating recommender systems: A survey of poisoning attacks and countermeasures},
  author={Nguyen, Thanh Toan and Quoc Viet Hung, Nguyen and Nguyen, Thanh Tam and Huynh, Thanh Trung and Nguyen, Thanh Thi and Weidlich, Matthias and Yin, Hongzhi},
  journal={ACM Computing Surveys},
  volume={57},
  number={1},
  pages={1--39},
  year={2024},
  publisher={ACM New York, NY}
}

@inproceedings{zhang2021causal,
  title={Causal intervention for leveraging popularity bias in recommendation},
  author={Zhang, Yang and Feng, Fuli and He, Xiangnan and Wei, Tianxin and Song, Chonggang and Ling, Guohui and Zhang, Yongdong},
  booktitle={Proceedings of the 44th international ACM SIGIR conference on research and development in information retrieval},
  pages={11--20},
  year={2021}
}

@article{dong2025memory,
  title={Memory Injection Attacks on LLM Agents via Query-Only Interaction},
  author={Dong, Shen and Xu, Shaochen and He, Pengfei and Li, Yige and Tang, Jiliang and Liu, Tianming and Liu, Hui and Xiang, Zhen},
  journal={arXiv preprint arXiv:2503.03704},
  year={2025}
}

@article{huang2023large,
  title={Large language models cannot self-correct reasoning yet},
  author={Huang, Jie and Chen, Xinyun and Mishra, Swaroop and Zheng, Huaixiu Steven and Yu, Adams Wei and Song, Xinying and Zhou, Denny},
  journal={arXiv preprint arXiv:2310.01798},
  year={2023}
}

@inproceedings{du2024improving,
  title={Improving factuality and reasoning in language models through multiagent debate},
  author={Du, Yilun and Li, Shuang and Torralba, Antonio and Tenenbaum, Joshua B and Mordatch, Igor},
  booktitle={Forty-first international conference on machine learning},
  year={2024}
}

@article{rotar2026can,
  title={Can Fairness Be Prompted? Prompt-Based Debiasing Strategies in High-Stakes Recommendations},
  author={Rotar, Mihaela and Rampisela, Theresia Veronika and Maistro, Maria},
  journal={arXiv preprint arXiv:2603.12935},
  year={2026}
}

@inproceedings{hayati2020inspired,
  title={Inspired: Toward sociable recommendation dialog systems},
  author={Hayati, Shirley Anugrah and Kang, Dongyeop and Zhu, Qingxiaoyang and Shi, Weiyan and Yu, Zhou},
  booktitle={Proceedings of the 2020 conference on empirical methods in natural language processing (EMNLP)},
  pages={8142--8152},
  year={2020}
}

@article{lu2023user,
  title={User perception of recommendation explanation: Are your explanations what users need?},
  author={Lu, Hongyu and Ma, Weizhi and Wang, Yifan and Zhang, Min and Wang, Xiang and Liu, Yiqun and Chua, Tat-Seng and Ma, Shaoping},
  journal={ACM Transactions on Information Systems},
  volume={41},
  number={2},
  pages={1--31},
  year={2023},
  publisher={ACM New York, NY}
}

@article{harper2015movielens,
  title={The movielens datasets: History and context},
  author={Harper, F Maxwell and Konstan, Joseph A},
  journal={Acm transactions on interactive intelligent systems (tiis)},
  volume={5},
  number={4},
  pages={1--19},
  year={2015},
  publisher={Acm New York, NY, USA}
}

@article{li2018towards,
  title={Towards deep conversational recommendations},
  author={Li, Raymond and Ebrahimi Kahou, Samira and Schulz, Hannes and Michalski, Vincent and Charlin, Laurent and Pal, Chris},
  journal={Advances in neural information processing systems},
  volume={31},
  year={2018}
}

@inproceedings{zhou2020towards,
  title={Towards topic-guided conversational recommender system},
  author={Zhou, Kun and Zhou, Yuanhang and Zhao, Wayne Xin and Wang, Xiaoke and Wen, Ji-Rong},
  booktitle={Proceedings of the 28th international conference on computational linguistics},
  pages={4128--4139},
  year={2020}
}

@inproceedings{chaney2018algorithmic,
  title={How algorithmic confounding in recommendation systems increases homogeneity and decreases utility},
  author={Chaney, Allison JB and Stewart, Brandon M and Engelhardt, Barbara E},
  booktitle={Proceedings of the 12th ACM conference on recommender systems},
  pages={224--232},
  year={2018}
}

@article{zhang2025robust,
  title={Robust recommender system: a survey and future directions},
  author={Zhang, Kaike and Cao, Qi and Sun, Fei and Wu, Yunfan and Tao, Shuchang and Shen, Huawei and Cheng, Xueqi},
  journal={ACM Computing Surveys},
  volume={58},
  number={1},
  pages={1--38},
  year={2025},
  publisher={ACM New York, NY}
}

@article{zhang2020explainable,
  title={Explainable recommendation: A survey and new perspectives},
  author={Zhang, Yongfeng and Chen, Xu},
  journal={Foundations and Trends{\textregistered} in Information Retrieval},
  volume={14},
  number={1},
  pages={1--101},
  year={2020},
  publisher={Emerald Publishing Limited}
}

@incollection{jeckmans2012privacy,
  title={Privacy in recommender systems},
  author={Jeckmans, Arjan JP and Beye, Michael and Erkin, Zekeriya and Hartel, Pieter and Lagendijk, Reginald L and Tang, Qiang},
  booktitle={Social media retrieval},
  pages={263--281},
  year={2012},
  publisher={Springer}
}

@inproceedings{koren2009collaborative,
  title={Collaborative filtering with temporal dynamics},
  author={Koren, Yehuda},
  booktitle={Proceedings of the 15th ACM SIGKDD international conference on Knowledge discovery and data mining},
  pages={447--456},
  year={2009}
}

@article{gunes2014shilling,
  title={Shilling attacks against recommender systems: a comprehensive survey},
  author={Gunes, Ihsan and Kaleli, Cihan and Bilge, Alper and Polat, Huseyin},
  journal={Artificial Intelligence Review},
  volume={42},
  number={4},
  pages={767--799},
  year={2014},
  publisher={Springer}
}

@article{resnick1997recommender,
  title={Recommender systems},
  author={Resnick, Paul and Varian, Hal R},
  journal={Communications of the ACM},
  volume={40},
  number={3},
  pages={56--58},
  year={1997},
  publisher={ACM New York, NY, USA}
}

@article{tran2021recommender,
  title={Recommender systems in the healthcare domain: state-of-the-art and research issues},
  author={Tran, Thi Ngoc Trang and Felfernig, Alexander and Trattner, Christoph and Holzinger, Andreas},
  journal={Journal of Intelligent Information Systems},
  volume={57},
  number={1},
  pages={171--201},
  year={2021},
  publisher={Springer}
}

@article{urdaneta2021recommendation,
  title={Recommendation systems for education: Systematic review},
  author={Urdaneta-Ponte, Mar{\'\i}a Cora and Mendez-Zorrilla, Amaia and Oleagordia-Ruiz, Ibon},
  journal={Electronics},
  volume={10},
  number={14},
  pages={1611},
  year={2021},
  publisher={MDPI}
}

@article{ke2025survey,
  title={A survey of frontiers in llm reasoning: Inference scaling, learning to reason, and agentic systems},
  author={Ke, Zixuan and Jiao, Fangkai and Ming, Yifei and Nguyen, Xuan-Phi and Xu, Austin and Long, Do Xuan and Li, Minzhi and Qin, Chengwei and Wang, Peifeng and Savarese, Silvio and others},
  journal={arXiv preprint arXiv:2504.09037},
  year={2025}
}

@inproceedings{wang2025empowering,
  title={Empowering large language model for sequential recommendation via multimodal embeddings and semantic ids},
  author={Wang, Yuhao and Pan, Junwei and Li, Xinhang and Wang, Maolin and Wang, Yuan and Liu, Yue and Liu, Dapeng and Jiang, Jie and Zhao, Xiangyu},
  booktitle={Proceedings of the 34th ACM International Conference on Information and Knowledge Management},
  pages={3209--3219},
  year={2025}
}

@article{chen2024hllm,
  title={Hllm: Enhancing sequential recommendations via hierarchical large language models for item and user modeling},
  author={Chen, Junyi and Chi, Lu and Peng, Bingyue and Yuan, Zehuan},
  journal={arXiv preprint arXiv:2409.12740},
  year={2024}
}

@inproceedings{xia2026multi,
  title={Multi-agent collaborative filtering: Orchestrating users and items for agentic recommendations},
  author={Xia, Yu and Kim, Sungchul and Yu, Tong and Rossi, Ryan A and McAuley, Julian},
  booktitle={Proceedings of the ACM Web Conference 2026},
  pages={8649--8652},
  year={2026}
}

@article{da2026agentic,
  title={Agentic Recommender Systems: A Systematic Literature Review},
  author={da Silva Portugal, Ivens and Alencar, Paulo and Cowan, Donald},
  journal={IEEE Transactions on Software Engineering},
  year={2026},
  publisher={IEEE}
}

@article{chen2026memrec,
  title={MemRec: Collaborative Memory-Augmented Agentic Recommender System},
  author={Chen, Weixin and Zhao, Yuhan and Huang, Jingyuan and Ye, Zihe and Ju, Clark Mingxuan and Zhao, Tong and Shah, Neil and Chen, Li and Zhang, Yongfeng},
  journal={arXiv preprint arXiv:2601.08816},
  year={2026}
}

@inproceedings{shokri2021privacy,
  title={On the privacy risks of model explanations},
  author={Shokri, Reza and Strobel, Martin and Zick, Yair},
  booktitle={Proceedings of the 2021 AAAI/ACM Conference on AI, Ethics, and Society},
  pages={231--241},
  year={2021}
}

@inproceedings{kumbam2025exploiting,
  title={Exploiting explainability to design adversarial attacks and evaluate attack resilience in hate-speech detection models},
  author={Kumbam, Pranath Reddy and Syed, Sohaib Uddin and Thamminedi, Prashanth and Harish, Suhas and Perera, Ian and Dorr, Bonnie J},
  booktitle={Proceedings of the International AAAI Conference on Web and Social Media},
  volume={19},
  pages={1038--1050},
  year={2025}
}

@article{wang2023decodingtrust,
  title={DecodingTrust: A Comprehensive Assessment of Trustworthiness in $\{$GPT$\}$ Models},
  author={Wang, Boxin and Chen, Weixin and Pei, Hengzhi and Xie, Chulin and Kang, Mintong and Zhang, Chenhui and Xu, Chejian and Xiong, Zidi and Dutta, Ritik and Schaeffer, Rylan and others},
  year={2023},
  publisher={Neural Information Processing Systems Datasets; Benchmarks Track}
}

@article{hu2022lora,
  title={Lora: Low-rank adaptation of large language models.},
  author={Hu, Edward J and Shen, Yelong and Wallis, Phillip and Allen-Zhu, Zeyuan and Li, Yuanzhi and Wang, Shean and Wang, Liang and Chen, Weizhu and others},
  journal={Iclr},
  volume={1},
  number={2},
  pages={3},
  year={2022}
}

@article{wang2024comprehensive,
  title={A comprehensive survey of continual learning: Theory, method and application},
  author={Wang, Liyuan and Zhang, Xingxing and Su, Hang and Zhu, Jun},
  journal={IEEE transactions on pattern analysis and machine intelligence},
  volume={46},
  number={8},
  pages={5362--5383},
  year={2024},
  publisher={IEEE}
}

@article{lampinen2000multiobjective,
  title={Multiobjective nonlinear pareto-optimization},
  author={Lampinen, Jouni},
  journal={Pre-investigation Report, Lappeenranta University of Technology},
  volume={114},
  pages={125},
  year={2000}
}

@article{rahimian2019distributionally,
  title={Distributionally robust optimization: A review},
  author={Rahimian, Hamed and Mehrotra, Sanjay},
  journal={arXiv preprint arXiv:1908.05659},
  year={2019}
}

@article{arjovsky2019invariant,
  title={Invariant risk minimization},
  author={Arjovsky, Martin and Bottou, L{\'e}on and Gulrajani, Ishaan and Lopez-Paz, David},
  journal={arXiv preprint arXiv:1907.02893},
  year={2019}
}

@article{ben2006analysis,
  title={Analysis of representations for domain adaptation},
  author={Ben-David, Shai and Blitzer, John and Crammer, Koby and Pereira, Fernando},
  journal={Advances in neural information processing systems},
  volume={19},
  year={2006}
}

@article{puterman1990markov,
  title={Markov decision processes},
  author={Puterman, Martin L},
  journal={Handbooks in operations research and management science},
  volume={2},
  pages={331--434},
  year={1990},
  publisher={Elsevier}
}

@inproceedings{hausknecht2015deep,
  title={Deep Recurrent Q-Learning for Partially Observable MDPs.},
  author={Hausknecht, Matthew J and Stone, Peter},
  booktitle={AAAI fall symposia},
  volume={45},
  pages={141},
  year={2015}
}

@article{li2017deep,
  title={Deep reinforcement learning: An overview},
  author={Li, Yuxi},
  journal={arXiv preprint arXiv:1701.07274},
  year={2017}
}

@inproceedings{wang2024macrec,
  title={Macrec: A multi-agent collaboration framework for recommendation},
  author={Wang, Zhefan and Yu, Yuanqing and Zheng, Wendi and Ma, Weizhi and Zhang, Min},
  booktitle={Proceedings of the 47th International ACM SIGIR Conference on Research and Development in Information Retrieval},
  pages={2760--2764},
  year={2024}
}

@book{mesterton2019introduction,
  title={An introduction to game-theoretic modelling},
  author={Mesterton-Gibbons, Mike},
  volume={37},
  year={2019},
  publisher={American Mathematical Soc.}
}



\end{document}